\documentclass[12pt]{article}
\usepackage{cite}
\usepackage{amsmath,amsfonts, amsthm, amssymb,slashed,slashed,url,upgreek,textgreek,url}
\usepackage[mathscr]{euscript}

\usepackage{ifpdf}
\ifpdf
  \usepackage[pdftex]{graphicx}
  \usepackage{epstopdf}
\else
  \usepackage[dvips]{graphicx}
\fi
\parskip=\baselineskip
\begin{document}
\def\x{{\mathbf x}}
\def\y{{\mathbf y}}
\def\z{{\mathbf z}}
\def\RP{{\mathbb{RP}}}
\def\Bbb{\mathbb}
\def\grav{{\mathrm{grav}}}
\def\etta{\upeta}
\def\Tr{{\rm Tr}}
\def\i{{\mathrm{i}}}
\def\Sp{{\mathrm{Sp}}}
\def\eD{{\Delta}}
\def\eps{\epsilon}
\def\P{{\mathcal P}}
\def\PF{{\mathit {Pf}}}
\def\pin{{\mathrm{pin}}}
\def\ch{{\mathrm{ch}}}
\def\16{{\bf 16}}
\def\1{{\bf 1}}
\def\K{{\mathcal K}}
\def\KB{{\mathrm{KB}}}
\def\frak{\sf}
\def\2{{\bf 2}}
\def\3{{\bf 3}}
\def\T{{\mathcal T}}
\def\sgn{{\mathrm{sign}}}
\def\4{{\bf 4}}
\def\hA{{\widehat A}}
\def\W{{\mathcal W}}
\def\P{{\mathcal P}}
\def\V{{\mathcal V}}
\def\S{{\mathcal S}}
\def\iota{{\mathfrak I}}
\def\I{{\mathcal I}}
\def\CS{{\mathrm{CS}}}
\def\OO{{\mathcal O}}
\def\i{{\mathrm i}}
\def\sCT{{\sf{CT}}}
\def\L{{\mathcal L}}
\def\UU{{\mathcal U}}
\def\D{{\mathcal D}}
\def\DD{{\sf D}}
\def\CP{{\sf{CP}}}
\def\sR{{\sf R}}
\def\Y{{\mathcal Y}}
\def\sP{{\sf P}}
\def\sT{{\sf T}}
\def\C{{\Bbb C}}
\def\sC{{\sf C}}
\def\bg{{\bar\gamma}}
\def\Pin{{\mathrm{Pin}}}
\def\sRC{{\sf{CR}}}
\def\sRT{{\sf{RT}}}
\def\RRR{{\mathcal R}}
\def\O{{\mathrm O}}
\def\U{{\mathrm U}}
\def\tr{{\mathrm {tr}}}
\def\RT{{\sf{RT}}}
\def\H{{\mathcal H}}
\def\t{\widetilde}
\def\h{\widehat}
\def\sCRT{{\sf{CRT}}}
\def\sCPT{{\sf{CPT}}}
\def\sC{{\sf{C}}}
\def\sCR{{\sf{CR}}}
\def\Spin{{\mathrm{Spin}}}
\def\bp{\begin{pmatrix}}
\def\ep{\end{pmatrix}}
\def\reg{{\mathrm{reg}}}
\def\diag{{\mathrm{diag}}}
\def\ind{{\mathrm{i\Tnd}}}
\def\diag{{\mathrm{diag}}}
\def\SU{{\mathrm{SU}}}
\def\veps{\varepsilon}
\def\bar{\overline}
\def\tilde{\widetilde}
\def\be{\begin{equation}}
\def\ee{\end{equation}}
\def\R{{\Bbb{R}}}
\def\sign{{\mathrm{sign}}}
\def\grav{{\mathrm{grav}}}
\def\coker{{\mathrm{coker}}}
\def\dim{{\mathrm{dim}}}
\def\Z{{\Bbb{Z}}}
\def\ZZ{{\mathcal Z}}
\def\N{{\mathcal N}}
\def\TT{\mathfrak T}
\def\U{{\mathrm U}}
\def\Sp{{\mathrm{Sp}}}
\def\Pf{{\mathrm{Pf}}}
\def\Pfaff{\Pf}
\def\g{{\gamma}}
\def\SO{{\mathrm{SO}}}
\def\SL{{\mathrm{SL}}}
\def\hat{\widehat}
\def\mfB{{\mathfrak B}}
\def\ICS{{\mathrm{CS}}}
\font\teneurm=eurm10 \font\seveneurm=eurm7 \font\fiveeurm=eurm5
\newfam\eurmfam
\textfont\eurmfam=\teneurm \scriptfont\eurmfam=\seveneurm
\scriptscriptfont\eurmfam=\fiveeurm
\def\eurm#1{{\fam\eurmfam\relax#1}}
\font\teneusm=eusm10 \font\seveneusm=eusm7 \font\fiveeusm=eusm5
\newfam\eusmfam
\textfont\eusmfam=\teneusm \scriptfont\eusmfam=\seveneusm
\scriptscriptfont\eusmfam=\fiveeusm
\def\eusm#1{{\fam\eusmfam\relax#1}}
\font\tencmmib=cmmib10 \skewchar\tencmmib='177
\font\sevencmmib=cmmib7 \skewchar\sevencmmib='177
\font\fivecmmib=cmmib5 \skewchar\fivecmmib='177
\newfam\cmmibfam
\textfont\cmmibfam=\tencmmib \scriptfont\cmmibfam=\sevencmmib
\scriptscriptfont\cmmibfam=\fivecmmib
\def\cmmib#1{{\fam\cmmibfam\relax#1}}
\numberwithin{equation}{section}
\def\neg{\negthinspace}
\def\mfB{{\mathfrak B}}
\def\sD{{\sf D}}
\def\mfT{{\mathfrak T}}
\def\mT{{\mathcal T}}
\def\veps{\varepsilon}
\def\d{{\mathrm d}}
\def\bp{\begin{pmatrix}}
\def\ep{\end{pmatrix}}

\begin{titlepage}
\begin{flushright}
hep-th/yymm.nnnn
\end{flushright}
\vskip 1.5in
\begin{center}
{\bf\Large{Fermion Path Integrals And Topological Phases}}
\vskip0.5cm 
{ Edward Witten} 
\vskip.5cm
 {\small{ School of Natural Sciences, Institute for Advanced Study, Princeton NJ 08540  USA}}\end{center}
\vskip.5cm
\baselineskip 16pt
\begin{abstract}  
Symmetry-protected topological (SPT) phases of matter have been interpreted in terms of anomalies,
and it has been expected that a similar picture should hold for SPT phases with fermions.  Here, we describe in 
detail what this picture means for phases of quantum matter that
can be understood via band theory and free fermions.  The main examples we consider are time-reversal invariant topological
insulators and superconductors in 2 or 3 space dimensions.  Along the way, we clarify the precise meaning of the statement that 
in the bulk of a 3d topological insulator, the electromagnetic $\theta$-angle is equal to $\pi$.
\end{abstract}
\date{August, 2015}
\end{titlepage}
\def\Hom{\mathrm{Hom}}

\tableofcontents
\section{Introduction}\label{intro}

\subsection{Goal Of This Paper}\label{goal}
Symmetry-protected topological (SPT) bosonic phases of matter can be characterized by anomalies \cite{WenEtAl}.  
 The meaning of this
statement is as follows.  

A $d$-dimensional SPT phase is gapped in bulk and has no topological order in the usual sense: on a compact
$d$-dimensional spatial manifold $M$ without boundary, it has a unique ground state.   But such a theory is not quite trivial.  On a compact
spacetime manifold $X$ of dimension $D=d+1$,  in the large volume limit, in general in the presence of a suitable
background gauge field, and after removing local, nonuniversal terms, the partition function of such a theory is a complex number of modulus 1,
$Z_X=e^{\i\Phi}$.   Here $e^{\i\Phi}$ is the partition function of a topological quantum field theory $\mfT$ that is ``invertible,'' its inverse being
the theory with complex conjugate partition function $Z_X^{-1}=e^{-\i\Phi}$.  

In this situation, typically $e^{\i\Phi}$ cannot be defined as a topological invariant in a satisfactory way -- consistent with all expected
symmetries and with physical principles such as unitarity -- if the spatial manifold $M$ (and therefore the spacetime $X$)
has a boundary.  To define the theory 
$\mfT$ in this situation, the boundary of $M$ (and $X$) must be endowed with 
additional degrees of freedom of some kind.  There are several
possibilities, but the case of interest in the present paper is that there are gapless degrees of freedom along the boundary of $M$ and $X$,
preserving the symmetries of theory $\mfT$.  

We  write $\mfB$ for the theory describing the gapless degrees of freedom on the boundary of $X$.  This theory has an ``anomaly'' of some
kind, and does not possess all of the desired physical properties, since if $\mfB$ were an entirely satisfactory and symmetry-preserving
theory by itself, then including it on the boundary $Y=\partial X$ of $X$ would not resolve whatever difficulty 
there is in defining theory $\mfT$ on  $X$ in the first place. The anomaly in defining theory $\mfB$ on $\partial X$ somehow 
cancels the difficulty in defining theory $\mfT$ on $X$.
Thus, this situation exemplifies -- usually in a more abstract way -- the idea of anomaly inflow from the bulk to the boundary \cite{CH}. 
For  a general mathematical treatment of this situation, see \cite{FreedTwo}. The
prototype of this situation in condensed matter physics is the integer quantum Hall effect on a two-manifold $M$ with boundary.  On
$M$ -- or rather on the three-manifold $X=\R\times M$, where $\R$ parametrizes the time -- there is a 
Chern-Simons coupling of the electromagnetic
vector potential $A$:
\begin{equation}\label{csm}I=k\ICS(A), ~~~\ICS(A)=\frac{e^2}{4\pi}\int_Y\,\d^3x \,\epsilon^{ijk}A_i\partial_jA_k,~~~k\in\Z. \end{equation}
This coupling is not gauge-invariant on a manifold with boundary, and the associated ``anomaly'' is canceled by the coupling of $A$ to chiral
edge modes that propagate on the boundary. (If these modes all have unit charge, then their multiplicity is $k$.)
  Theory $\mfB$ describes those chiral edge modes, and its anomaly is the two (spacetime)
dimensional
version of the Adler-Bell-Jackiw anomaly for chiral fermions \cite{Adler,BJ}.

Some of the most interesting SPT phases studied in recent years -- and observed experimentally in several important cases -- are
the topological insulators and superconductors
 that can be constructed out of free fermions via band theory.  For  introductions and reviews, see \cite{KH,ZQ,MZ}.  The crucial
symmetries in these examples are time-reversal symmetry $\sT$ and the $\U(1)$ symmetry associated to conservation of electric charge.
The purpose of the present paper is  to understand these free fermion topological phases of matter from the standpoint
of anomalies and anomaly inflow.  Several results in this direction were obtained previously in  \cite{RML}, 
where it was conjectured -- correctly,
as will become clear -- that a fuller picture would result from consideration of global anomalies. 
In analyzing this problem, we make use of index theory, applied to fermion path integrals.  
This is not surprising, since the free fermion phases
can be classified via K-theory \cite{Kitaev} (or by classes of matrices, which are related \cite{RSFL}), and index theory was invented \cite{ASfirst,AS} as the analysis associated to K-theory.

We should say at the outset that a few of our considerations are somewhat fanciful from the point of view of condensed matter physics.  We work
in a relativistic framework in which one can consider a theory on an arbitrary $D$-manifold $X$, with spatial boundary $Y$.  As in most studies
of anomalies, it is technically convenient to take $X$ to have Euclidean signature, and we do so.  The relativistic framework  is available because the boundary fermions
of the standard topological phases are governed by Dirac-like equations with an emergent relativistic symmetry, but relying on it 
is unnatural from the point
of view of condensed matter physics.  Perhaps it is possible to interpret our results in a Hamiltonian framework more natural for condensed
matter physics, but we will not try to do this. Experience seems to show that gapped condensed matter systems are frequently governed
by fully relativistic theories, but this is not fully understood.  

Also, because of their emergent relativistic symmetry, the boundary fermions we study
inevitably possess an emergent symmetry that in  $D=4$ is usually called $\sCPT$ (where $\sC$ is charge conjugation and
$\sP$ is parity or spatial inversion). However, this formulation is not
valid for odd $D$ (even spatial dimension $d$), since in that case $\sP$ is contained in the connected part of the spatial rotation group
and should be replaced by an operation that reverses the orientation of space.  
To use a language that is equivalent to the usual $\sCPT$ statement for even $D$ but is uniformly valid for all $D$,
we will refer instead to $\sCRT$ symmetry, where $\sR$ is  a reflection of one spatial coordinate. $\sCRT$ symmetry  is always 
valid in a relativistic theory of
any dimension, so, though not
very natural in condensed matter physics, it is an emergent symmetry 
of the usual gapless fermions on $\partial X$ for all of the usual free fermion phases.  This will be used at some points, notably in section
\ref{welfin} to describe the topological quantum field theory  associated a topological insulator in 2 space dimensions.

Actually we have been a little imprecise in this introduction in speaking of topological quantum field theory (TQFT). 
Theories studied in this paper contain fermions, so they require a 
 spin structure on spacetime (or a pin structure in the unorientable case). Hence the structure of interest is analogous
to a TQFT except that spacetime is required to have a chosen spin (or pin) structure. 
We will use the name sTQFT (spin  topological quantum 
field theory)
to refer to a theory that is like a TQFT except that it includes fermions.  For early study of such theories in the context
of $D=3$ Chern-Simons theory, see \cite{DW}.  The ``s'' in sTQFT can refer to either a spin structure or a pin
structure, depending on context.  The reader may want to consult the appendices for explanations of spin and pin structures,
and related matters.

\subsection{Some Generalities About Fermions}\label{generalities}

Now we will explain a few generalities about fermions and anomalies that will be useful background for this paper.  First of all,
anomalies always come from massless fermions only; massive fermions never contribute to anomalies.  A technical explanation is that
whatever symmetries a massive fermion $\psi$ may possess are also possessed by a Pauli-Villars regulator field (a field of opposite
statistics, obeying the same Dirac equation as $\psi$ but with a much larger mass).  So massive fermions can always be regularized in 
a way that preserves their symmetries, and they never contribute anomalies.  

Hence, as one would expect intuitively, in analyzing  fermion anomalies, one can ignore
gapped degrees of freedom.  In fact, we can ignore gappable degrees of freedom -- fermions that {\it could} have a symmetry-preserving
bare mass, even if they do not -- since again a symmetry-preserving regulator is possible for such fermions.  
A corollary is that in any relativistic theory of fermions $\psi$ in any dimension,
 there is no problem in defining the absolute value $|Z_\psi|$ of the fermion path
integral $Z_\psi$, consistent with all symmetries and physical principles.  This is because a conjugate set of fermions $\t\psi$, transforming
under any symmetries as the complex conjugate  of $\psi$, and with an action that is the complex conjugate of the action of $\psi$,
would have partition function $\bar Z_\psi$.  The combined system of fermions
$\psi\oplus \t\psi$ is always gappable.  This combined system would have partition function $Z_\psi \bar Z_\psi=|Z_\psi|^2$.  Thus there is no anomaly in $|Z_\psi|^2$ or
equivalently in $|Z_\psi|$.

Implicit in the last paragraph is that in Euclidean signature, and in contrast to 
Lorentz signature, fermions do not necessarily transform in a real
or self-conjugate representation of the relevant symmetry group (so $\psi$ and $\t\psi$ may transform differently).  
For example, the  edge modes in the integer quantum Hall effect are described
by a fermion field $\psi$ in spacetime dimension 2 that has definite chirality, meaning in Euclidean signature that $\psi$ transforms with spin 1/2 (or spin $-1/2$) under the rotation
group $\SO(2)$.  The chiral asymmetry of the edge modes mean that there is no corresponding field of the opposite chirality, transforming
as the complex conjugate of $\psi$.  It is only in Lorentz signature that fermions -- like bose fields -- are real.  For instance, continuing
with the example of
the chiral edge states,  the spin $1/2$ representation of $SO(2)$, in which a rotation by an angle $\varphi$ acts as $e^{i\varphi/2}$,
 becomes after  continuation to
Lorentz signature a 1-dimensional representation of $SO(1,1)$,
in which a Lorentz transformation $\begin{pmatrix}\cosh w & \sinh w\cr \sinh w & \cosh w\end{pmatrix} $ acts by $e^{w/2}$, which is real.

In general, relativistic  fermions fit in three broad classes, according to how they transform under the appropriate
symmetry group $K$ in Euclidean signature.  In $K$, we include the rotation group $SO(D)$, possible discrete symmetries such as\footnote{In
Euclidean signature, all symmetries act on the fermions in a linear fashion.  The antilinear nature of $\sT$ arises in analytic
continuation back to Lorentz signature.  The $\sCRT$ theorem means that one Euclidean symmetry can be continued to two distinct
symmetries in Lorentz signature; for example, a Euclidean signature symmetry $\sCR$ 
continues to either the antilinear symmetry $\sT$
or the linear symmetry $\sCR$ in Lorentz signature, depending on whether the reflection $\sR$ acts on the spacetime coordinate that is being analytically continued.} $\sT$ or $\sR$, and gauge
and/or global symmetries.  Fermions may transform in (1) a pseudoreal representation, (2) a real representation, or (3) a complex representation.
Of course, mixtures of these cases are also possible; for example, a theory may have some fermions transforming in a representation of one
type and some in a representation of the other type. (Also, a reducible representation can be of more than one type; see section \ref{orch}.)  We will encounter all three types in this paper.

  An irreducible representation $R$ of the group $K$ is said
to be pseudoreal if it admits an invariant, antisymmetric bilinear form $\omega$.  In this case, fermions transforming in the representation
$R$, which we denote as $\psi^\alpha$, $\alpha=1,\dots,\dim\,R$, can have a $K$-invariant bare mass $\omega_{\alpha\beta}\psi^\alpha\psi^\beta$.
(This is consistent with fermi statistics, since $\omega$ is antisymmetric.)
Hence, fermions transforming in a pseudoreal representation will never contribute to anomalies, and strictly speaking case (1) is not
relevant to our considerations.  However, there is a variant of this case
that we might call case (1$'$).  This is the case of a $\sT$-conserving theory with 
fermions that transform in a representation $R$ that is pseudoreal if one omits $\sT$ 
in the definition of $K$, but not if one includes $\sT$.  Since the fermions are pseudoreal if $\sT$ is ignored, such a theory can always be
quantized in a consistent way, but there may be an anomaly in $\sT$ symmetry.   This anomaly comes from a problem involving
 the sign of the fermion path integral $Z_\psi$, as explained in section \ref{pseudo}.
 Basic examples are the boundary fermions of a 3d topological
insulator or superconductor, formulated on an orientable manifold only. We will begin with this example, because  the
topological invariants involved are relatively familiar.  
Also, our results in the case of the 3d topological insulator may be particularly interesting, as we will be able to get a more precise description
of the sense in which \cite{ZQH} a 3d topological insulator is characterized by an electromagnetic $\theta$-angle equal to $\pi$.

Concerning case (2), an  irreducible representation $R$ of $K$ is said to be real if it admits an invariant, symmetric bilinear form $h$.  For a fermion $\psi$ transforming
in such a representation, a 
bare mass is forbidden by fermi statistics, since $h_{\alpha\beta}\psi^\alpha\psi^\beta =0$ because of fermi statistics.
However, if we double the fermion spectrum, adding a second multiplet $\t\psi$ also transforming in the representation of $R$, then
a $K$-invariant 
bare mass $h_{\alpha\beta}\psi^\alpha\t\psi^\beta$ becomes possible.  Doubling the spectrum replaces the fermion path integral $Z_\psi$
with its square $Z_\psi^2$, so $Z_\psi^2$ is anomaly-free and for real fermions, an anomaly can only affect the sign of $Z_\psi$.  
The basic examples that we will consider are edge states of a 2d topological superconductor or insulator.

Finally, concerning case (3), $R$ is said to be complex if it admits no invariant bilinear form.  
In this case, the fermion path integral $Z_\psi$ is complex in general. Unlike the other cases, fermions
that transform in a complex representation can have perturbative anomalies -- the standard Adler-Bell-Jackiw anomaly, and its analogs in other dimension. 
In this context, a ``perturbative'' anomaly is one that can be seen when fermions are quantized in a weak gauge or gravitational
field, as opposed to ``global'' or ``nonperturbative'' anomalies that are invisible in weak background fields and involve global considerations.  
For example, the edge modes in the integer quantum Hall effect
transform in a complex representation, and they do have a perturbative anomaly (which compensates for the lack of gauge invariance
of the Chern-Simons coupling (\ref{csm}) on a manifold with boundary).  
 However, in the present paper, we consider only cases in which 
perturbative anomalies are absent, leaving  only the more subtle global anomalies.
Our basic example of complex fermions  will be edge modes of 
 a 3d topological superconductor or insulator, now on a possibly unorientable manifold.  In particular, in the case of the topological
superconductor, it turns out that the anomaly involves a $16^{th}$ root of unity.  (A specific computation for this system
exhibiting an anomaly involving
an $8^{th}$ root of unity was performed in \cite{HCR}; the sense in which the anomaly is of order 16 rather than 8 is rather subtle,
as we explain in section \ref{zeb}.)   The fact that the anomaly involves a $16^{th}$ root of unity means that although at the free fermion level,
topological superconductors are classified by a $\Z$-valued invariant, anomalies see only the reduction of this invariant mod 16.   In fact,
it is known \cite{FCV,WS,MFCV,KitTwo} that with interactions included, the classification of 3d topological supeconductors is reduced from $\Z$ to $\Z_{16}$,
so anomalies precisely detect the 16 different classes.
For a 3d topological insulator, the anomaly even on an unorientable manifold involves only a sign and  the anomaly
detects precisely the expected $\Z_2$-valued invariant.

Anomalies for the case that the representation $R$ in Euclidean signature is real  were originally studied in \cite{Witold,Witoldtwo}. 
Such anomalies can be understood topologically in a fairly direct way, in terms of the mod 2 index of the Dirac operator. The
case of pseudoreal fermions can be understood somewhat similarly, using the ordinary Dirac index
rather than the mod 2 index,  and was originally studied in \cite{R , Semenoff , ADM}.  Finally, global anomalies for complex fermions
were analyzed in \cite{Global} and were expressed in terms of the Atiyah-Patodi-Singer (APS)
$\eta$-invariant \cite{APS}.  This relation was later refined in a result that we will call the Dai-Freed theorem \cite{DF}.
The Dai-Freed theorem is useful in a variety of problems involving complex fermions, such as the worldsheet path integral of the 
heterotic string \cite{Dinstanton}.  We will make use of it in section \ref{comb}.

Part of the motivation for the present work was the suggestion  \cite{KTTW}
 that the partition function of a 3d topological superconductor on a 4-manifold would be the exponential of 
an $\eta$-invariant.  We will see that this proposal
follows from a standard characterization of the phase transition between a trivial and topological superconductor.  We will show by
direct computation of partition functions that,  as suggested by previous work,\footnote{As briefly explained in the discussion of
eqn. (5.4) in \cite{Freed}, in general $\U(1)$-valued cobordism invariants are partition functions of unitary TQFT's or sTQFT's.  It was shown in \cite{KTTW} that in the examples we consider
in this paper, the possible cobordism invariants are in 1-1 correspondence with free fermion phases of matter.  (In higher dimensions, there are cobordism invariants that do not have an obvious
connection to free fermion phases, so they may be associated to interacting phases of matter.)  
 In sections \ref{cobo} and \ref{rpo}, we give
some very partial indications of how cobordism invariance is related to physical principles such as unitarity and the behavior under cutting and gluing. }   sTQFT's associated to free fermion states of matter have cobordism-invariant partition functions.
the proposal in \cite{KTTW} that sTQFT's associated to free fermion states of matter have cobordism-invariant partition functions  follows from direct calculation of partition functions for the bulk gapped fermions. 
The relation of these bulk calculations to the anomalies of boundary fermions follows from 
the Dai-Freed theorem in general, and from more elementary considerations when the gapless boundary
fermions are real or pseudoreal.

The  literature on anomalies is too vast to be properly summarized here.  However,  several additional
papers in which various problems are treated in a similar spirit to the approach here are \cite{Bunke}, in which considerations of \cite{Dinstanton} concerning
the heterotic string are formulated rigorously; \cite{DMW} with (in section 2) a precise definition of the path integral of M-theory on an 11-manifold\footnote{Because of the mod 8 periodicity of real K-theory, this 11-dimensional problem actually has much in common with the
case of 3-dimensional boundary fermions treated in section \ref{pseudo}.  But the details are different and the anomaly cancels with
no need for inflow from 12 dimensions.}; \cite{FM}, in which a more general version of that problem is treated very precisely; 
\cite{MW},  in which (in section 5.3) an analog of the topological insulator appears in a problem involving D-branes in string theory; 
and \cite{Monnier}, on the anomaly field theory associated to chiral fermions.  As this paper neared completion, the author became
aware of related work involving the $\eta$-invariant in the context of SPT phases \cite{Metlitski}.

\subsection{Organization and General Remarks}\label{zorg}

The contents of this paper can be summarized as follows.
In section \ref{pseudo}, we discuss pseudoreal fermions or more precisely fermions that are pseudoreal if one ignores $\sT$ symmetry.
In section \ref{realf}, we discuss real fermions, and in section \ref{complex}, we discuss complex fermions.  In section
\ref{maj}, we briefly consider another example with real fermions: the Majorana chain \cite{FK} in $1+1$ dimensions.  The overall idea is to start
with simple examples and gradually introduce more complicated ones.  We give a sort of overview of the 
different cases in section \ref{overview}.

We have tried to make the body of this paper relatively self-contained, but some
 details are contained in appendices.  For basic definitions about spinors in Riemannian geometry,  see Appendix \ref{dirram}.  For
details in low dimension, see Appendix \ref{examples}. The $\eta$-invariant in 4 (and 2) dimensions is analyzed in Appendix \ref{feta}.

In this paper, with the sole exception of section \ref{maj},
we only consider time-reversal symmetries with $\sT^2=(-1)^F$ (here $(-1)^F$ is the operator that counts fermions
mod 2).  This also means that the square of a spatial reflection $\sR$ (or $\sCR$ in the presence of a $\U(1)$ symmetry)
is $+1$, and that when we work on an unorientable manifold,
we use a $\Pin^+$ structure, not a $\Pin^-$ structure.  For an explanation of these concepts, see  Appendix \ref{dirram}.  In section
\ref{maj}, all statements are reversed: $\sT^2=+1$, $\sR^2=(-1)^F$, and the spacetime is endowed with a $\Pin^-$ structure.

We frequently introduce a Pauli-Villars regulator field as an aid in defining precisely a continuum theory of fermions.  The sign of the regulator
mass is then an important variable.  In condensed
matter physics, there would be no regulator but a physical cutoff.  For example, in band theory, gapless relativistic fermions arise at a particular
point (or points) in a band, and the topology of the rest of the band plays a role analogous to the sign of the regulator mass in the Pauli-Villars
approach.

We usually phrase our arguments in terms of partition functions.  There is much more to a physical theory than its partition function,
but in the case of free fermions, if there is a satisfactory
definition of the partition function, there is always a satisfactory definition of correlation functions.  So focusing on the partition function is
a convenient shorthand for determining if a theory is well-defined.  Occasionally the partition function vanishes and one must
consider the path integral measure instead.

\section{Pseudoreal Fermions}\label{pseudo}
\subsection{Topological Insulator in $d=3$}\label{topthree}

\subsubsection{Basics}

Our first example will be a topological insulator in $d=3$, studied in this section on an  oriented $3+1$-dimensional spacetime $X$.
$X$ has a spatial boundary, which is a $2+1$-dimensional orientable manifold $Y$.
 The boundary mode of the topological insulator is a $2+1$-dimensional massless two-component Dirac fermion $\psi$,
with action
\be\label{forx}I=\int_Y\d^3x \bar\psi \i\slashed{D}\psi.\ee
Here $\slashed{D}=\sum_{\mu=0}^2\gamma^\mu D_\mu$ is the usual Dirac operator.  $\psi$ couples to the vector potential $A$ of 
electromagnetism, just like the electron; indeed, in band theory, $\psi$ originates as a mode of the electron.  

It is possible to endow $\psi$ with a Lorentz-invariant mass term.  The Dirac equation then becomes\footnote{In local Lorentz coordinates,
our  Dirac
matrices are real matrices obeying $\{\g_a,\g_b\}=2\eta_{ab}$ where $\eta=\mathrm{diag}(-1,1,1)$.  These conventions are convenient
for $\sT$-invariant systems with $\sT^2=(-1)^F$, and 
 make it straightforward to compare Majorana and Dirac fermions. }
\be\label{orx}\left( \slashed{D}-m\right)\psi=0,\ee
with real $m$.  This describes a particle of mass $|m|$ and spin $\frac{1}{2}\sign(m)$.  Time-reversal and reflection symmetry reverse the sign
of the particle spin, so the perturbation to $m\not=0$ is $\sT$- and $\sR$-violating.  Concretely, a $\sT$-transformation\footnote{For
a Majorana fermion, we will have to consider two possible signs in this equation: $\sT\psi(-t,x_1,x_2)=\pm\g_0\psi(-t,x_1,x_2)$.  For a
Dirac fermion with $\U(1)$ symmetry, we can eliminate this sign by transforming, if necessary, $\psi\to\i\psi$.}
\be\label{rox}\sT\psi(t,x_1,x_2)=\g_0 \psi(-t,x_1,x_2) \ee
is easily seen to reverse the sign of $m$ in the Dirac equation.  If, however, we have two identical Dirac fermions $\psi_1,\,\psi_2$,
both transforming under $\sT$ as in eqn. (\ref{rox}), then one can add a $\sT$-invariant\footnote{The mass term in eqn. (\ref{worx}) is 
$\sR$-violating, assuming that $\sR$ -- a spatial reflection --
is supposed to act the same way on both $\psi_1$ and $\psi_2$ by $\sR\psi_i(t,x_1,x_2)=\pm
\g_1\psi_i(t,-x_1,x_2)$.  (The mass term  conserves a different $\sR$ symmetry.) The only general symmetry in relativistic QFT is $\sCRT$, which means
that $\sT$ symmetry is equivalent to $\sCR$ symmetry.  Given a $\sT$ symmetry, there is a canonical  $\sCR$ symmetry,
but there is no canonical $\sR$ symmetry.} mass term to the Dirac equation:
\be\label{worx}\left(\slashed{D}-\bp 0 & \i m \cr -\i m & 0\ep\right)\bp \psi_1\cr \psi_2\ep=0 . \ee
Diagonalizing the mass term, one finds that $\sT$ exchanges two modes of equal mass (and charge) and opposite spin.   Since a single Dirac fermion
cannot acquire  a $\sT$-conserving mass, it follows that in  dimension $3+1$, if a $\sT$-invariant material has a single massless
Dirac fermion on its boundary, then this state of affairs is protected by $\sT$-invariance.  We have written this paragraph in Lorentz signature,
because this makes the action of $\sT$ most transparent, but in our analysis below of the path integral for $\psi$ and its anomaly,
we take $Y$ to have Euclidean signature.

The path integral $Z_\psi$ of the $\psi$ field is formally the determinant of the operator $\D=\i\slashed{D}$:
\be\label{deto}Z_\psi=\det\,\D.\ee
The operator $\D$ is hermitian, so its eigenvalues are real:
\be\label{meto}\D\psi_i=\lambda_i\psi_i,~~\lambda_i\in\R.\ee
Formally, the determinant is the product of eigenvalues:
\be\label{peto}\det\,\D=\prod_i\lambda_i.\ee

Formally, this is real, since all the factors are real. 
This  reality has a simple physical meaning: it reflects the $\sT$-invariance of the classical field theory that we are trying to quantize.
 In general, in any unitary QFT, on an orientable spacetime $X$
 of Euclidean signature, the partition function is complex-conjugated if one reverses the orientation of $X$.
 In a $\sT$-invariant theory, this reversal of orientation is a symmetry, so the partition function is always real.  All this is true only for orientable $X$;
 for unorientable $X$, $\sT$-invariance does not require the partition function to be real.  See section \ref{complex} for examples.

For {\it two} massless Dirac fermions, the path integral would be
\be\label{reto}Z_\psi^2=\prod_i\lambda_i^2. \ee
Every factor is not just real but positive, so formally one certainly expects $Z_\psi^2$ to be positive,
and indeed the path integral for two massless Dirac fermions can be naturally defined
to be positive.  Being positive, it is equal to $|Z_\psi|^2$, and thus is completely anomaly-free,
since as we explained in section \ref{generalities}, the absolute value $|Z_\psi|$ of a relativistic fermion
path integral can always be defined in a completely anomaly-free way, preserving all symmetries. 
In particular, the path integral for two massless Dirac fermions on $Y$ can be defined in a $\sT$-invariant
fashion.  This should come as no surprise, since we have seen above that a $\sT$-invariant mass and
therefore a $\sT$-invariant Pauli-Villars regulator are possible.  (The $\sT$-invariant
Pauli-Villars regularization is based on the $\sT$-invariant massive Dirac equation (\ref{worx}), so it
uses a regulator of positive mass for one linear combination of the  two Dirac fermions and  of negative
mass for the other.)

With only one Dirac fermion, however, we have a problem, because the formal expression
\be\label{wink}Z_\psi=\prod_i\lambda_i \ee
is naturally real but not naturally positive.  Its sign is roughly speaking the number of $\lambda_i$
that are negative, mod 2, but the number of negative $\lambda_i$ is infinite, and there is no natural way to decide
if this infinite number is even or odd. 

\subsubsection{Global Anomaly And Spectral Flow}\label{trada}

In fact, if we try to define $Z_\psi$ to be positive and gauge-invariant, we run into a contradiction.   We could pick an arbitrary metric and gauge
field, say $g=g_0$ and $A=A_0$, and define $Z_\psi$ to be, say, positive at $(A,g)=(A_0,g_0)$.  Then letting $A$ and $g$ vary, we could
follow the sign of $Z_\psi$ continuously, saying that this sign changes whenever an eigenvalue of $\D$ passes through 0.  However, this
procedure leads to a conflict with gauge-invariance.  Let $\phi$ be a gauge transformation or the combination of a gauge transformation
and a diffeomorphism.  Let $(A_0^\phi,g_0^\phi)$ be whatever $A_0$ and $g_0$ transform into under $\phi$.  It is always possible
to continuously interpolate from $(A_0,g_0)$ to $(A_0^\phi,g_0^\phi)$.  One introduces a real parameter $s$, $0\leq s\leq 1$, and then one
sets
\be\label{zink}A_s=(1-s)A_0+sA_0^\phi,~~~ g_s=(1-s)g_0+sg_0^\phi. \ee
These formulas have been chosen so that $(A_s,g_s)$ coincides with $(A_0,\phi_0)$ at $s=0$ and with $(A_0^\phi,g_0^\phi)$ at $s=1$.
Before accepting this interpolation from $(A_0,\phi_0)$ to $(A_0^\phi,g_0^\phi)$, we should make sure that $A_s$ and $g_s$ are well-defined
for all $s$.  A metric $g_s$ is supposed to be positive-definite.  There is no problem because $g_s$ for $0< s< 1$ is a linear combination
with positive coefficients of the positive metrics $g_0$ and $g_0^\phi$.  For $A_s$, there is actually no analogous condition to be checked.

\begin{figure}
 \begin{center}
   \includegraphics[width=3in]{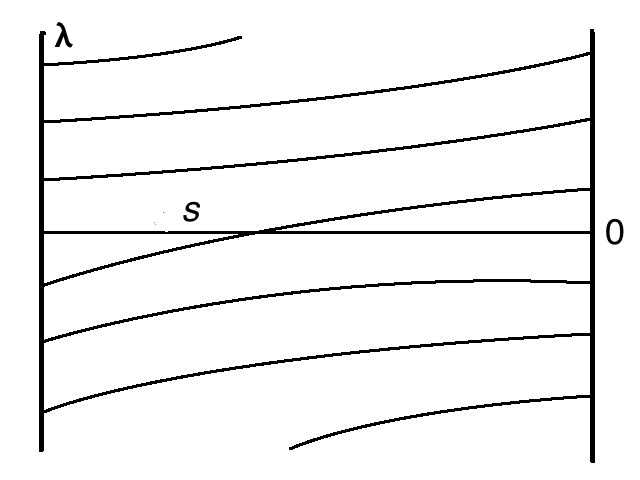}
 \end{center}
\caption{\small  Spectral flow for a  Dirac operator. The vertical axis parametrizes an eigenvalue $\lambda$ and the horizontal axis
parametrizes a parameter $s$ on which the eigenvalues depend.  In the case shown, the spectrum is the same at $s=1$ as at $s=0$,
but there is a net upward flow of one eigenvalue through
$\lambda=0$ between $s=0$ and $s=1$.  This leads to a sign change of the fermion path integral $Z_\psi$.}
 \label{flow}
 \end{figure}
Now we evolve $(A_s,g_s)$ continuously from $s=0$ to $s=1$, and we check how many times $Z_\psi$ changes sign.  Gauge invariance
implies that $Z_\psi$ should have the same sign at $s=1$ as at $s=0$, since $(A_s,g_s)$ at $s=1$ is gauge-equivalent to what it is
at $s=0$.  However, in general there is a problem. Gauge-invariance implies that the spectrum of the Dirac operator at $s=1$ 
is the same as it is at $s=0$, but between $s=0$ and $s=1$, there can be a net ``spectral flow'' of the Dirac eigenvalues, as shown in
fig. \ref{flow}.  Such a spectral flow is only possible because $\D$ has infinitely many positive and negative eigenvalues.

In the case of the boundary fermions of the 3d topological insulator, there definitely is such an inconsistency in trying to define the sign of 
$Z_\psi$, assuming that one expects it to be real.  There actually is a topological formula for the net number $\Delta$ of
eigenvalues flowing through $\lambda=0$.  It is
\be\label{worf}\Delta=\iota,\ee
where as described momentarily, $\iota$ is a certain Dirac index in $D=4$.
The corresponding formula for 
the sign change of $Z_\psi$ between $s=0$ and
$s=1$ is
\be\label{torf}Z_\psi\to Z_\psi (-1)^\iota.\ee
 
$\iota$ will be a Dirac index on a certain four-manifold $X$, known as the mapping torus.  This manifold 
 will be  a calculational tool to study anomalies
in the fermion theory on $Y$.   In our discussion of the topological insulator, a four-manifold will generically be called $X$, whether it is a tool to study anomalies
or is the worldvolume of a physical system.   This reason is that the same mathematical considerations will enter in either case.  
This is not a coincidence but represents the importance of the boundary
anomalies in the physics of the topological insulator.  

To define $X$, 
let $I$ be the unit interval $0\leq s\leq 1$. We glue together the ends of $I\times Y$ 
via the diffeormorphism $\phi$ to build a four-manifold $X$ without boundary.\footnote{One can slightly modify the $s$-dependence of the
metric so that the metric on $X$ is smooth at the endpoints $s=0$ and 1.}  A manifold constructed in this way is called a mapping torus. 
 Using also the gauge transformation part of $\phi$, we glue
together the gauge bundles over the ends of $I\times Y$ to make a $U(1)$ gauge bundle $\L\to X$.  Then $\iota$ is the index of the
4d Dirac operator on\footnote{\label{helpful} Instead of working on the compact mapping torus $X$, we could set $u=1/(1-s)-1/s=(2s-1)/s(1-s)$, so $0<s<1$
corresponds to $-\infty<u<\infty$, and work on the noncompact four-manifold $X'\cong\R\times Y$, with metric $\d u^2+ g_{s(u)}$,
and the corresponding $u$-dependent 
gauge field $A_{u(s)}$.  The index of the Dirac operator on $X'$, in the space of square-integrable wavefunctions, is the same
as the index on $X$.}
$X$, coupled to $\L$.    (See section \ref{why} for the definition of the Dirac index.)
 This relation between spectral flow in three dimensions and the Dirac index in four dimensions,
originally obtained in \cite{APS},
has applications in QCD and has been described in the physics literature \cite{DCG,Kiskis}.

What has just been described, in which one looks for a possible inconsistency in the sign or phase of $Z_\psi$ as a function of  a parameter $s$,
is the framework of the 1980's 
\cite{Witold,Witoldtwo,Global}   for studying global anomalies.   At one time it was assumed to be the whole story.  Nowadays it is clear that this is not
the case, and in the present paper we will need a more complete treatment.  The reason that we have begun with the traditional framework is not that we want to delve into
history but that this is the most elementary way to introduce the important topological invariants -- notably the index $\iota$ in the present example, and its analogs in examples
that we introduce later.

What is missing in the traditional framework  can be described as follows.  The absence of an anomaly in the sign of $Z_\psi$
means that on  a particular $Y$, $Z_\psi$ is well-defined as a function of $(A,g)$, up to an overall sign that depends on $Y$ but not on $(A,g)$.
(In the above reasoning, this sign was introduced as the sign of  $Z_\psi$ at some arbitrary starting point $(A_0,g_0)$.)   But we need a way to determine
this overall sign, and 
we should certainly not expect to get a satisfactory theory if we define the sign of $Z_\psi$ independently for each $Y$.
A definite physical theory produces definite answers for these signs.  Physically, there must be a sensible behavior under various cutting
and pasting operations in which various three-manifolds $Y_i$ are cut
in pieces and glued together in different ways.

The prototype of such a cutting relation arises in ordinary quantum mechanics, with Hamiltonian $H$,
if a time interval of length $t$ is decomposed in successive intervals of length $t_1$ and $t_2$, where $t=t_1+t_2$.  One
has $\exp(-\i Ht)=\exp(-\i Ht_2)\exp(-\i Ht_1)$, leading to a well-known result for transition amplitudes 
\be\label{cutting}\langle f|\exp(-\i Ht)|i\rangle
=\sum_k \langle f|\exp(-\i H t_2)|k\rangle \langle k|\exp(-\i Ht_1)|i\rangle,\ee
 with a sum over intermediate states. This has an analog in quantum field theory
in any dimension, where one cuts a manifold into pieces and claims that the path integral on the whole
manifold is the product of path integrals on individual pieces (with a sum over physical states where gluing occurs).  
The quantum field theory story is much richer than the quantum mechanical case because in higher dimensions
there are many more ways to cut a manifold into pieces. 
Compatibility with cutting and pasting is the essence of the locality of quantum field theory.

In cases relevant to topological states of matter -- a good
example being the 3d topological superconductor -- even if there is
no anomaly in the traditional sense of an inconsistency in 
defining the path integral on a specific $Y$, there can be an anomaly in the
more subtle sense that there is no satisfactory way to define overall
signs or phases of the path integral on different  $Y$'s. One needs to
take these more subtle anomalies into account as part of the
paradigm ``Anomalies in $d$ dimensions $\longleftrightarrow$  SPT phases in $d+1$ dimensions.''

Taking these anomalies into account means giving an absolute definition of the sign of the fermion path integral $Z_\psi$ for each $Y$
and each $(A,g)$.  We will do this next, following \cite{ADM}.

\subsubsection{$\sT$ Anomaly}\label{zelfan}

The fact that there is a problem in defining the sign of $Z_\psi$ for a $2+1$-dimensional Dirac fermion $\psi$ does not mean that this
theory is inconsistent.  It only means that the theory cannot be quantized in a $\sT$-invariant (or $\sR$-invariant) way.   It is possible
to define $Z_\psi$ consistently if we do not try to make it real.

After all,
$\psi$ could have a gauge-invariant bare mass, which violates $\sT$ symmetry but otherwise is perfectly physically acceptable.
The possibility of a $\sT$-violating mass means that, at the cost of losing $\sT$ symmetry, we can regularize this theory by adding a Pauli-Villars
regulator field $\chi$, which one can think of as a bosonic field that obeys a massive Dirac equation $(\i\slashed{D}+\i\mu)\chi=0$, for very
large $\mu$.  
In Euclidean signature, the regularized path integral is
\be\label{pv} Z_{\psi,\,\reg}=\prod_k\frac{\lambda_k}{\lambda_k+\i \mu}. \ee
Actually, what we have written, though good enough for our purposes, is only an approximation to the Pauli-Villars procedure.
In general one introduces a variety of massive bose and fermi fields, with different large masses $\mu_a, \,a=1,\dots, t$, chosen
so that the regularized path integral converges for fixed $\mu_a$.  Then one takes the limit $\mu_a\to\infty$, adding local counterterms $W(\mu_a,
A,g)$ 
 to the action so that the limit of $Z_{\psi,\,\reg}\exp(-W(\mu_a,A,g))$ exists.  The limit is the renormalized fermion 
 path integral.  We do not need to follow this procedure in detail, because we are only interested in the phase of
 $Z_\psi$, and eqn. (\ref{pv}) is good enough to motivate the correct formula for this phase.  (The counterterms that are needed
 in 3 spacetime dimensions are all real and do not affect the discussion of the phase.)
 
 Going back to eqn. (\ref{pv}), we see that for large $\mu>0$, each eigenvalue $\lambda_k$ contributes a phase $\i^{-1}$ or $\i$ to $Z_\psi$,
 depending on the sign of $\lambda_k$.   So formally
 \be\label{nv}Z_\psi=|Z_\psi|\exp\left(-\frac{\i\pi}{2}\sum_k \sign(\lambda_k)\right). \ee
 Thus
 \be\label{mv} Z_\psi=|Z_\psi|\exp\left(-\i\pi\etta/2\right), \ee
 where $\etta$ (the Atiyah-Patodi-Singer or APS $\etta$-invariant) is a regularized version of the difference between the number of positive
 and negative eigenvalues\footnote{\label{later} In the present derivation, the $\lambda_k$ are the eigenvalues of the Dirac
 operator $\D$ acting on the positively charged Dirac fermion $\psi$, without including charge conjugate eigenmodes of $\D$ acting on the negatively charged field $\bar\psi$.
We can also define a similar invariant $\eta$ in which one sums over modes of both kind.
 By charge conjugation symmetry, the relation between the two is just $\eta=2\etta$.  Later on, when we consider Majorana fermions (with no
 $\U(1)$ symmetry and so no distinction between modes of positive or negative charge), we will want to express all formulas in terms of $\eta$.  } of $\D$.
 
 As usual, the precise regularization does not matter.  The original APS definition was\footnote{Some analytic continuation is required in this
 definition.  One first defines the sum on the right hand side of eqn. (\ref{rv}) for large $\mathrm{Re}\,s$, where it converges.  Then one
 analytically continues to $s=0$.  Note that the alternative definition in eq. (\ref{rov}) does not require such analytic continuation.}
 \be\label{rv}\etta=\lim_{s\to 0}\sum_k\sign(\lambda_k)|\lambda_k|^{-s}. \ee
 An equivalent definition would be
  \be\label{rov}\etta=\lim_{\veps\to 0^+}\sum_k\sign(\lambda_k)\exp(-\veps\lambda_k^2). \ee
 For our present application, we do not need to know what is meant by $\sign(\lambda)$ if $\lambda=0$, because if one of the $\lambda_k$
 is 0, then $|Z_\psi|=0$ and it does not matter what value we assign to $\etta$.  However,  the formulas
 of index theory work  smoothly if in the definition of $\etta$, we set 
  \be\label{uv}\sign(\lambda)=\begin{cases} 1 &  \text{if}~\lambda\geq 0\cr -1&\text{if}~\lambda<0. \end{cases}.\ee
  It would work just as well to take $\sign(\lambda)=-1$ for $\lambda=0$; the important thing is to treat all zero-modes in the same way.
  
  The formula 
  \be\label{sv}Z_\psi=|Z_\psi|\exp(-\i\pi\etta/2), \ee
  together with any standard procedure to define $|Z_\psi|$, gives a satisfactory definition of $Z_\psi$ for all $Y$ and all $A,g$, with all desireable
  properties except $\sT$ and $\sR$ symmetry.  $\sT$ and $\sR$ symmetry do not hold, since they would require $Z_\psi$ to be real. $\sT$
  and $\sR$ symmetry have been violated by the choice of sign of the regulator mass.
  An equally good regularization with the opposite sign would have given the opposite phase to $Z_\psi$, so we really have two equally good
  definitions with
  \be\label{yv}Z_\psi=|Z_\psi|\exp(\mp \i\pi\etta/2). \ee
  Either of these formulas gives the partition function, in a background field $(A,g)$,  of a gapless free fermion QFT that is perfectly unitary
  and Poincar\'e invariant -- and even conformally invariant -- and otherwise physically sensible,  but is not $\sT$- or $\sR$-invariant. 
  
  To conclude this section, we will explain some aspects of the statement that the formula (\ref{yv}) for the path integral is physically
  sensible.  The fermion path integral $Z_\psi$ is supposed to change sign when a fermion eigenvalue passes through 0.  The formula
  (\ref{sv}) does have this property.  While the first factor $|Z_\psi|$ of course does not change in sign, the second factor $\exp(\mp \i\pi\etta/2)$
  does change sign, because $\etta$ jumps by $\pm 2$ when an eigenvalue passes through 0.
  
\begin{figure}
 \begin{center}
   \includegraphics[width=3in]{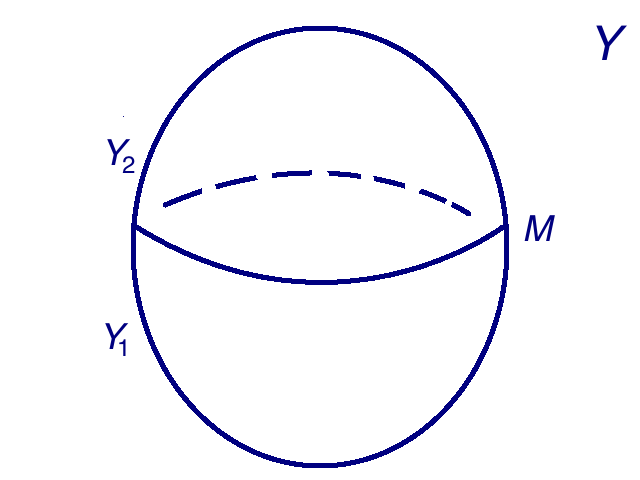}
 \end{center}
\caption{\small  A mirror-symmetric construction of a 3-manifold $Y$, by gluing together identical pieces $Y_1$ and $Y_2$ along their common boundary $M$,
with reversed orientation for $Y_2$.}
 \label{dodo}
 \end{figure}

  We may also want to verify that our definition of the partition function is consistent with unitarity.  In Euclidean signature,
  unitarity corresponds to reflection positivity.   Let $Y_1$ and $Y_2$ be  two identical three-manifolds, endowed with the same gauge fields and spin structures,
  and let $M$ be their common boundary.  
  Build a three-manifold $Y$ by gluing $Y_1$ and $Y_2$  along  $M$ (after reversing the orientation of $Y_2$ so that the orientations
   match).  This construction is illustrated 
  in fig. \ref{dodo}.  $Z_\psi$ should be real and nonnegative in this situation for the following reason.  The path integral
  on $Y_1$ constructs a state $|\Phi\rangle$ -- a ``ket'' -- in the Hilbert space of quantum states on $M$.  The path integral on $Y_2$
  constructs the corresponding ``bra'' $\langle\Phi|$ in the same Hilbert space.  The full path integral $Z_\psi$ on $Y$ is the 
  inner product $\langle\Phi|\Phi\rangle$.  In a unitary quantum field theory, this of course must be real and nonnegative.
  
  To show that $Z_\psi\geq 0$, we need to show that $\exp(\mp \i\pi \etta/2)$ is equal to 1 in this situation, as long as there are no zero-modes
  (that is, as long as $|Z_\psi|\not=0$).  This is true because the reflection that exchanges $Y_1$ and $Y_2$, and reverses the orientation of $Y$,
   anticommutes with the Dirac
  operator $\D=\i\slashed{D}=\i\sum_{\mu=1}^3\g^\mu D_\mu.$ For example, if $Y=\R^3$ with flat metric, a reflection 
  \be\label{refact}\sR\psi (x_1,x_2,x_3)=\g^1\psi(-x_1,x_2,x_3) \ee
  clearly anticommutes with $\slashed{D}$. This is actually a universal result for any orientation-reversing symmetry on a manifold
  of odd dimension.
   So  reflection symmetry in fig. \ref{dodo} implies that the eigenvalues of the Dirac operator are invariant
  under $\lambda\leftrightarrow -\lambda$, and this in turn implies  (in the absence of zero-modes) that $\etta=0$ and $\exp(\mp \i\pi\etta/2)=1$.
  
  The fact that a symmetry reversing the orientation of spacetime
   anticommutes with the Dirac operator is the reason that a mass term added to the Dirac equation violates
  such symmetries.  Obviously, if $\sR$ commuted with $\i\slashed{D}$, then a massive Dirac equation, which in Euclidean signature
  would be $\left(\i\slashed{D}+\i m\right)\psi=0$, would be $\sR$-invariant.

 \subsubsection{Canceling The Anomaly From The Bulk}\label{canbulk}
 
The anomaly in $\sT$-invariance that we have just discussed for the boundary fermions of a 3d topological insulator is related
to the fact \cite{ZQH} that in bulk, such a material has an electromagnetic $\theta$-angle of $\pi$.
 
 The instanton number of a $\U(1)$ gauge field in four spacetime dimensions is defined as
 \begin{equation}\label{zelm}P=\frac{e^2}{32\pi^2}\int_X \d^4x\,\epsilon^{\mu\nu\alpha\beta}F_{\mu\nu}F_{\alpha\beta},\ee
 where $F_{\mu\nu}=\partial_\mu A_\nu -\partial_\nu A_\mu$ is the electromagnetic field strength.  If $X$ is a compact manifold
 with boundary, then $P$ is always an integer.  A typical example with $P\not=0$ is the case that $X=S^2\times S^2$ with one unit
 of flux (that is, $\int_{S^2}F=2\pi/e$) on each factor.  This example has $P=1$.
 
 The electromagnetic $\theta$-angle is defined by saying that the $P$-dependent part of the effective action is
 \be \label{zpart}I_\theta=\theta P.\ee
 If $P$ can be assumed to be integer-valued, then  physics will be invariant under $\theta\to\theta+2\pi$,
 since in quantum mechanics, we only care about the value of the action mod $2\pi\Z$.  
 
 $P$ is odd under reflection or time-reversal, so $\theta$ is likewise odd.   In a $\sT$- or $\sR$-invariant theory, $\theta$ must
 equal 0 or $\pi$ ($\sT$ and $\sR$ map $\theta=\pi$ to $\theta=-\pi$, which is equivalent to $\theta=\pi$ mod $2\pi$).  It is shown
 in \cite{ZQH} that, assuming that $\theta=0$ in vacuum, $\theta=\pi$ inside a 3d topological insulator.  
 \
 
 However, a real topological insulator has a boundary and its worldvolume $X$ does 
 not fill all of spacetime. In this case, $P$ (defined as an integral over $X$ and not over all of spacetime) is generically
 not an integer. So it is not $\sT$-invariant to merely include in the path integral a factor
 $\exp(\pm \i\pi P)$.  On an orientable spacetime,
 $\sT$ invariance requires that the argument of the path integral should be real, and $\exp(\pm \i\pi P)$ certainly does not have
 this property when $P$ is not $\Z$-valued.  Acting with $\sT$ reverses the sign in the exponent $\exp(\pm\i\pi P)$.
 
 More concretely, if we simply include in the functional integral in the presence of a topological insulator a factor $\exp(\pm \i \pi P)$,
 then by a well-known argument,
  this produces on the surface of the topological insulator
  a Hall conductivity with $\nu=\pm 1/2$ (as usual, $\nu$ is the Hall conductivity in units of $e^2/2\pi\hbar$).  This is certainly
 not $\sT$-conserving.  
 
 In the context of a topological insulator, the assertion that $\theta=\pi$ 
  is $\sT$-conserving really means that it is $\sT$-conserving in bulk, and that we may be able to
 maintain $\sT$ invariance along the boundary if we find the right boundary state.  A trivial gapped boundary
 state is not suitable.
 
 However, the standard boundary state with massless Dirac fermions does combine with the bulk system of $\theta=\pi$
 to maintain $\sT$ symmetry.   We recall that the partition function of the boundary
 fermions is
 \be\label{dont} Z_\psi=|Z_\psi|\exp(\mp\i\pi\etta/2),\ee
 where the sign depends on the choice of regulator.   It turns out that if one subtracts from $P$ a gravitational correction that for the moment
 we will just denote\footnote{We write $\hA(R)$ as an abbreviation for the more usual $\int_X\hA(R)$.} as $\hA(R)$, then  the bulk contribution to the path integral with $\theta=\pi$  combines with the partition
 function of the boundary fermions to give a $\sT$-conserving result.
 In fact, according to    Atiyah, Patodi, and Singer (APS) \cite{APS}, 
 \be\label{wont}\exp(\mp\i\pi\etta/2)\exp\left(\pm \i\pi(P-\hA(R))\right)=(-1)^\iota,\ee
 where $\iota$ is an integer.    As we explain in section \ref{aps}, $\iota$ is the index of the Dirac operator computed
 with APS boundary conditions.
Hence the complete path integral measure after integrating out the boundary fermions is
\be\label{ont}|Z_\psi|\exp(\mp\i\pi\etta/2)\exp\left(\pm \i\pi(P-\hA(R))\right)=|Z_\psi|(-1)^\iota. \ee
(This formula was essentially found in \cite{MW}, section 5.3, in a related context involving D-branes
in string theory.)  This is real and so $\sT$-conserving.

In the rest of this section, we discuss the physical interpretation of the formula  (\ref{ont}), and then review the APS formula (\ref{wont}).

\subsubsection{Physical Meaning of $\theta=\pi$}\label{physical}

 \begin{figure}
 \begin{center}
   \includegraphics[width=2.5in]{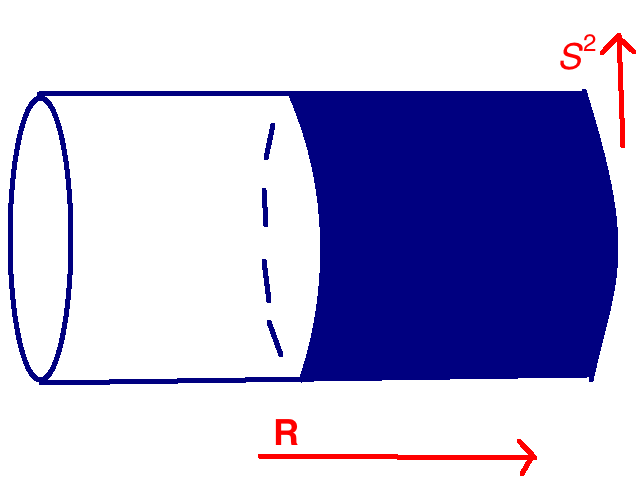}
 \end{center}
\caption{\small  A topological insulator  supported on the right half of $\R\times S^2$ (solid color).}
 \label{support}
 \end{figure}

To have $\theta=\pi$ in a topological insulator 
should mean, in some sense, that the path integral measure changes sign when a $\U(1)$ instanton
moves from outside to inside a topological insulator.  To make this concrete, we should set up a thought experiment that makes some 
sense in condensed 
matter physics in which it is possible to move an instanton in this way.

We take space to be not $\R^3$ but $\R\times S^2$, with two dimensions compactified to a two-sphere $S^2$.  (We could just as well
replace $S^2$ with $T^2$, a two-torus corresponding to periodic boundary conditions in two spatial directions.) We consider a topological
insulator supported on $\R_+\times S^2\subset \R\times S^2$ (fig. \ref{support}) where $\R_+\subset \R$ is a half-line $y\geq 0$.  Including the time
direction, which we parametrize by $\R'$ (another copy of $\R$), the full spacetime is $\h X= \R'\times \R\times S^2$ and the worldvolume 
of the topological insulator is $X=\R'\times \R_+\times S^2$.   

We consider a gauge field that has one unit of magnetic flux on $S^2$, and also one unit of electric flux on $\R'\times \R$:
\be\label{zelb} \int_{\R'\times \R}\d t \d x \, F_{01}=\frac{2\pi}{e}. \ee
Such a gauge field has the value 1 of 
\be\label{elb}\h P=\frac{e^2}{32\pi^2}\int_{\h X}\d^4x\,\epsilon^{\mu\nu\alpha\beta}F_{\mu\nu}F_{\alpha\beta}. \ee
Note that this is an integral over the whole spacetime $\h X=\R'\times \R\times S^2$. By contrast, we define $P$ (eqn. (\ref{zelm}))
as a similar integral over only the worldvolume $\R'\times \R_+\times S^2$ of the topological insulator.  So $\h P$ is a topological invariant
but $P$ is not.  If the unit of electric flux in eqn. (\ref{zelb}) is localized far to the left in fig. \ref{support}, then $\h P=1$ but $P=0$.
If it is localized far to the right, then $\h P=P=1$.   

To consider in this context the formula
\be\label{zont}|Z_\psi|(-1)^\iota \ee
for the path integral measure, we should compactify the time direction and go to Euclidean signature, since the formula was derived in that case.
For example, to compute a thermal partition function $\Tr\, \exp(-\beta H)$, with $H$ the Hamiltonian of the system, we should take the 
time direction to be a circle $S$ of circumference $\beta$.
  As long as the electric flux is deep inside or outside the topological insulator, but at any rate far from its boundary,
the boundary fermions do not ``see'' this flux.   Generically (for a generic value of $\oint_SA$, to be precise) the boundary
fermions have no zero-mode, so $|Z_\psi|>0$.  On the other hand, $\iota=0$ if the instanton is outside the topological insulator -- that
is if $P=0$ -- and $\iota=1$ if the instanton is inside the topological insulator -- that is if $P=1$.  The path integral measure is always real
and so $\sT$-conserving, but it passes smoothly from positive to negative values as the instanton is brought inside the topological insulator.
 The jump in $\iota$ from 0 to 1 -- and thus the change in sign of the path integral measure -- occurs
precisely when an eigenvalue of the boundary Dirac operator passes through zero, or in other words when $|Z_\psi|=0$.   

The fact that the path integral
is negative when $P=1$ can be described by saying that the path integral measure of the topological insulator contains
a factor $(-1)^P=\exp(\i\pi P)$.  This gives a precise meaning to the statement that $\theta=\pi$ inside the topological insulator.

Another  and perhaps less technical way to give a precise meaning to the statement that  $\theta=\pi$ 
has already been described in the literature \cite{RF}. Let us go back to the Lorentz signature picture in which the full spacetime
is $\hat X=\R'\times \R\times S^2$,  the topological insulator lives on $X=\R'\times \R_+\times S^2$,
and the boundary Dirac fermion lives on $\partial X\cong \R'\times\{0\}\times S^2$. To find the quantum states of the boundary fermions, we have
to find the eigenmodes of the 2d Dirac operator on $S^2$, and then quantize them.  In the presence of a unit magnetic flux on $S^2$,
the 2d Dirac operator has a single charge $-e$ zero-mode of one 2d chirality, and a single charge $+e$ zero-mode of the opposite 2d chirality.
(The existence of these
zero-modes is guaranteed by the 2d version of the Atiyah-Singer index theorem.  The 
two zero-modes are complex conjugate; complex conjugation reverses both the charge and the 2d chirality.)   
As in \cite{JR}, quantization of these modes gives a pair
of states with electric charge $\pm e/2$.     The fact that the boundary of the topological insulator, when pierced by a unit of flux, supports
half-integral electric charge can be understood if we imagine creating this unit of magnetic flux by dragging a magnetic monopole of unit
charge into the topological insulator.  (We assume that the monopole starts far to the left of the topological insulator in fig.
\ref{support}, with its flux in the initial state going
off to the left and not entering the topological insulator.)   When the monopole enters the material, a unit of flux appears on its surface and accordingly 
the charge on its surface jumps by a half-integral amount. But the electric charge of the monopole also jumps by a half-integral amount
 when it enters the
topological insulator, because a magnetic monopole in a world with $\theta=\pi$ has half-integral electric charge \cite{Dyons}.  
It has indeed been shown
in a lattice model of a topological insulator that when a magnetic monopole enters a topological insulator, half-integral charge appears on the surface of the topological insulator and the monopole itself acquires half-integral charge \cite{RF}.
This gives a direct interpretation of what it means to say that the topological insulator has $\theta=\pi$.

\subsubsection{Why is The Partition Function Equal to $(-1)^\iota$?}\label{why}

Part of the above story is that  on a compact four-manifold $X$ without boundary, with a background $U(1)$ gauge field $A$, the topological field theory
associated to a topological insulator has
partition function  $(-1)^\iota$.   The reader may wonder how this is related to other known descriptions of a topological insulator.

We use the following standard characterization of the phase transition between a topological insulator and an ordinary one.  This transition
occurs when the mass parameter $m$ of a $D=4$ Dirac fermion $\psi$ passes through zero and changes sign.  

First we recall the definition of the index of the Dirac operator.  On an oriented 4-manifold, we define the chirality operator 
\be\label{chiro}\bg=\frac{1}{4!}\eps^{ijkl}\g_i\g_j\g_k\g_l,\ee where $\eps^{ijkl}$ is the Levi-Civita tensor.  We write $\psi=\psi_++\psi_-$ where 
$\psi_\pm=\frac{1}{2}(1\pm \bg)\psi$ satisfy  $\bg\psi_\pm=\pm\psi_\pm$; we say that $\psi_+$ and $\psi_-$ have positive or negative
chirality.    $\bg$ anticommutes
with the Dirac operator $\D$, so $\D$ maps $\psi_+$ to $\psi_-$ and vice-versa.  We write $n_+$ for the dimension of the space of zero-modes
of $\D$ acting on $\psi_+$ and $n_-$ for the dimension of the space of zero-modes of $\D$ acting on $\psi_-$.
The index of the Dirac operator is $\iota=n_+-n_-$.  It is a basic topological invariant.

To prove topological invariance of $\iota$,
observe that $\bg$ anticommutes with $\D$, so that if $\D\psi=\lambda\psi$ with $\lambda\not=0$, then $\D(\bg\psi)=-\lambda\bg\psi$.  $\psi$ and $\bg \psi$ must therefore
be linearly independent if $\lambda\not=0$, and hence the chiral projections $\psi_\pm=\frac{1}{2}(1\pm\bg)\psi$ are nonzero.
Consider the ``Hamitlonian'' $H=\D^2$. Zero-modes of $\D$ are the same as zero-modes of $H$, and more generally  
eigenstates of $H$, obeying $H\psi=E\psi$, are linear combinations of solutions of $\D\psi=\pm\sqrt{E}\psi$.
If $E\not=0$, a solution of $\D\psi=\pm\sqrt{E}\psi$ has nonzero chiral projections $\psi_\pm$ with the same $E$.  Hence nonzero energy levels
of $H$ occur in pairs with positive and negative chirality.  When we vary the gauge field or metric on $X$, all that can happen to the space
of zero energy states is that a pair of states of opposite chirality can move to or from zero energy.  In the process, $n_+$ and $n_-$
change, but $\iota=n_+-n_-$ is unchanged. 

Let us assume that $X$ and $A$ are such that $\iota\not=0$ and for definiteness let us take $\iota>0$.    Generically, for $\iota>0$, $\psi_+$ will have $\iota$ zero-modes and $\psi_-$ will have none.  Let us assume that we are in this situation.
What about zero-modes of the oppositely charged fermion field $\bar\psi$?  Writing $\bar\psi_-$ and $\bar\psi_+$ for its components
of definite chirality, $\bar\psi_-$ will have $\iota$ zero-modes and $\bar\psi_+$ will have none.  (The zero-modes of $\bar\psi_-$ are complex
conjugates of the zero-modes of $\psi_+$.)   

Because of these $2\iota$ zero-modes, the fermion path integral vanishes if $m=0$.   However, for $m\not=0$, the action has a term
\be\label{delfo}I=\int\d^4x \left(\dots -m\bar\psi_-\psi_++\dots\right),\ee
and accordingly the integrand $e^{-I}$ of the path integral has a factor
\be\label{elfo} \exp\left(m\int\d^4x \bar\psi_-\psi_+\right). \ee
Expanding this factor, a $\bar\psi_-$ zero-mode and a $\psi_+$ zero-mode can be lifted at the cost of a factor of $m$.

Accordingly, for small $m$, the path integral is proportional to $m^\iota$.  Therefore, if the path integral is positive-definite for, say, $m>0$,
then its sign for $m<0$ is $(-1)^\iota$.  (Zero-modes beyond the ones we assumed would contribute an even power of $m$ and thus would
not affect this sign.)  This sign is the universal part of the answer and is the partition function of the topological field theory that one extracts
as the long distance limit of the topological insulator.

We treat several similar problems later in this paper in a slightly more precise way including an explanation of the regularization.
For example, see section \ref{bulkboundary}.

\subsubsection{The Index And Cobordism Invariance}\label{cobo}

We defined the Dirac index $\iota$ in section \ref{why} by using the fact that on an oriented 4-dimensional spin manifold $X$ without
boundary, the equation $\D\psi=0$ for a fermion zero-mode splits as separate equations for the chiral projections $\psi_+$ and $\psi_-$.

\begin{figure}
 \begin{center}
   \includegraphics[width=3in]{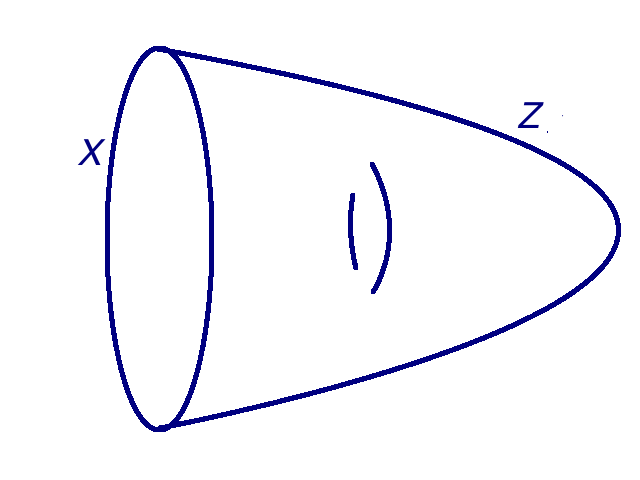}
 \end{center}
\caption{\small A topological invariant $\iota$ defined for a manifold $X$ is 
said to be a cobordism invariant if it vanishes whenever $X$ is the boundary
of a manifold $Z$ of one dimension more.  If $X$ has some structure (such as an orientation, a spin structure, or a $\U(1)$ gauge field)
that is required in defining $\iota$, then this structure is required to extend over $Z$.}
 \label{coob}
 \end{figure}
The same is true in any even dimension, and the original Atiyah-Singer index theorem \cite{ASfirst,AS} gives a cohomological formula for $\iota$.
For a $\U(1)$ gauge field in $D=4$, the formula is 
\be\label{erfo}\iota=\hA-P.\ee
 Here  $P$ is the instanton number defined in eqn. (\ref{zelm}), and 
\be\label{ziffy}\hA=-\frac{1}{48}\int_X\,\frac{\tr\,R\wedge R}{(2\pi)^2},\ee
with $R$ being the Riemann tensor regarded as a matrix-valued 2-form.
The formula implies immediately\footnote{Here we are reversing the original logic.  In the original proof of the index theorem \cite{ASfirst},
it was shown first that the index is a cobordism invariant, and this knowledge was used to deduce a formula for it.}
 that $\iota$ is a cobordism invariant: it vanishes if $X$ is the boundary of an oriented 5-dimensional spin manifold 
$Z$ over which the $\U(1)$ gauge field extends (fig. \ref{coob}).  For in that situation,  $\hA=P=0$, so $\iota=0$.  For example, one has 
\be\label{wiffy}\hA=-\frac{1}{48}\int_X \,\frac{\tr\,R\wedge R}{(2\pi)^2}=-\frac{1}{48}\int_Z\d \,\frac{\tr\,R\wedge R }{(2\pi)^2}=0\ee
(here one  uses Stokes's theorem to express an integral over $X$ as an integral over $Z$, and then one
uses the Bianchi identity $DR=0$ to prove that $\d\,\tr\, R\wedge R=0$). A similar argument shows that $P=0$.

Since $\iota$ is a cobordism invariant, so is $(-1)^\iota$,
which we have seen to be the partition function (on an orientable manifold) of the bulk sTQFT associated to a topological insulator.
As we proceed in this paper, it will become clear that partition functions of sTQFT's associated to free fermion states of matter are always
cobordism invariants. This was conjectured in \cite{KTTW}.

In 
general,\footnote{For a brief explanation, see the discussion of eqn. (5.4) in \cite{Freed}.}
 starting with such a $\U(1)$-valued  cobordism invariant (a homomorphism from an appropriate cobordism 
 group to $\U(1)$), one can always construct an invertible topological
field theory with the given invariant as its partition function.  The rough idea behind this statement is that cobordism invariance implies
relations among partition functions that one would expect on physical grounds in a topological field theory.  The simplest example
of such a statement is illustrated in fig. \ref{cob}. Another example of how cobordism invariance implies a physical property (unitarity) is described at the
end of section \ref{rpo}. 

 It appears to be unknown whether all unitary invertible TQFT's or sTQFT's 
are  associated to cobordism invariants.   Since this is unknown, we do not really know if the fact that the usual free fermion phases are associated
to cobordism invariants is a consequence just of the fact that they are gapped and have no topological order in bulk, or is more special.  At any rate,
cobordism invariance is much stronger than topological invariance.

\begin{figure}
 \begin{center}
   \includegraphics[width=3in]{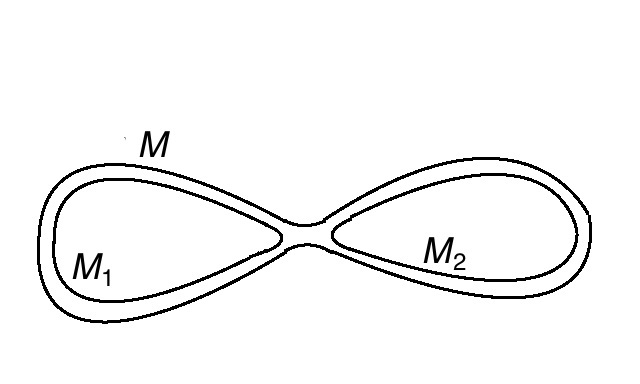}
 \end{center}
\caption{\small  Here we give the simplest illustration of the statement that cobordism invariance leads
to relations that are natural in topological field theory.  The {\it connected sum} $M$ of two $D$-manifolds $M_1$ and $M_2$ is made
by cutting small holes out of $M_1$ and $M_2$ and gluing them together along their boundaries.  If the space of
physical states on a sphere $S^{D-1}$ is 1-dimensional (as expected in a unitary topological field theory), one can deduce a universal 
relation between partition functions: $Z_M = g Z_{M_1}Z_{M_2}$, where $g$ is a constant characteristic of the theory.  Cobordism
invariance implies such a relation with $g=1$, because $M$ is cobordant to the disjoint union of $M_1$ and $M_2$.  This cobordism
is sketched in this figure (for the case $D=1$ with all manifolds being circles).   Topological field theories derived from cobordism invariants
always satisfy the condition that the space of physical states on any $D-1$-manifold has dimension 1, which is why they lead to a relation of the given
kind.}
 \label{cob}
 \end{figure}

\subsubsection{The APS Index Theorem}\label{aps}

To define the Dirac index $\iota$ in section \ref{why}, we used the fact that on an oriented 4-manifold, the Dirac equation $\D\psi=0$
splits as separate equations for the chiral projections $\psi_+$ and $\psi_-$ of $\psi$.

\begin{figure}
 \begin{center}
   \includegraphics[width=3in]{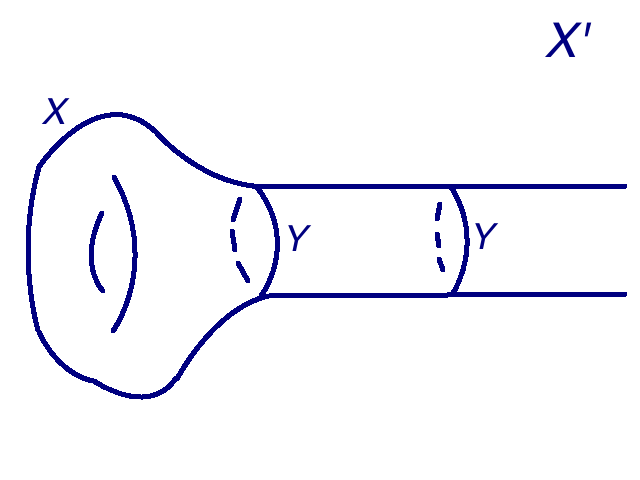}
 \end{center}
\caption{\small  The complete but not compact manifold $X'$ is built by gluing onto the manifold $X$, which has boundary $Y$,
a semi-infinite tube $Y\times \R_+$.}
 \label{longer}
 \end{figure}
The APS index theorem \cite{APS} is a generalization of the original Atiyah-Singer index theorem \cite{ASfirst,AS} to a manifold $X$ with boundary.
In trying to find such a generalization, one immediately finds that there is no convenient local boundary condition along $Y=\partial X$
that can be used in defining an index.  The Dirac operator admits a local boundary condition
\be\label{torz}\gamma\cdot n\,\psi=\pm \psi,\ee
where $n$ is the unit normal to $Y$ and $\gamma\cdot n=\g^\mu n_\mu$.  This boundary condition is the most general one that
preserves the local rotation symmetry of the boundary
and it is physically sensible in the sense that with this boundary condition the Dirac operator is hermitian.  However, because $\g\cdot n$
anticommutes with the chirality operator $\bg$, this boundary condition mixes the two fermion chiralities and so does not allow one to define
a Dirac index. 

What can we do to define a Dirac index in this situation?  The  APS construction starts by replacing a manifold  $X$ that has a boundary
 with a manifold
$X'$ without boundary that is not compact but has a complete Riemannian metric.  This is done by picking a Riemannian metric on $X$, and in
particular on $Y$, and building $X'$ by gluing
to the boundary of $X$ a semi-infinite tube $Y\times \R_+$, where $\R_+$ is the half-line $y\geq 0$ (fig. \ref{longer}). The Riemannian
metric in the tube is taken to be the obvious one $\d y^2+g_Y$, where $g_Y$ is the chosen metric on $Y$.  

Because $X'$ has no boundary, the Dirac equation on $X'$ does not mix positive and negative chirality spinors.  Therefore, one can
define spaces $H_+$ and $H_-$ of {\it square-integrable} solutions of the Dirac equation on $X'$ of positive and negative chirality.  Their
dimensions are integers $n_+$ and $n_-$.  Assume for the moment that the metric $g_Y$ of $Y$ is sufficiently generic so that the Dirac operator
$\D_Y$ of $Y$ has no zero-modes.  Then the index $\iota$ is defined in the usual way as $\iota=n_+-n_-$.

The APS index theorem gives a formula for $\iota$:
\be\label{trof}\iota=\hA-P-\frac{\etta}{2}. \ee
Here $\etta$ is the $\etta$-invariant (\ref{rv}) of the Dirac operator $\D_Y$ on $Y$. 
Thus $-\etta/2$ appears as a sort of boundary correction to the index theorem.

On a manifold $X$ without boundary, the index of the Dirac operator is an integer and moreover is a topological invariant.  
When $X$ has a boundary and the index  is defined with APS boundary conditions, it is still an integer but it is no longer a topological
invariant.
The APS index formula makes this clear.  As the metric and gauge field on $X$ and $Y$ are varied so that an eigenvalue of $\D_Y$ passes through 0,
$\etta$ jumps by $\pm 2$, so $\iota $ must jump by $\mp 1$.  To understand explicitly how this happens, we look at the Dirac equation in
the tube $Y'=Y\times \R_+$.   The Dirac operator $\D_{Y'}$ on $Y'$ is defined using gamma matrices $\g_y$ and $\g_i$, $i=1,2,3$ with
\be \label{drowsy}\D_{Y'}=\i\left(\g^y\frac{\partial}{\partial y} +\sum_{i=1}^3 \g^i\frac{D}{D x^i}\right), \ee
where $x^i$, $i=1,\dots,3$ are local coordinates on $Y$.   So the equation $\D_{Y'}\psi=0$ is
\be\label{rowsy}\left(\frac{\partial}{\partial y}-\i\sum_{i=1}^3\i\g^y\g^i\frac{D}{D x^i}\right)\psi=0. \ee
We want to express this in terms of the Dirac operator on $Y$, which would be
\be\label{owsy}\D_Y=\i\sum_{i=1}^3\h\g^i \frac{D}{D x^i}. \ee
Here $\h \g_i$ are purely 3-dimensional gamma matrices obeying $\{\h\g_i,\h\g_j\}=2g_{ij}$.  In a locally Euclidean frame, we can take them to be $\h \g_i=\sigma_i$ (where
$\sigma_i$ are the Pauli matrices) and so to obey \be\label{blg}\h\g_1\h\g_2\h\g_3=\i\epsilon_{123}\ee (and cyclic
permutations) where $\epsilon_{ijk}$ is the Levi-Civita tensor
of $Y$.

  In the definition (\ref{chiro}) of the chirality operator of $X$, we orient $X$ so that
$\epsilon_{yijk}=\epsilon_{ijk}$.  The matrices $\g'_i=\i\g_y \g_i$ appearing in eqn. (\ref{rowsy}) obey the same Clifford algebra
$\{\g'_i,\g'_j\}=2g_{ij}$ as the $\h \g_i$.  They also satisfy in a local Euclidean frame
$\g'_1\g'_2\g'_3=\bg \cdot \i \epsilon_{123}$.  From this one can deduce
that  the matrices $\bg \g'_i$
obey the same Clifford algebra and the same relation (\ref{blg})
 as the purely 3d Dirac matrices $\h\g_i$, and so we can set $\bg\g'_i=\h\g_i$.  Then we can rewrite the
$D=4$ Dirac equation (\ref{rowsy}) in the form
\be\label{wsy}\left(\frac{\partial}{\partial y}-\bg\D_Y\right)\psi=0.\ee
So if $\psi_0$ is a spinor field on $Y$ satisfying $\D_Y\psi_0=\lambda\psi_0$,  and moreover $\bg \psi_0=a\psi_0$, $a=\pm 1$,
then we can solve the $D=4$ Dirac equation with
\be\label{syw}\psi =\exp(a\lambda y)\psi_0.\ee
To get a square-integrable solution of this equation, we need $\lambda<0$ if $a=1$ or $\lambda>0$ if $a=-1$.

Now we can complete the construction by explaining
 APS boundary conditions on the original manifold with boundary $X$.  Let $\H$ be the space of all spinor fields
on $Y$, and decompose $\H=\H_+\oplus \H_-$ where $\H_+$ and $\H_-$ are spanned, respectively, by the eigenstates of $\D_Y$ with
$\lambda>0$ or with $\lambda<0$.  Let $\psi_+$ and $\psi_-$ be as usual the positive and negative chirality parts of a 4d spinor
$\psi$.  Then APS boundary conditions on $X$ say that the restrictions of $\psi_\pm$ to $Y=\partial X$ obey
\be\label{yw}\psi_+|_Y\in \H_-, ~~\psi_-|_Y\in\H_+. \ee
The point  is that solutions to the Dirac equation on $X$ that can be extended to square-integrable solutions
on $X'$ are precisely those that obey this condition.  So we can compute the index $\iota$ either  in terms of solutions of the Dirac equation on
$X$ that obey this APS boundary condition, or in terms of square-integrable solutions on $X'$.

This should make it clear how $\iota$ can fail to be a topological invariant.  When an eigenvalue of $\D_Y$ passes through 0, the spaces
$\H_\pm$ change discontinuously, the APS boundary conditions change discontinuously, and the index $\iota$ also changes discontinuously.

As long
as $\D_Y$ has no zero-modes, the spectrum of the Dirac operator on $X$ with APS boundary conditions varies continuously with the metric
and gauge fields on $X$ (and in particular on $Y$) and $\iota$ is constant.   In the above, we defined $\iota$, $\H_+$, and $\H_-$ only in the
generic case that $\D_Y$ has no zero-modes.  The definitions can be slightly modified so that the APS index formula (\ref{trof}) is valid
in general, with $\etta$ defined using eqn. (\ref{uv}).

We will make a few more remarks on the APS index theorem.  First of all, in eqn.  (\ref{trof}), we have written the APS index theorem for $D=4$.
However, this theorem holds in any dimension $D$.  In general, $\hA-P$ has to be replaced by the bulk contribution to the index that is
given by the original Atiyah-Singer index theorem \cite{ASfirst,AS}  (this bulk contribution can be written in general as $\int_X\hA(R)\,\tr \,e^F$,
where $F$ is the Yang-Mills field strength or 
curvature and the trace is taken in the representation that the fermions live in.  The boundary contribution is always $-\etta/2$.  However,
for odd $D$, the bulk contribution vanishes\footnote{The reason for this is just that $R$ and $F$ are both two-forms, so $\hA(R)\,\tr\,e^F$ is a linear combination
of forms of even degree.  If $D$ is odd, then $\hA(R)\,\tr\,e^F$ has no degree $D$ piece that could be a bulk contribution in the index theorem.}
 and the APS index theorem reduces to
\be\label{zolg} \iota=-\frac{\etta}{2}. \ee
Since $\iota$ is always an integer, this shows that for odd $D$, if a $D-1$-manifold $Y$ is the boundary of some $D$-dimensional $X$ over which the relevant
structures extend (so that an APS index theorem on $X$ is available),  then the $\etta$-invariant on $Y$ obeys $\etta/2\in\Z$.  
In later applications to Majorana fermions, we will want to write this formula in terms of $\I=2\iota$, $\eta=2\etta$:
\be\label{molg}\I=-\frac{\eta}{2}. \ee

Another important consequence of the APS index formula is a relationship between $\etta$ and the Chern-Simons function 
\be\label{zerf}\ICS(A)=\frac{e^2}{4\pi}\int_Y\,\d^3x \,\epsilon^{ijk}A_i\partial_jA_k. \ee
$\ICS(A)$ is only gauge-invariant mod $2\pi\Z$ (which is why the coupling $k$ in eqn. (\ref{csm}) is an integer in any material
that can be described by an effective action for $A$ only), but its variation $\delta\ICS(A)$ in a continuous change in $A$ is completely gauge-invariant.  
If the spin manifold $Y$ on which we are trying to define $\CS(A)$ is the boundary of some $X$ (over which $A$ and the spin
structure extend), then 
\be\label{xerf}\ICS(A)=2\pi P~{\mathrm{mod}}~2\pi\Z, \ee
where $P$ is 
the instanton number, integrated over $X$, that appears in the index formula (\ref{trof}).      This formula depends on the choice of a 4-dimensional spin manifold $X$ with boundary $Y$, along with an extension of $A$ over $X$,
but it is independent of these choices mod $2\pi\Z$. 
To prove this, suppose that $X$ and $X'$ are spin manifolds with the same boundary $Y$; let $P$ and $P'$ be the instanton
numbers integrated over $X$ and $X'$, respectively.  Then the difference $P-P'$ is an integer, because it is the instanton
number integrated on a closed oriented
 four-dimensional spin manifold $\h X$ built by gluing $X'$ to $X$ with a reversal of orientation of $X'$ (fig. \ref{gluing}). 

 \begin{figure}
 \begin{center}
   \includegraphics[width=3in]{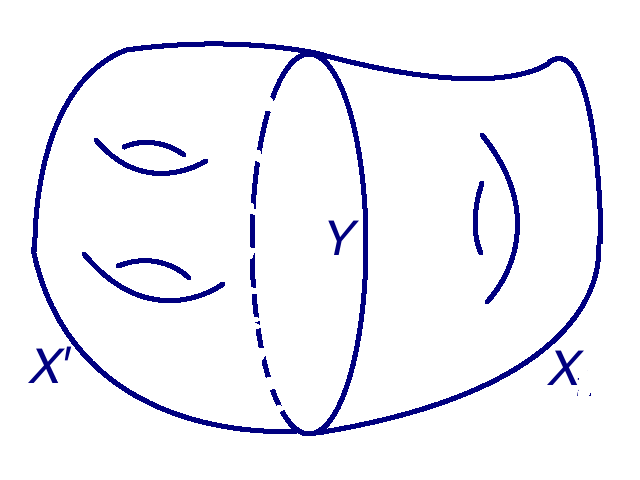}
 \end{center}
\caption{\small  By gluing together two oriented four-manifolds $X$ and $X'$ that have the same boundary $Y$ (after reversing
the orientation of $X'$ so that the orientations are compatible), we build a closed oriented four-manifold $\h X$.  $\h X$ is a spin
manifold if $X$ and $X'$ have spin structures that agree along $Y$.}
 \label{gluing}
 \end{figure}

 Similarly, one can define a gravitational Chern-Simons term $\ICS_\grav$, also defined mod $2\pi\Z$,
 that satisfies
\be\label{nerf}\ICS_\grav=2\pi \frac{\hA}{2}. \ee
(It is understood that $\hA$ is integrated over $X$.  The formula
is written in terms of $\hA/2$ because $\hA$ is even on a four-dimensional spin manifold without boundary.  This fact
will be important in section \ref{topsup}.)  
Given these definitions, and assuming a suitable $X$ exists, the APS index theorem for $X$ implies
\be\label{moj}\frac{\etta}{2}=\frac{\ICS(A)}{2\pi}-2\frac{\ICS_\grav}{2\pi}~~{\mathrm {mod}}~\Z. \ee
The statement is true only mod $\Z$ because it does not take into account the $\iota$ term in the index formula.
(Also, $\ICS(A)/2\pi$ is only well-defined mod $\Z$ so the statement really only makes sense mod $\Z$.)

In deducing this relation between $\etta$ and Chern-Simons, we have assumed that the manifold $Y$ on
which we are studying these objects is that boundary of some $X$ on which we can use the APS index theorem.
This is true if $Y$ is a 3-dimensional  spin manifold with a $\U(1)$ gauge field, but in a more general problem to which we might
apply the APS index theorem, it will not always be true.
Even  then, a somewhat weaker statement is true.  In general, the difference between $\etta/2$
and the corresponding Chern-Simons invariant is constant (but is not necessarily an integer) as one varies the metric and gauge field on $Y$,
keeping away from jumps in $\etta$. To state the relation in generality, we assume that $Y$ has dimension $d$,
and we  generically write $\ICS(g,A)/2\pi$ for the Chern-Simons
invariant that is related to the curvature polynomial\footnote{A curvature polynomial is just a polynomial in the Riemann curvature $R$ and the gauge theory field strength
or curvature $F$.}
 $\hA\, \tr\,e^F$  that appears in the index theorem in $D=d+1$ dimensions 
(generalizing
$\hA-P$ for $\U(1)$ gauge theory in 4 dimensions).  
Then  under any continuous variation of the metric and gauge field $(g,A)$ on $Y$, modulo even integer
jumps in $\etta$, we have
\be\label{noj}\frac{\delta\etta}{2}=\frac{\delta \ICS(g,A)}{2\pi}. \ee
This is proved by applying the APS index theorem to the $D$-manifold $X=Y\times [0,1]$, with one pair of fields $(g,A)$
at one end, and a slightly different pair $(g',A')=(g+\delta g,A+\delta A)$ at the other end.    Including the jumps in $\etta$, the statement is
\be\label{oj} \frac{\etta}{2} =\frac{\ICS(g,A)}{2\pi}+{\mathrm{constant}}~~{\mathrm{mod}}~\Z.\ee The constant, which is valued in 
$\R/\Z$, is an important topological invariant in some situations.  It can appear in global anomalies, because of the relation of global
anomalies to $\etta$ (section \ref{complex}).

In a context in which one only cares about $\etta/2$ mod $\Z$, and assuming $X$ exists (as it always does for $\U(1)$ gauge
theory on a 3-dimensional spin manifold),
eqn. (\ref{moj}) can be used to replace $\etta$ with the more
elementary invariants $\ICS(A)$ and $\ICS_\grav$.  However, 
the boundary fermions of the 3d topological insulator have partition function $Z_\psi=|Z_\psi|
\exp(-\i\pi\etta/2)$, which depends on $\etta/2$ mod $ 2\Z$, not mod $\Z$.  So the replacement (\ref{moj}) is not adequate for describing
those boundary fermions.  If we try to make that replacement, we will lose control of the overall sign of $Z_\psi$, which as we have seen is
very important in describing the topological insulator.

\subsubsection{Reflection Positivity}\label{rpo}

Another question is why the formula $|Z_\psi|(-1)^\iota$ for the partition function is consistent with unitarity,
that is with reflection positivity. 

Actually, in the specific case at hand, we can give a trivial answer to this question.  We have built 
$|Z_\psi|(-1)^\iota$ as a product of two factors $|Z_\psi|\exp(\mp \i\pi\etta/2)$ and $\exp(\pm \i\pi( P-\h A(R)))$.  We showed
the first factor to be reflection positive in the discussion of fig. \ref{dodo}, and the second factor is also compatible
with reflection-positivity; otherwise
 gauge  and gravitational $\theta$-angles would in general not be physically sensible.  (Concretely, $\exp(\pm\i\theta (P-\h A(R)))$
is consistent with reflection positivity for any real $\theta$ because $P$ and $\h A(R)$ are odd under reflection and vanish for
any reflection-symmetric gauge field and metric.)

 \begin{figure}
 \begin{center}
   \includegraphics[width=2.5in]{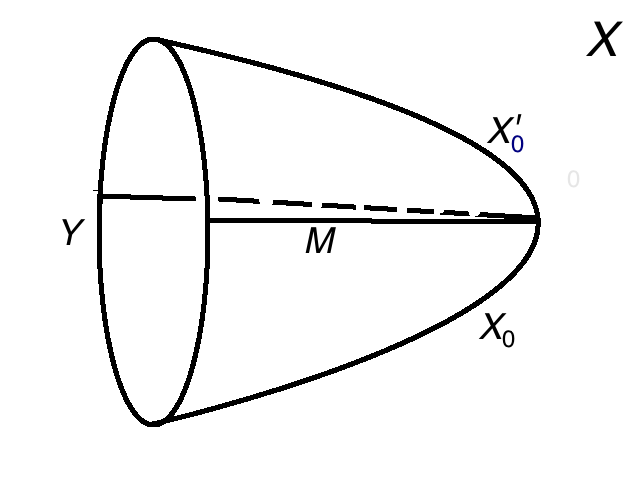}
 \end{center}
\caption{\small  A spacetime $X$ with boundary $Y$ that is built by gluing together mirror image copies
of some manifold $X_0$,  The copies
 are denoted $X_0$ and $X_0'$ in the figure.  
$Y$ is likewise the union of mirror image pieces.  $X_0$ has two types of boundaries: $M$, along
which it is glued to $X_0'$, and also its intersection with $Y=\partial X$.  $X_0$ has corners
along $M\cap Y$.}
 \label{dodotwo}
 \end{figure}

However, with a view to other cases, we will establish directly
the reflection-positivity of $|Z_\psi|(-1)^\iota$.  We want to show that $|Z_\psi|(-1)^\iota$ is real and nonnegative
whenever a spacetime $X$, with boundary $Y$,
is built by gluing together, along their common boundary $M$, two mirror image copies of some manifold
$X_0$, as in fig. \ref{dodotwo}.   If $X$ has boundary $Y$, they this boundary is likewise built by gluing together mirror
image pieces. In that case, $X_0$ is really a manifold with corners, as in the figure.
 We assume that the gauge field, metric, and spin structure
 on $X$ are invariant under the reflection that exchanges
the two copies of $X_0$.

Since $|Z_\psi|$ is trivially real and nonnegative, to show reflection positivity of $|Z_\psi|(-1)^\iota$ it suffices to show that $\iota=0$ in 
a reflection-symmetric situation.
 A reflection reverses the chirality of a $D=4$ fermion,
so in any reflection-symmetric situation, the index $\iota$ computed with
APS boundary conditions vanishes.  This completes the proof.

Alternatively, we  could prove the vanishing of $(-1)^\iota$ in the situation of fig. \ref{dodotwo} by using
cobordism invariance.  This explanation has the advantage that it applies almost 
directly\footnote{To do this properly requires some important details about the gluing of the spin or pin
structures of $X_0$ and $X_0'$ that have been described to me by D. Freed.  These details have to be used
in establishing that the spin or pin structure of $X$ extends over the manifold $Z$ constructed below.} to the other examples
that we will study in the rest of this paper, since they all involve cobordism-invariant sTQFT's.  If $X$
is constructed as in fig. \ref{dodotwo} by ``doubling'' some manifold $X_0$, then $X$ is always the boundary
of a manifold $Z$ that can be constructed as follows.  Let $M$ be the boundary of $X_0$ along
which the gluing occurs, as in the figure. Let $Z_0=X_0\times I$,
where $I$ is the unit interval $[0,1]$, and in $X_0$, collapse $M\times I$ to $M$.   This gives a manifold
$Z$ with boundary $X$.  The local picture, near $M$, is that $Y$ looks like $M\times \R_+$ and $Z$ looks
like $M\times \R^2_+$, where $\R_+$ and $\R^2_+$ are halfspaces in $\R $ or $\R^2$.  The existence
of $Z$ ensures vanishing of a cobordism invariant of $X$ such as $(-1)^\iota$.

\subsubsection{Running The Story In Reverse}\label{reverse}

Finally, it may be illuminating to briefly run the above story in reverse.  We do this by first postulating in bulk a spin
topological quantum field theory or sTQFT, and then determining what sort of symmetry-preserving boundary state it can have.

We consider a $D=4$  sTQFT, with $\U(1)$ symmetry, that is characterized by saying that its partition function on a compact $D=4$ spin
manifold $X$, in the presence of a background $\U(1)$ gauge field $A$, is $(-1)^\iota$, $\iota$ being the index of the Dirac operator. 
This is the partition function of an sTQFT because the Atiyah-Singer index theorem shows that $\iota$ is a cobordism invariant
and in general such a $\U(1)$-valued cobordism invariant is the partition function of a topological field theory.

A topological insulator can have a spatial boundary, and in fact in the real world -- as opposed to thought experiments -- spatial boundaries
are unavoidable.  So we would like to extend our sTQFT to make sense on a manifold $X$ with boundary.
This is not straightforward.
The index $\iota$ of the Dirac operator cannot be defined with local boundary conditions on $\partial X$, but it can be defined with global
APS boundary conditions.  But  $\iota$ defined this way is not a topological invariant, and moreover as the background data (the metric
and gauge field on $X$) are varied, $\iota$ jumps in an unphysical way.

The standard boundary state of the topological insulator has massless Dirac fermions living on $\partial X$.  In the presence of these modes,
the jumps in $\iota$ make sense physically.  The formula $|Z_\psi|(-1)^\iota$ that we have described is a physically sensible formula
for the partition function of a combined system consisting of an sTQFT in bulk coupled to Dirac fermions on the boundary.
The jumps in $\iota$ compensate for the problem in defining the sign of the fermion partition function.

\subsection{Topological Superconductor In $d=3$}\label{topsup}

We now turn to the topological superconductor in $3+1$ dimensions, formulated for now on an orientable manifold only.

\subsubsection{Fermions And Pfaffians}\label{ferp}

The topological superconductor is gapped in bulk (at least for fermionic excitations), but on the boundary it supports a two-component
Majorana fermion $\psi$.   The main difference from the analysis of the 
topological insulator is that $\psi$ does not carry any conserved charge
(except $(-1)^F$, the operator that counts fermions mod 2).  In particular, and in contrast to the topological insulator,
there is no distinction between fermion modes of positive and negative charge.  This necessitates a few changes in the details
of our explanations, although the main ideas are the same.  In the absence of a conserved additive charge carried by the fermions,
we have to use fermion Pfaffians rather than determinants.  Moreover,  we will  need a more careful
argument to explain a doubling of the spectrum that is rather trivial in the $\U(1)$-invariant case.

A 3d Majorana fermion $\psi$ in Lorentz signature has 2 real components -- they transform in the real 2-dimensional representation
of $\Spin(1,2)\cong \SL(2,\R)$.  After continuing to Euclidean signature, $\psi$ still has 2 independent components, although
they are no longer real fields; they transform in the 2-dimensional pseudoreal representation of $\Spin(3)\cong \SU(2)$.
(This is simply the spin 1/2 representation of $\SU(2)$.)  We cannot claim that $\psi$ is real, because this would not be
compatible with $\SU(2)$ symmetry, but nonetheless
the complex conjugate of $\psi$ does not appear in the formalism.\footnote{This leads to no contradiction because fermion
integration is really an algebraic procedure.  We define it for an odd variable $\psi$
by saying $\int\d \psi\cdot 1=0$, $\int \d\psi\cdot\psi=1$, without
ever having to claim that $\psi$ is real or to mention its complex conjugate.}  So when we write
the Euclidean version of the action for $\psi$
\begin{equation}\label{zorb}I_\psi=\int\d^3x\,\bar\psi \DD \psi,  \end{equation}
despite appearances, there is actually no complex conjugation involved in the definition of
 $\bar\psi$.  Indeed, for a Majorana fermion continued to Euclidean space, the definition is just $\bar\psi_\alpha=\veps_{\alpha\beta}\psi^\beta$,
 where $\veps_{\alpha\beta}$ is the $\Spin(3)=\SU(2)$-invariant antisymmetric tensor. Thus $I_\psi$ is simply
an antisymmetric bilinear expression in the fermion field $\psi$.  It is antisymmetric because of fermi statistics.  

It is useful to distinguish carefully between operators and bilinear forms. The usual hermitian
Dirac operator $\D=\i\slashed{D}=\i\sum_{i=1}^3\g^iD_i$ is an ``operator,'' schematically 
$\psi^\alpha\to \sum_\beta\i\slashed{D}^\alpha_\beta\psi^\beta$.  
A spinor index of $\psi^\alpha$ can be raised or lowered using the  antisymmetric tensor $\veps_{\alpha\beta}$,
and if we do this we get the antisymmetric bilinear form $\DD_{\gamma\beta}=\veps_{\gamma\alpha}\i\slashed{D}^\alpha_\beta$.

The path integral of the $\psi$ field is, formally, the Pfaffian of this antisymmetric form:
\begin{equation}\label{turmo}Z_\psi=\Pf(\DD). \ee
Whenever one has a theory of fermions, the quadratic part of the fermion action is always antisymmetric by virtue of fermi
statistics and the corresponding fermion path integral is the Pfaffian of the antisymmetric bilinear form that appears in the action.

When can one speak of a fermion ``determinant'' rather than a Pfaffian?  This is possible if there is a conserved $\U(1)$
charge\footnote{$\U(1)$ can be replaced by some other group as long as there is a splitting of $\psi$ that leads to a structure like that of 
eqn. (\ref{wurmo}).} and a splitting of $\psi$ into fields $\psi_+$ and $\psi_-$ of positive and negative charge, respectively.
$\U(1)$ symmetry then ensures that in a basis $\bp \psi_+\cr \psi_-\ep$, the form $\DD$ will be block off-diagonal
\begin{equation}\label{wurmo}\DD=\bp 0 & \D\cr -\D^\tr & 0 \ep. \ee  Here $\D$ is not subject to any condition of symmetry
or antisymmetry, and $\D^\tr$ is its transpose.  In such a situation, one has
\begin{equation}\label{zurmo}\Pf(\DD)=\det(\D),\ee
and so the fermion path integral can be written as a determinant.

Going back to the hermitian Dirac operator $\D$ for the Majorana fermion, 
its eigenvalues have even multiplicity because of a Euclidean analog of
Kramers doubling.\footnote{The mathematical facts that lead to this Euclidean analog of 
Kramers doubling for $c$-number modes in 3 spacetime dimensions
are the same ones that lead to conventional Kramers doubling for quantum states in $3+1$ dimensions.   
A definition of $\mathcal T$ that is less intrinsic but may look more familiar is given
in Appendix \ref{wwod}; see eqn. (\ref{zork}).}  If a $c$-number spinor $\chi$ is an eigenvector of the Dirac operator,
\be\label{dolf}\D\chi=\lambda\chi,~~\lambda\in\R,\ee
then $\t\chi$ defined by $\t\chi^\alpha=\veps^{\alpha\beta}\chi^*_\beta$ is an eigenvector with the same eigenvalue
\be\label{old}\D\t\chi =\lambda\t\chi.\ee
Here we write $\chi^*_\beta$ for the complex conjugate of $\chi^\beta$.  Although there is no field in the Euclidean field
theory that corresponds to the complex conjugate of $\psi$,  when we expand $\psi$ in eigenmodes of $\D$ as a step
toward performing a functional integral, it makes sense to consider the behavior of these eigenmodes under complex conjugation.
If we define an antilinear operation $\T$ on spinors by \be\label{norg}(\T\chi)^\alpha =\veps^{\alpha\beta}\chi^*_\beta,\ee
then $[\T,\D]=0$ and $\T^2=-1$.   Just as in the study of Kramers doubling in $3+1$ dimensions, this implies that the
eigenvalues of $\D$ have even multiplicity.   The doubling of the spectrum always occurs, for the same reason, for
pseudoreal fermions.  One simply has to replace $\veps^{\alpha\beta}$ by the invariant antisymmetric bilinear form
relevant to a given case.

Let $\chi$ and $\T\chi$ be a pair of eigenmodes of $\D$ with the same eigenvalue $\lambda$.  Because of the definition
$\DD_{\g\beta}=\veps_{\g\alpha}\D^\alpha_\beta$, in a $2\times 2$ block with basis $\bp \chi\cr\T\chi\ep$, $\DD$ takes
the form 
\be\label{tf}\DD=\bp 0 & -\lambda \cr \lambda & 0 \ep.\ee
More generally,, consider a basis of eigenmodes of $\D$ of the form $\chi_i,\T\chi_i$ (where $\chi_i,\,\T\chi_i$ are
orthogonal to $\chi_j,\,\T\chi_j$ for $i\not=j$).  In such a basis, $\DD$ is block-diagonal with $2\times 2$ blocks of the form
(\ref{tf}).  Accordingly, the Pfaffian of $\DD$ is formally
\be\label{rf} \Pf(\DD)=\prod_i{}'\,\lambda_i, \ee
where  the product is over eigenvalues of $\D$ and the symbol $\prod{}'$ means that in the product, we take only one
eigenvalue from each pair.  

Let us compare this to the corresponding formula (\ref{peto}) for a Dirac fermion.  On the left hand side of (\ref{peto}), we wrote
$\det\,\D$ instead of $\Pf(\DD)$, reflecting the fact that $\Pf(\DD)$ can be written as $\det\,\D$ in the $\U(1)$-invariant case.
  On the right hand side of eqn. (\ref{peto}), the product over eigenvalues runs only over fermion
eigenvalues of  positive charge. (In that derivation, $\D$ was defined as the hermitian Dirac operator acting on a Dirac
fermion $\psi$, meaning a fermion of definite charge, which we may as well consider to be positive charge.)  
By charge conjugation symmetry of the Dirac equation, the corresponding
negatively charged fermion has exactly the same eigenvalues, but they are not considered in eqn. (\ref{peto}).  The doubling
of Dirac eigenvalues that we got in the Majorana case from a relatively subtle argument using the fact that $\T^2=-1$
is more trivial in the Dirac case: it just means that the eigenvalues are the same for fermions of positive or negative charge.
A formula like (\ref{rf}) can be written for any system of pseudoreal fermions; it reduces to something along the lines of (\ref{peto})
when there is a suitable  symmetry.

\subsubsection{Vanishing Of The Spectral Flow}\label{specvan}

Now let us consider the sign of $\DD$.  As in the discussion of the topological insulator, there is potentially a problem in 
defining the sign of the formal expression (\ref{rf}), since there are infinitely many eigenvector pairs with $\lambda<0$.
On a given 3-manifold $Y$, we can start with a particular metric $g_0$, define the sign of $\Pf(\DD)$ as we wish,
and then evolve this sign as the metric is varied by counting how many times an eigenvalue pair flows through $\lambda=0$.
To decide if this procedure gives an answer that is invariant under diffeomorphisms not connected to the identity, one follows
the procedure involving the mapping torus that was described in section \ref{trada}.
We consider the same spectral flow problem as in fig. \ref{flow},
and ask if the number $\Delta$ of eigenvalue pairs flowing through $\lambda=0$ is even or odd.

It turns out that $\Delta$ always vanishes.  The steps in proving this are as follows.  First of all, the spectral flow arguments
\cite{APS,DCG,Kiskis} that we mentioned in explaining eqn. (\ref{worf}) are also valid for Majorana fermions.  The only
differences are these: (1) when there is no $\U(1)$ symmetry, the spectral flow $\Delta$ should be defined as the number of
eigenvalue pairs flowing through $\lambda=0$, rather than the number of positive charge eigenvalues flowing through $\lambda=0$;
(2) similarly, when there is no $\U(1)$ symmetry, we have to express $\Delta$ in terms of a $D=4$ Dirac index $\I$ computed
in the space of all fermion fields, not an index $\iota$ computed in the space of positively charged fermions only.  The
relation between the two types of index (when they are both defined) is just
\be\label{reft} \I=2\iota,\ee
since charge conjugation implies that the $D=4$ index for negatively charged fermions is the same as that for positively
charged ones.  So the generalization of eqn. (\ref{worf}) that is valid without assuming a $\U(1)$ symmetry
 is
\be\label{neft}\Delta =\frac{\I}{2}. \ee  This formula makes sense because in $D=4$, $\I$ is always even,\footnote{
This assertion is proved by an argument similar
to the one we used to show the doubling of the Dirac  spectrum for $d=3$. The spinor representation of $\Spin(4)$ is pseudoreal,
and therefore the eigenmodes of a  Majorana fermion in $D=4$ are doubled because of an antilinear symmetry 
$(\T\chi)^\alpha=\veps^{\alpha\beta}\chi^*_\beta$,
$\T^2=-1$, $[\T,\D]=0$.  The zero-modes of positive or negative chirality are both doubled by this
argument, and therefore $\I$ is even.}  a fact that
will be important in what follows.
Similarly the analog of (\ref{torf}) is that the transformation of the fermion path integral under a diffeomorphism $\phi$
is
\be\label{weft}\Pf(\DD)\to \Pf(\DD)(-1)^{\I/2}. \ee
Here $\I$ is the index of the Dirac operator on the mapping torus built from $\phi$, taking all fermions into account.

This formula is not limited to the case of a single Majorana fermion coupled to gravity only. It applies (on an orientable
3-manifold; we postpone the unorientable case to section \ref{complex}) to any system of 3d Majorana
fermions transforming in a real representation of any compact gauge group.  In many cases (such as the topological insulator,
where eqn. (\ref{weft}) is equivalent to eqn. (\ref{torf})), the formula implies an anomaly in the
sign of $\Pf(\DD)$.  However, there is no such anomaly for a single Majorana fermion coupled to gravity only.
Even though $\I/2$ can be odd in general on a compact $D=4$ spin manifold without boundary,\footnote{For example,
$\I/2=1$ on a K3 surface.}  it always vanishes on a mapping torus.\footnote{One way to prove this is to note that in four
dimensions, the Dirac index for a Majorana fermion coupled to gravity only is related to the signature $\sigma$ by $\I=\sigma/8$.
But for a mapping torus of any dimension,
a relatively elementary topological argument shows that $\sigma=0$. Let $X\to S^1$ be a mapping
torus, of dimension $4n$ for some $n$, with fiber $Y$ of dimension $4n-1$.  Consider the subspace $\Gamma\subset H_{2n}(X;\Z)$ generated by cycles of the
form $W\times p$, with $W\subset Y$ and $p$ a point in $S^1$.   Modulo torsion, $\Gamma$ is a null subspace for the intersection pairing on
$H_{2n}(X;\Z)$.  (For any $W,W'\subset Y$, the intersection number of $W\times p$ and $W'\times p$ is 0 because $W'\times p$ is homologous
to $W'\times p'$, for some $p'\not= p \in S^1$, and $W'\times p'$ does not intersect $W\times p$.)   Poincar\'e duality can be used to show that 
$\Gamma$ is middle-dimensional.  The existence of a middle-dimensional null subspace of $H_{2n}(X;\Z)$ mod torsion  shows that the signature of $X$ is 0.}

\subsubsection{The Sign Of The Path Integral}\label{signp}

The last statement means that in the traditional sense, there is no anomaly in $\Pf(\DD)$ for a 3d Majorana fermion coupled to gravity
only on an oriented 3-manifold $Y$: after 
picking its sign in an arbitrary fashion at some starting metric $g_0$, the sign can be defined
in a consistent way for all metrics in a way invariant under orientation-preserving diffeomorphisms. However, it is not physically satisfactory to allow an 
arbitrary independent 
sign in the path integral for every $Y$.  As already remarked in section \ref{trada}, a satisfactory definition of the theory will
determine the phase of the path integral for all $Y$'s, compatibly with physical requirements of cutting and pasting.  

We will now explain how, even though there is no anomaly in the traditional sense, a physically natural procedure to define
the fermion path integral of the $2+1$-dimensional Majorana fermion 
in a consistent way for all $Y$'s leads to a violation of $\sT$- and $\sR$-symmetry, just as happens for the topological insulator.
Moreover, in the context of this procedure, it will be clear that, even though $(-1)^{\I/2}=1$ for mapping tori, the fact that
$(-1)^{\I/2}\not=1$  for some other (compact, closed, orientable spin) manifolds represents an obstruction
to defining the $2+1$-dimensional Majorana fermion by itself in a $\sT$-invariant way, without coupling to a bulk topological
superconductor.  With coupling to the bulk included, we will get a well-defined and $\sT$-invariant answer.

 Though I do not claim to have a formal proof, one would expect that any other procedure to define precisely the partition
function of the $2+1$-dimensional Majorana fermion for all $Y$ will lead to the same conclusions.  In general, one expects that two reasonable
regularizations of the same theory differ by a local counterterm, but there is no possible local counterterm that would change our conclusions.

To define the $2+1$-dimensional Majorana fermion, we will use the same Pauli-Villars regularization that we used
for the Dirac fermion.   The analog for the Pfaffian of eqn. (\ref{pv}) for the fermion determinant is
\be\label{polk}\Pf(\DD)_\reg=\prod_k{}'\frac{\lambda_k}{\lambda_k\pm\i\mu},\ee
where we allow both signs of the regulator mass.
The same steps that led to (\ref{yv}) now give for the Majorana fermion path integral
\be\label{oldo}Z_\psi=|Z_\psi| \exp(\mp\i\pi\etta/2),\ee
where we can now   define $\etta$ as a restricted sum 
 \be\label{riv}\etta=\lim_{s\to 0}\sum_k{}'\sign(\lambda_k)|\lambda_k|^{-s} \ee
 in which one only includes one eigenvalue from each pair.  However, it is more customary to define
 an $\eta$-invariant as a sum over all eigenvalues
 \be\label{roov}\eta=\lim_{s\to 0}\sum_k\sign(\lambda_k)|\lambda_k|^{-s} ,\ee
 so that \be\label{mold}\eta=2\etta,\ee
 and then we should write the fermion path integral as
 \be\label{zold}Z_\psi=|Z_\psi|\exp(\mp \i\pi \eta/4).\ee

 This is $\sT$-violating because it is not real, but as in section \ref{topthree}, we can restore $\sT$ invariance
 if we assume that the 3-manifold $Y$ that supports the massless Majorana fermion is actually the boundary of the worldvolume
 $X$ of a topological superconductor.   The salient property of the topological superconductor is that in the large volume
 limit, its partition function
 on a closed (orientable) four-manifold without boundary is $(-1)^{\I/2}$.  (This is always $\pm 1$, since $\I$ is even.)  This
 claim about the partition function of a topological superconductor can be justified by precisely
 the same reasoning that we gave in section \ref{why} for the topological insulator, starting with the fact that the phase
 transition between an ordinary and topological $D=4$ superconductor occurs  when the mass $m$ of  a $D=4$ Majorana fermion  passes
 through 0.  What happens to the sign of the path integral when this mass changes sign? To absorb $\I$ fermion
 zero modes requires $\I/2$ mass insertions, so the partition function
 is proportional for small $m$ to $m^{\I/2}$.  So if it is positive for one sign of $m$, then for the opposite sign of $m$ it is proportional
 to $(-1)^{\I/2}$.
 
 The partition function $(-1)^{\I/2}$ can be interpreted in terms of a gravitational $\theta$-angle.
In general, such an angle would give
 a factor $\exp(i\theta\I/2)$ in the path integral measure, so the fact that the topological
superconductor partition function is $(-1)^{\I/2}$ means that its gravitational $\theta$-angle is
$\pi$. (See \cite{RML} for a previous explanation of this fact.)  On a manifold without boundary, the Atiyah-Singer index theorem
gives $(-1)^{\I/2}=\exp(\pm \i\hA(R)/2)$, but on a manifold with boundary, with $\I$ defined using
APS boundary conditions, the APS index theorem gives a boundary
correction in this statement:
  \be\label{wizont}\exp(\mp\i\pi\eta/4)\exp\left(\pm \i\pi\hA(R)/2\right)=(-1)^{\I/2}.\ee
This formula is the analog of eqn. (\ref{wont}).

The combined path integral of the Majorana fermion on $Y=\partial X$ and the topological superconductor in bulk is then
\be\label{izont}|Z_\psi|  \exp(\mp \i\pi \eta/4) \exp\left(\pm \i\pi\hA(R)/2\right)=|Z_\psi|(-1)^{\I/2}.\ee
This is quite analogous to the topological insulator, with the sole difference $\iota\to\I/2$.  Since
$|Z_\psi|(-1)^{\I/2}$ is real, $\sT$ invariance has been recovered.

Now let us re-examine the partition function of the boundary Majorana fermions on $Y$.  The combined path integral
that describes these fermions and the bulk sTQFT on $X$ is supposed to give $|Z_\psi|(-1)^{\I_X/2}$, where we write $\I_X$
(and not just $\I$) to emphasize the role of $X$.   But does this formula depend on $X$?  To decide this question, suppose
that $X$ and $X'$ are orientable spin manifolds with the same boundary $Y$.  By gluing them together with a reversal of orientation
on one of them, we build an orientable spin manifold $\h X$ without boundary (see fig. \ref{gluing} of section \ref{aps}).  
The gluing formula for the index
with APS boundary conditions gives\footnote{The theorem says that $\I_{\h X}=\I_X+\I_{X'}$, which clearly implies eqn. (\ref{zely}).     It follows from the Atiyah-Singer
index formula $\I_{\h X}=\int_X \hA(R)\,\tr\,e^F$, together with the APS index theorem, which says that $\I_{X}=\int_X\hA(R)\,\tr\,e^F-\eta/2$, with a similar formula
for $\I_{X'}$.  When we take the sum $\I_X+\I_{X'}$, the integrals over $X$ and $X'$ add up to the integral over $\h X$ that gives $\I(\h X)$, and the boundary terms $-\eta/2$ cancel
because $\eta$ is odd under reversal of orientation.}
\be\label{zely}(-1)^{\I_X/2}(-1)^{\I_{X'}/2}=(-1)^{\I_{\h X}/2}.\ee
Thus in general $(-1)^{\I_X/2}$ and $(-1)^{\I_{X'}/2}$ are different precisely because $(-1)^{\I_{\h X}/2}$ can be nontrivial
for a suitable $\h X$ without boundary.   Because $(-1)^{\I_X/2}$ and $(-1)^{\I_{X'}/2}$ are different in general,
$|Z_\psi|(-1)^{\I_X/2}$ does depend on $X$.  This $X$-dependence is the obstruction to defining the $2+1$-dimensional Majorana
fermion by itself in a $\sT$-invariant way.

That is the basis for saying that the nontriviality of $(-1)^{\I_X/2}$ for some $X$, even though it is trivial for mapping tori,
represents an anomaly for the $2+1$-dimensional Majorana fermion.  

Viewed from the bulk point of view, the significance of the fact that $(-1)^{\I_X/2}$ can be nontrivial  may seem more
obvious.  If $(-1)^{\I_X/2}$ were equal to 1 for all $X$ without boundary, then the corresponding sTQFT would be completely
trivial, and we could define it to be still trivial if $X$ has a boundary.  Boundary fermions would play no essential role.

In section \ref{complex}, we will reconsider the 3d topological insulator and superconductor on unorientable manifolds.
We will see that this does not change much for the topological insulator, but for the topological superconductor it does.  What we have
said here about the topological superconductor  is only a small part of the full story, but treating this part by itself gave us
the chance to explain some important points in a simple context.

\subsection{Overview Of The Remaining Cases}\label{overview}

Before getting into any details, here we will give an overview of the rest of this paper.
We will discuss the remaining cases of real and complex fermions in a way very similar to the way that we have analyzed
pseudoreal fermions.

In each case, there will be a ``bulk'' invariant that will play the role of the Dirac index $\iota$ or $\I$ in the cases analyzed in
section \ref{pseudo}.  For real fermions, this invariant will be $(-1)^\zeta$, where $\zeta$ is the mod 2 index of a  
Dirac operator in dimension $D$.  (This concept will be explained in section \ref{inter}.)
For complex fermions, the analog will be $\exp(-\i\pi\eta/2)$, where $\eta$ is the APS $\eta$-invariant\footnote{Typically $\exp(-\i\pi\eta/2)$
is  not a topological invariant, but if the original complex fermions have no perturbative anomaly, then the particular invariant $\exp(-\i\pi\eta/2)$ that
controls their global anomaly is a topological invariant and in fact a cobordism invariant.} of a Dirac operator in dimension $D$. In each case,
$(-1)^\zeta$ or $\exp(-\i\pi\eta/2)$ will be the partition function of a suitable sTQFT in $D$ dimensions.

However, just like $(-1)^{\I/2}$ (or its special case $(-1)^\iota$), 
the invariants $(-1)^\zeta$ and $\exp(-\i\pi\eta/2)$ are not straightforwardly defined on a manifold
$X$ with a spatial boundary.  When $X$ has a spatial boundary, something has to happen on $Y=\partial X$.  Though there are other possibilities,
the standard boundary states of topological insulators and superconductors have gapless fermions supported on the boundary.  In each case,
the fermion partition function $Z_\psi$ is not well-defined by itself.  But the product $|Z_\psi|(-1)^\zeta$ or $|Z_\psi|\exp(-\i\pi\eta/2)$ is well-defined
and compatible with all physical principles.  

Here $|Z_\psi|(-1)^\zeta$ or $|Z_\psi|\exp(-\i\pi\eta/2)$ is the partition function of the combined system consisting of gapless fermions on $\partial X$
and an sTQFT in bulk.  These combined partition functions cannot be usefully factored as the product of a partition function for the boundary
fermions and one for the bulk sTQFT.  In each case, the boundary fermions have an anomaly that compensates for the difficulty in defining
the bulk sTQFT on a manifold with boundary.

In a certain sense, the invariant $\exp(-\i\pi\eta/2)$ is universal and the more elementary invariants
$(-1)^{\I/2}$ and $(-1)^\zeta$ 
can be viewed as special cases to which $\exp(-\i\pi\eta/2)$ reduces under favorable circumstances.  It logically would have been possible
to begin this paper with a general analysis leading to $\exp(-\i\pi\eta/2)$ and then to deduce as corollaries the more easily understood formulas involving
$(-1)^{\I/2}$ or $(-1)^\zeta$, but this might have made the presentation rather opaque. 

\section{Real Fermions}\label{realf}

In this section, we consider topological states of matter in which the boundary modes, in Euclidean signature, are real
fermions.  Our basic example will be the $2+1$-dimensional $\sT$-invariant topological superconductor.  
However, we will also consider
the $2+1$-dimensional topological insulator.

\subsection{Symmetry Of The Spectrum}\label{rsf}

In two spacetime dimensions, we need only two gamma matrices.  In Euclidean signature, they should obey
$\{\g_a,\g_b\}=2\delta_{ab}$, and they can be $2\times 2$ real matrices, for example $\g_1=\sigma_1$, $\g_2=\sigma_3$.

The operator $\slashed{D}=\sum_{i=1}^2 \gamma^i D_i$ is therefore a real antisymmetric operator.  The corresponding hermitian
Dirac operator $\D=\i\slashed{D}$ is imaginary and antisymmetric. This is true even on on unorientable 2-manifold.

Consider the eigenvalue problem $\D\chi=\lambda\chi$ in this situation.  The spectrum is invariant under $\lambda\to
-\lambda$, because if $\D\chi=\lambda\chi$, then $\D\chi^*=-\lambda\chi^*$.   Note that we have made use of an antilinear operation $\T:\chi\to\chi^*$ that obeys $\T^2=1$.
By contrast, for pseudoreal fermions,  to prove
that the spectrum is doubled for each value of $\lambda$, we used in section \ref{topsup} an antilinear symmetry with $\T^2=-1$.    
 In the real case, the symmetry $\lambda\to-\lambda$
gives a pairing of eigenvalues with $\lambda\not=0$, but there may be unpaired zero-modes.  (Indeed, these would
represent a mod 2 index, which we come to in section \ref{inter}.)  For the moment we assume that generically
there are no zero-modes.

The edge modes of a $2+1$-dimensional $\sT$-invariant topological superconductor are a $1+1$-dimensional  Majorana fermion (coupled
to gravity only).  There is no $\U(1)$ symmetry and the fermion path integral is best understood as the Pfaffian of an antisymmetric bilinear form $\DD$.
The definition of $\DD$ in terms of the hermitian Dirac operator $\D$ is slightly different from what it is for pseudoreal
fermions, which were discussed in section \ref{topsup}.  Instead of $\DD_{\gamma\beta}=\veps_{\gamma\alpha}\D^\alpha_\beta$
as in that case, with $\veps_{\alpha\beta}$ an invariant antisymmetric tensor in the space of fermion fields, we now
have $\DD_{\gamma\beta}=\delta_{\gamma\alpha}\D^\alpha_\beta$, with $\delta_{\gamma\alpha}$ an invariant symmetric
tensor.

The result of this is that $\DD$ is the direct sum of $2\times 2$ blocks, with one such block for each eigenvalue pair
$\lambda_i,-\lambda_i$ of $\D$:
\be\label{molfo}\DD=\bigoplus_i \bp 0&\lambda_i\cr -\lambda_i & 0 \ep. \ee
This contrasts with the pseudoreal case, in which $\DD$ has a $2\times 2$ block for each pair of states with the {\it same}
eigenvalue $\lambda$.  

An important detail is that if we are given an antisymmetric bilinear form $\DD$, then its ``skew eigenvalues'' $\lambda_i$
are uniquely determined only up to sign.  An orthogonal transformation by the $2\times 2$ matrix $\mathrm{diag}(1,-1)$ acting
in the $i^{th}$ block 
would reverse the sign of $\lambda_i$ in eqn. (\ref{molfo}).
So {\it up to sign} (and modulo the need for regularization), the fermion path integral is
\be\label{wolfo}\Pf(\DD)=\prod_i\lambda_i.\ee

Broadly speaking, we are in a familiar situation.  To decide on the sign of $\Pfaff(\DD)$, we need to know, mod 2,
how many of the $\lambda_i$ should be considered negative. In general, such a question does not have a natural answer.
 However, we observe that as a function of the metric tensor $g$ of a 2-manifold $Y$,
$\Pfaff(\DD)$ should change sign whenever a pair of eigenvalues $\lambda_i,-\lambda_i$ passes through 0.  
So to get a partial answer, we can pick the sign of $\Pfaff(\DD)$ for some chosen
metric $g_0$ on a 2-manifold $Y$, and then determine the sign as a function of $g$ by counting how many times
an eigenvalue pair passes through 0 in interpolating from $g_0$ to $g$.

\begin{figure}
 \begin{center}
   \includegraphics[width=3in]{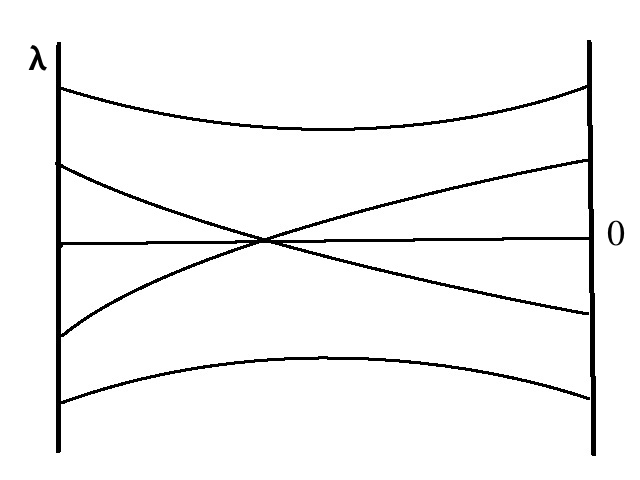}
 \end{center}
\caption{\small  Spectral flow for real fermions.  The spectrum is symmetric under $\lambda\leftrightarrow-\lambda$.
The fermion path integral changes sign whenever an eigenvalue pair flows through 0, so we count mod 2 how many
times this happens in interpolating from an initial metric $g_0$ to its transform $g_0^\phi$ under some diffeomorphism $\phi$. In
the example shown, this occurs once, so $\Pfaff(\DD)$ is odd under $\phi$.}
 \label{real}
 \end{figure}

To understand the behavior of $\Pfaff(\DD)$ under a diffeomorphism $\phi$, 
we have to count mod 2 how many times
the sign of $\Pfaff(\DD)$ changes when we interpolate, via a family of metrics $g_s$ parametrized by $s\in[0,1]$,
 from a starting metric $g_0$ to its conjugate $g_0^\phi$ under 
$\phi$ (fig. \ref{real}).   The difference from the pseudoreal case of fig. \ref{flow} is that instead of counting the net flow of eigenvalues
(or eigenvalue pairs) from $\lambda<0$ to $\lambda>0$ as $s$ increases from 0 to 1, which is an integer-valued invariant, 
now we are interested
in counting mod 2 how many times a pair of eigenvalues $\pm\lambda$ passes through 0.  It is easily seen that this is only
a mod 2 topological invariant.  

In section \ref{pseudo}, we exploited the fact that the
 integer-valued spectral flow invariant that arises for pseudoreal fermions in $d$ spacetime dimensions
can be expressed in terms of the index of the Dirac
operator in $D=d+1$ dimensions.  
The general version of this formula without assuming a $\U(1)$ symmetry is in eqn. (\ref{neft}).   The $\Z_2$-valued spectral
flow invariant for real fermions is likewise a Dirac index of a sort, but a less familiar sort.
It is equal to the mod 2 index \cite{AStwo} of the Dirac operator in $D$ dimensions, which we will denote as $\zeta$.  
This is a rather subtle invariant.   In the original papers on global anomalies in fermion path integrals \cite{Witold,Witoldtwo}, 
it was shown that the mod 2 spectral flow invariant for real fermions in $d$ spacetime
dimensions equals the corresponding mod 2 Dirac index
$\zeta$ in $D=d+1$ dimensions.  Thus the transformation law
of the fermion path integral is
\be\label{zott}\Pf(\DD)\to \Pf(\DD)(-1)^\zeta, \ee
where $\zeta$ is the mod 2 index of the Dirac operator on the mapping torus.
The proof is similar to the proof of eqns.
(\ref{worf}) and (\ref{neft}), which we recall reflect a relationship between spectral flow for pseudoreal fermions in $d$ spacetime
dimensions and an ordinary Dirac index in $D=d+1$ dimensions \cite{APS,DCG,Kiskis}.

\subsection{An Interlude On The Mod 2 Index}\label{inter}

The most straightforward way to define the mod 2 index physically is just to observe the following.  Suppose in general that in $D$ spacetime
dimensions, we can write an action for some system of fermions:
\be\label{facte}I=\int\d^Dx \bar\psi\DD\psi. \ee
The only universal property of $\DD$ is that it can be viewed as an antisymmetric bilinear form -- antisymmetric because of fermi statistics.
The canonical form of an antisymmetric bilinear form  $\DD$ is a direct sum of $2\times 2$ blocks with some zero eigenvalues:
\be\label{conform}\bp 0&a_1& && && \cr
                                   -a_1 & 0 &&&& \cr
                                    && 0 & a_2 &&&\cr
                                     && -a_2&0 &&&\cr
                                     &&&& \ddots && \cr
                                     &&&&&0 & \cr
                                      &&&&&&0 \ep. \ee
The number of zero-modes of $\DD$ is a topological invariant mod 2, since as one varies the parameters $a_1,a_2,\dots$ that appear in the nonzero
blocks, zero-modes can only appear or disappear in pairs.  The number of zero-modes mod 2
is called the mod 2 index.  (As an example of this definition, the reader can see that
if $\Omega$ is  an antisymmetric bilinear form on a finite-dimensional
vector space $V$, then its mod 2 index is simply the dimension of $V$ mod 2.)

Here are the simplest examples of the mod 2 index in low dimension. (In all these examples, we consider fermions coupled to gravity only.)
 In 1 dimension, there is only 1 gamma matrix $\g_1$ obeying $\g_1^2=1$,
so we represent it as $\g_1=1$ and consider a 1-component fermion field $\psi$.  A compact 1-manifold without boundary is a circle $S^1$.
There are two possible spin structures, depending whether $\psi$ obeys periodic or antiperiodic boundary conditions.  The Dirac equation is
$\d\psi/\d t=0$, and there is 1 zero-mode in the periodic case and none in the antiperiodic case.  So the mod 2 index is 1 or 0 for periodic
or antiperiodic fermions.  

In 2 dimensions, we can consider a 1-component chiral fermion $\psi$ coupled to gravity only
(the edge mode of a chiral superconductor in $2+1$ dimensions),
which can only be formulated on an {\it orientable} 2-manifold.
An example of a compact orientable 2-manifold is 
a torus $T^2$, for example the quotient of the complex $z$-plane by 
\be\label{zp}z\cong z+1\cong z+\i.\ee The Dirac equation
for a chiral fermion is $\partial \psi/\partial \bar z=0$.  There are 4 spin structures depending on whether $\psi$ is periodic or antiperiodic
under $z\to z+1$ and under $z\to z+\i$.  In the completely periodic case, there is 1 zero-mode and in the other cases there are none.
So the mod 2 index is nonzero precisely in the completely periodic spin structure.

What about a Majorana fermion $\psi$ in 2 dimensions?    On an orientable 2-manifold, $\psi$ can be decomposed in components $\psi_+$ and $\psi_-$
of positive and negative chirality, and the Dirac equation $\slashed{D}\psi=0$ for a zero-mode splits as separate equations for $\psi_+ $ and $\psi_-$.
The complex conjugate of a $\psi_+$ zero-mode is a $\psi_-$ zero-mode, so the number of zero-modes of $\psi$ is trivially even and the mod 2
index inevitably vanishes.

This is not so if we formulate the Majorana fermion on an unorientable 2-manifold.  For a concrete example, we consider the Klein bottle KB
constructed as the quotient of the two-torus (\ref{zp}) by $z\to \bar z+1/2$.  Equivalently, introducing real coordinates
by setting $z=x_1+\i x_2$, the relations are
\be\label{rp}(x_1,x_2)\cong (x_1+1,x_2)\cong (x_1,x_2+1)\cong (x_1+1/2,-x_2).  \ee
We define a $\pin^+$  structure\footnote{ This concept, which is the analog of a spin structure on an unorientable manifold,
 is described in Appendix \ref{unor}.}
 on KB by requiring
\be\label{tp}\psi(x_1,x_2)=\psi(x_1+1,x_2)=\psi(x_1,x_2+1)=\g_2\psi(x_1+1/2,-x_2).\ee
One of the two zero-modes that $\psi$ would have on $T^2$ is projected out when we go to KB, so with this $\pin^+$ structure, the mod 2 index of $\psi$ is
1.

For our last example, we consider a two-component Majorana fermion $\psi$ in $D=3$ spacetime dimensions.  
There is a Dirac action for such a
system, so it is possible to define a mod 2 index.  This mod 2 index always vanishes on an orientable 
3-manifold $X$.  Indeed, on an orientable
3-manifold, the eigenvalues of the Dirac operator always have even multiplicity, as we showed earlier using the antilinear 
operation defined in eqn.
(\ref{norg}).  However, on an unorientable 3-manifold, the mod 2 index can be nontrivial.  A simple example is $X=\KB\times S^1$,
where $\KB$ and the spin structure on it were defined in eqns. (\ref{rp}), (\ref{tp}), and we take the periodic spin structure on $S^1$.  
The only zero-mode of $\psi$ is a mode that is constant in the $S^1$ direction and whose restriction to KB is the zero-mode found in the last
paragraph.  So the mod 2 index is 1.  

We conclude this introduction to the mod 2 index by stressing that despite its name, the mod 2 
index is not the mod 2 reduction of an ordinary index
or indeed of any integer-valued invariant.  In fact, we have given examples in 1 and 3 dimensions 
of a nontrivial mod 2 index, but the ordinary index
always vanishes in odd dimensions.    The mod 2 version of the Atiyah-Singer index theorem \cite{AStwo} does not give
a formula for the mod 2 index,\footnote{To be more precise, the formula expresses the mod 2 index in terms of K-theory but does not give
a formula in a conventional sense as an integral.} analogous to the familiar Atiyah-Singer index formula for the ordinary index.  It does imply
that the mod 2 index is a cobordism invariant, so in particular 
there is a 3d sTQFT (defined on spin or more precisely $\pin^+$ manifolds)
with partition function $(-1)^\zeta$.   Before discussing what happens when this theory is considered on a manifold
with boundary, we pause for some simple examples of the global anomaly of a 2d Majorana fermion.

\subsection{Simple Examples Of The Global Anomaly}\label{simpex}

We learned in section \ref{inter} that in $D=3$, the mod 2 index of the Dirac operator is always zero on an orientable manifold, 
but that on  an unorientable manifold the mod 2
index of a Majorana fermion coupled to gravity only can be nonzero.

This should mean that in 2 spacetime dimensions, a Majorana fermion $\psi$ is consistent as 
long as we formulate it only on an orientable 2-manifold and only allow orientation-preserving diffeomorphisms,
but   becomes anomalous otherwise.

The anomaly if we formulate the Majorana fermion $\psi$ on an unorientable 2-manifold or if we allow orientation-reversing
symmetries can be seen by elementary examples that do not really require the general formalism.  First 
let us consider the $\psi$-field on the Klein bottle KB, which
we take as an example of an unorientable 2-manifold.  As we have seen above, with a suitable pin structure on KB, the mod
2 index is nonzero.  This means that generically the Dirac operator has a single $c$-number zero-mode $\chi_0$.  Other modes
are paired in a natural way under the symmetry $\lambda\leftrightarrow-\lambda$ of the spectrum.  In such a situation, the measure
in the fermion path integral is odd under the symmetry $(-1)^F$ that counts fermions mod 2.  Concretely, the fermion partition
function on KB vanishes because of the zero-mode, but the (unnormalized)
one-point function $\langle \int\d^2x\sqrt g\bar\chi_0\psi\rangle$
-- with an operator insertion that removes the zero-mode -- is nonzero.  Since this correlation function has an odd number of fermion insertions,
it is odd under $(-1)^F$.

Clearly, we can understand this example without thinking about the mapping torus, but how would we understand it in terms of the mapping torus construction?
Any diffeomorphism of a manifold $Y$ that can act on
fermions at all can act in two different ways, differing by an overall sign.  In particular, the trivial diffeomorphism of $Y$ can be taken to act trivially on fermions
or to act as multiplication by $-1$.  If we lift the trivial diffeomorphism of $Y$ to act trivially on fermions, we get the trivial symmetry 1, but if we take it to act
as $-1$ on fermions, we get the symmetry $(-1)^F$.   So we can think of the trivial symmetry 1 or the nontrivial symmetry $(-1)^F$ as special cases of
a diffeomorphism, and apply the mapping torus construction.

The mapping torus associated to the trivial diffeomorphism of $\KB$ is just the product $\KB\times S^1$.  The two lifts of the trivial diffeomorphism
to act on fermions correspond to two different $\pin^+$ structures in which the fermions are periodic or antiperiodic around the $S^1$ direction.  (The $\pin^+$
structure on $\KB$ is arbitrary and is kept fixed in this discussion.)  The trivial symmetry 1 and the symmetry $(-1)^F$ correspond
to the antiperiodic and periodic $\pin^+$ structures, respectively.  This is analogous to the fact that in statistical mechanics, with Hamiltonian $H$, 
to compute a trace
$\Tr\,e^{-\beta H}$ or $\Tr\,(-1)^Fe^{-\beta H}$, we do a path integral on a circle of circumference $\beta$ with fermions that are respectively antiperiodic
or periodic around the circle. Now let us compute the mod 2 index on $\KB\times S^1$ with our two $\pin^+$ structures.
 If the fermions are antiperiodic on $S^1$, corresponding to the trivial symmetry 1, then the lowest eigenvalue of the Dirac operator
on $\KB\times S^1$ is $\pi/\beta$ (or larger), where again $\beta$ is the circumference of the circle.  There are no zero-modes, so
the mod 2 index of the mapping cylinder vanishes, implying that the path integral measure on $\KB$ is invariant
under the trivial symmetry.  But if the fermions  are periodic on $S^1$, corresponding to the symmetry $(-1)^F$, then as we have seen in section \ref{inter},
the mod 2 index on $\KB\times S^1$ is the same as it is on $\KB$.  This means that precisely when the mod 2 index on $\KB$ is nonzero, the path integral
measure on $\KB$ is odd under $(-1)^F$.   

  In these statements, 
 $\KB$ could be replaced with any two-manifold $Y$.  Whenever there is a nontrivial mod 2 index on $Y$,
 and thus an anomaly in $(-1)^F$, this anomaly shows up in the mod 2 index on the mapping torus $Y\times S^1$.   
  
 For another simple example, let us consider the Majorana fermion $\psi$ on the orientable 2-manifold $T^2$.  But now we consider
 a diffeomorphism $\phi$ that acts as a reflection of one coordinate, reversing the orientation of $T^2$.  Taking on $T^2$ the
 spin structure that is periodic in both directions, $\psi$ has 2 zero-modes, say $\chi_1$ and $\chi_2$.  Because of these
 zero-modes, the partition function of the $\psi$-field on $T^2$ vanishes, but the (unnormalized) 2-point function
 \be\label{tolf}\left\langle\int_Y\d^2x \sqrt g \bar\chi_1\psi\cdot \int_Y\d^2x\sqrt g\bar\chi_2\psi\right\rangle\ee
 is nonzero.  Of the two zero-modes, one linear combination is even under $\phi$ and one is odd, so the matrix element
 in eqn. (\ref{tolf}) is odd under $\phi$, showing that the path integral measure is odd under $\phi$.
 
 Let us again try to understand this global anomaly in the mapping torus construction.  We take for $T^2$ the quotient
 of the $x_2-x_3$ plane by $(x_2,x_3)\cong (x_2+1,x_3)\cong (x_2,x_3+1)$, with the Majorana fermion field $\psi$
 assumed to be periodic in both directions.  We consider $\phi$ to act by
 $(x_2,x_3)\to (-x_2,x_3)$.  To construct the mapping torus associated to $\phi$, 
 we introduce a copy of $\R$ parametrized by another variable $x_1$ and consider the quotient of $\R\times T^2$ 
 by $(x_1,x_2,x_3)\to (x_1+1/2,-x_2,x_3)$.  
 But this is just another way to construct the familiar 3-manifold $\KB\times S^1$.  We must further say how we want $\phi$
 to act on fermions. We do this by saying that a fermion field $\psi $ on $\KB\times S^1$ is a fermion field on $\R\times T^2$, periodic in the $T^2$
 directions,
 that obeys $\psi(x_1+1/2,-x_2,x_3)=\g_2\psi(x_1,x_2,x_3)$.  But this gives the $\pin^+$ structure on $\KB\times S^1$ that was
 already considered in section \ref{inter}.  As explained there, the mod 2 index with this $\pin^+$ structure is nonzero, so again
 the mapping torus construction detects the global anomaly in reflection symmetry of $T^2$.
 
\subsection{Bulk sTQFT Associated To the $\sT$-Invariant Topological Superconductor}\label{bulkboundary}

Now let us consider the sTQFT whose partition function on a closed 3-manifold $X$ with  a $\pin^+$ structure is $(-1)^\zeta$.
This theory is trivial if $X$ is orientable, but in general not otherwise.

Our first task is to explain why this sTQFT is the bulk state of a $\sT$-conserving topological superconductor in 
$D=3$ spacetime dimensions.
The style of the argument should be familiar from section \ref{why}.

We start in Lorentz signature with two 3d fermions $\psi_1$, $\psi_2$ that transform oppositely\footnote{In eqn. (\ref{worx}), we wrote
a massive Dirac equation for a pair of 3d fermions $\psi_1$, $\psi_2$ assumed to transform the same way under $\sT$.   These,
however, were complex Dirac fermions, and the factor of $\i$ multiplying the mass term in eqn. (\ref{worx}) was necessary for
$\sT$-invariance.  For Majorana fermions, the Dirac equation must be real, so this factor of $\i$ is not possible; a $\sT$-invariant
mass term linking two fermions $\psi_1,\psi_2$ is possible only if they transform with opposite signs under $\sT$. }
  under $\sT$:
\be\label{molf}\sT\psi_1(t,x_1,x_2)=\g_0\psi_1(-t,x_1,x_2),~~~\sT\psi_2(t,x_1,x_2)=-\g_0\psi_2(-t,x_1,x_2).\ee
The opposite signs are chosen so that it is possible to have a $\sT$-conserving off-diagonal mass term $m\bar\psi_1\psi_2$.
(Diagonal mass terms $m'\bar\psi_1\psi_1$ or $m''\bar\psi_2\psi_2$ are $\sT$-violating.)   The phase transition between a topologically
trivial  and a topologically nontrivial
 insulator is achieved when $m$ changes sign.  Now we study as a function of $m$ the partition function 
of this system on a compact manifold $X$ of Euclidean signature.  Suppose that $X$ and the $\pin^+$ structure on $X$ are such
that $\zeta\not=0$.  Then setting $m=0$, both $\psi_1$ and $\psi_2$ have an odd number of zero-modes (generically 1 each).  It
takes an odd number of insertions of $m\bar\psi_1\psi_2$ to lift these zero-modes, and therefore the partition function is proportional
to an odd power of $m$; thus, it changes sign in passing through $m=0$ whenever $\zeta\not=0$.  There is no such sign change
if $\zeta=0$.   So if the partition function is positive-definite for, say, $m>0$ then its sign is $(-1)^\zeta$ for $m<0$.

A more precise explanation is as follows.  Since $\psi_1$ and $\psi_2$ transform oppositely under
an orientation-reversing symmetry, they combine together to a single Majorana fermion $\h\psi$ on an oriented three-manifold $\h X$
that arises as the oriented double cover of $X$.  The partition function of the $\psi_1-\psi_2$ system on $X$ is the same as the
$\h \psi$ partition function on $\h X$.  To be precise about this partition function, we add (rather as in section \ref{zelfan})
Pauli-Villars regulator fields $\chi_1,\chi_2$ with opposite statistics to $\psi_1,\psi_2$ and a $\sT$-conserving mass term 
$\mu\bar\chi_1\chi_2$.  On $\h X$, these combine to a single regulator field $\h\chi$.  $\h X$ has an orientation-reversing symmetry
(the quotient of $\h X$ by this symmetry is the original $X$).  The Dirac operator on $\h X$ anticommutes with such a symmetry
(see the discussion of eqn. (\ref{refact})),  
and hence the non-zero eigenvalues of the Dirac operator on $\h X$ are paired under
$\lambda\to-\lambda$.  (There is also a 
doubling of the spectrum on $\hat X$ for each value of $\lambda$, because of the antilinear symmetry (\ref{norg}), but as
in eqn. (\ref{rf}), a Majorana fermion partition function is computed by taking only one eigenvalue from each such pair.)
With $\h\psi$ having mass $m$ and $\h\chi$ having mass $\mu$, the contribution of a pair $\lambda,-\lambda$ to the fermion path
integral is
\be\label{zeffo}\frac{\lambda+\i m}{\lambda+\i \mu}~~\frac{-\lambda+\i m}{-\lambda+\i \mu}. \ee
This is always positive, regardless of the signs of $\mu$ and $m$, so the contribution of nonzero modes to the regularized
path integral is always positive.  Now let us suppose that the massless Dirac operator on $X$
has $\upiota$ zero-modes (with $\upiota\cong\zeta$ mod 2).  Then the field $\h\psi$ on $\h X$ has $2\upiota$ zero modes.  Since a mass
insertion  can lift two zero modes of $\h\psi$ or $\h\chi$, every  pair of such zero-modes
gives a factor of $m$ in the $\h\psi$ path integral and a factor of $1/\mu$ in the regulator path integral.  So the path integral
is proportional to $(m/\mu)^\upiota$; it is positive-definite if $m$ and $\mu$ have the same sign, and its sign is $(-1)^\zeta$     
otherwise.

The case that $m$ and $\mu$ have the same sign corresponds to a topologically trivial $\sT$-invariant superconductor.
The topologically non-trivial case is the case that $m$ and $\mu$ have opposite signs.  
 This second case is the topologically interesting one and $(-1)^\zeta$ is the partition function of the
bulk sTQFT associated to it.

\subsection{Extending the sTQFT To A Manifold With Boundary}\label{extending}

We would like to extend this sTQFT to the case that $X$ has a spatial boundary $Y$.  For this, we need to extend the definition of
the mod 2 index $\zeta$ to the case that $X$ has such a boundary.  However, there is no  symmetry-preserving local boundary
condition that will let us define $\zeta$.

Let us look carefully at this crucial point.  If $X$ is orientable, then its boundary $Y$ is also orientable, and
it is possible to define a local boundary condition on the Dirac equation along the lines of eqn. (\ref{torz}).  Let $\vec n$ be the unit normal vector
to $Y$.  Then we can impose the boundary condition
\begin{equation}\label{tumf}\gamma\cdot n\,\psi|_{Y}=\pm \psi|_Y,\end{equation}
with some choice of sign, and these are the most general possible covariant boundary conditions.\footnote{A local boundary condition
must set to zero one-half of the fermion components along $Y=\partial X$, and eqn. (\ref{tumf}) is the most general way to do this
that is invariant under rotation of the normal plane at a point $p\in Y$.  Any other matrix that we might use instead of
$\gamma\cdot n$ would violate the local rotation symmetry.}   With either choice of sign,
this is a physically sensible classical boundary condition,\footnote{Quantum mechanically,
a $D=3$ fermion with such a boundary condition will have a perturbative gravitational anomaly along $\partial X=Y$ \cite{HW}.  That is
not important here because we are interested in the mod 2 index of the $D=3$ fermion, not its partition function.} and with
this boundary condition, we can define the mod 2 index of the Dirac operator.  However, this boundary condition does not make
sense globally if $Y$ is unorientable.  A quick way to see this is as follows.  Starting with a two-component fermion field $\psi$,
the boundary condition (\ref{tumf}) selects one component that can be nonzero along $Y$, and it is not difficult to see\footnote{In a local
Euclidean frame, we can take the three gamma-matrices to be $\g_a=\sigma_a$, $a=1,\dots,3$, obeying $\g_1\g_2\g_3=\i$.  
Let us assume that the normal
direction to the boundary is the 3 direction. The 2d chirality along the boundary is measured by $\i\g_1\g_2=-\g_3$, so the condition
(\ref{tumf}) on the eigenvalue of $\slashed{n}=\g_3$ selects one 2d chirality.} that this
component has a definite chirality in the 2-dimensional sense.  But the chirality cannot be defined globally along $Y$ if $Y$ is unorientable,
so the boundary condition (\ref{tumf}) does not make sense globally.  

If $Y$ is orientable and we pick an orientation, then the boundary condition (\ref{tumf}) with some choice of sign
makes sense.  Defining $\zeta$ to be the mod 2 index of the Dirac operator on $X$ with some choice of sign, we can define a 3d sTQFT
on a 3-manifold with boundary whose partition function is $(-1)^\zeta$.  This gives a gapped and topologically unordered
but {\it symmetry-breaking} boundary
condition for the 3d sTQFT.  It is symmetry-breaking because the choice of sign in eqn. (\ref{tumf}), which involves a choice of
one  2d chirality, breaks reflection or time-reversal symmetry along the boundary.

We can define $\zeta$ in a $\sT$-invariant way if we use the global APS boundary conditions. 
Concretely this amounts to replacing $X$ by the noncompact manifold $X'$ of fig. \ref{longer}, and defining $\zeta$ to be the dimension
mod 2 of the space of {\it square-integrable} zero-modes of the Dirac operator.
This gives a $\sT$-invariant definition of
$\zeta$, but $\zeta$ defined this way is not a topological invariant.  When the metric on $Y=\partial X$ is varied so that
a pair of eigenvalues of the 2d Dirac operator on $Y$ passes through 0, the APS boundary conditions on $X$ jump and $\zeta$ also
jumps.  

At this point $(-1)^\zeta$ jumps in sign, so a bulk theory with partition function $(-1)^\zeta$ is not physically sensible by itself
in this situation.  However, hopefully it is clear at this point what we should say.  The bulk theory with partition function $(-1)^\zeta$
has to be combined with a boundary system consisting of Majorana fermions on the boundary.   The partition function of the combined
system is
\be\label{overdo}|Z_\psi|(-1)^\zeta.\ee
Here $(-1)^\zeta$ jumps whenever a pair of eigenvalues 
$\pm \lambda$ of the boundary fermions passes through $\lambda=0$, that is,
whenever the path integral of the boundary fermions would be expected to change sign. So the product is physically sensible.

\subsection{The Orientable Case And Chiral Symmetry}\label{orch}

Let us now look more closely at the case that $Y$ is orientable.  In this case, subject to one restriction that we mention
in a moment,  we can assume that $Y$ is the boundary of an orientable spin manifold $X$.
This manifold has $\zeta=0$.  Indeed, the argument given in section \ref{inter} showing that $\zeta=0$ if $X$ is orientable remains valid
with $\partial X\not=0$, as long as we use APS boundary conditions.   

Since $(-1)^\zeta$ is always 1 in this situation, we conclude that if $Y$ is orientable, we can sensibly define
the Majorana fermion partition function on $Y$ to be positive definite. A more direct explanation of this is to observe that if $Y$ is
orientable, then the Majorana fermion $\psi$ on $Y$ can be decomposed as a sum of fields $\psi_+$ and $\psi_-$ of positive
and negative chirality; moreover, the action and path integrals of $\psi_+$ and $\psi_-$ are complex conjugates.  So the overall
path integral is positive. 

As long as $Y$ and $X$ are orientable, the formula (\ref{overdo}) for the partition function does not depend on $X$ -- since $\zeta=0$ 
for any  $X$ -- so it seems
that we can forget about $X$ and we get a definition of the massless Majorana fermion theory on $Y$ as a purely 2-dimensional
theory.   However, there is a detail that requires some care.  Until this point, we have only assumed that $Y$ is orientable, but we have
not actually had to pick an orientation.  Momentarily, we will see that to give a complete definition of the Majorana fermion theory on $Y$
requires picking an orientation of $Y$, and therefore allowing only orientation-preserving diffeomorphisms as symmetries.  This should
not come as a surprise, because to analyze an orientation-reversing symmetry of $Y$, the mapping torus construction
will tell us to think about the mod 2 index on an unorientable 3-manifold $X$, and this can be nontrivial, as we discussed in an example in section \ref{inter}.

The reason that some care is needed is that, 
for given $Y$, with a given spin structure, a suitable $X$ may not
exist.  The obstruction to existence of $X$ arises precisely if $Y$ is such that the mod 2 index $\zeta_Y$ for a positive chirality fermion $\psi_+$
on $Y$ 
is not zero.  (By complex conjugation, $\zeta_Y$ is also the mod 2 index of a negative chirality fermion $\psi_-$ on $Y$.)   This mod 2 index
 is a cobordism invariant, so if it is nonzero on $Y$, then $Y$ with its given spin structure is not the boundary
of an orientable spin manifold $X$.

Suppose
that this index is nonzero on $Y$.  Then $\psi_+$ and $\psi_-$ each have an odd number of zero-modes; for simplicity, let us assume
that this number is 1, the generic value, and write $\psi_{+,0}$ and $\psi_{-,0}$ for the coefficients of the zero-modes in a mode
expansion of $\psi$.
The path integral over nonzero fermion modes on $Y$ is naturally positive, as the $\psi_-$
path integral is the complex conjugate of the $\psi_+$ path integral.  Since  $\psi_{+,0}$ and
$\psi_{-,0}$ are coefficients of zero-modes, their path integral vanishes in the absence of operator insertions, so the question
we should ask is how to define the path integral measure.  But there is no natural way to define the sign of this measure, unless
an orientation of $Y$ is chosen:
whether we write $\d\psi_{+,0}\d\psi_{-,0}$ or $\d\psi_{-,0}\d\psi_{+,0}$ is tantamount to a choice of orientation of $Y$.
We actually considered this situation in section \ref{simpex} for the case $Y=T^2$ and interpreted the fact that the path integral
measure on $Y$ is odd under an orientation-reversing symmetry in terms of the mapping torus construction.

We conclude this discussion with the following remark.  Although we have taken the 2d Majorana fermion as our basic
example of real fermions, the special behavior for orientable $Y$ that we have just found reflects the fact that this  example is
somewhat special.  The classification of representations of a compact group $K$ as being real, pseudoreal, or complex properly refers
to irreducible representations, which either are real (they admit an invariant symmetric bilinear form), pseudoreal (they admit
an invariant antisymmetric form), or complex (they admit no invariant bilinear form).  Reducible representation do not have
this classification and a reducible representation may admit both a symmetric bilinear form and an antisymmetric one.
That is the case with the $1+1$-dimensional Majorana fermion on an orientable spacetime $Y$.  The spinor representation of
$\Spin(2)$ is not irreducible but decomposes as the sum of components of positive or negative chirality.  If $b$ is a bosonic
spinor of $\Spin(2)$ with chiral components $b_+$ and $b_-$, then there exist both an invariant symmetric form $\d b_+\otimes \d b_-$
and an invariant antisymmetric form $\d b_+\wedge \d b_-$.  By contrast, if $Y$ is unorientable, then we have to consider the Majorana
fermion as a representation of $\Pin^+(2)$.  Only the symmetric form and not the antisymmetric one is invariant under the reflections in $\Pin^+(2)$
(which exchange $b_+$ and $b_-$), so as a representation of $\Pin^+(2)$, the Majorana fermion is real and not pseudoreal.
  So that case
is a better example of the general paradigm of real fermions.

There are many interesting examples of real fermions in which this issue does not arise (the Euclidean space fermions have a real structure
and no pseudoreal one). 
One such case is the original example in which the mod 2 index was interpreted as a global anomaly for real fermions \cite{Witold}, namely\footnote{The model was pointed out to the author by S. Coleman in Aspen in the summer of 1976.  It is difficult to recall what
his view  was at the time.}
an $\SU(2)$ gauge theory in four dimensions with a single multiplet of fermions transforming in the spin $1/2$ representation of $\SU(2)$.  A
more elementary example, which we discuss in section \ref{maj},
 is given by a one-component fermion on the boundary of a $D=2$ Majorana chain \cite{FK}.

\subsection{$2+1$-Dimensional Topological Insulator}\label{welfin}

We now turn to the topological insulator in $2+1$ dimensions.\footnote{Some of the following issues were treated previously in \cite{Hsieh}.}
  This is a system with $\sT$ symmetry and also with $\U(1)$ symmetry.
Crucially, $\sT$ commutes with the electric charge operator $Q$, the generator of $\U(1)$.  (At the end of this section, we consider
the contrary case of a conserved charge that is odd under $\sT$.)

The boundary modes of the $2+1$-dimensional topological insulator are a $1+1$-dimensional massless
charged Dirac fermion $\psi$, say of charge 1.  Its charge conjugate, of course, has charge $-1$.  
The Dirac fermion $\psi$ can be expanded in terms of two Majorana fermions $\psi_1$ and $\psi_2$ as
$\psi=(\psi_1+\i\psi_2)/\sqrt 2$.  The $\U(1)$ symmetry of electric charge rotations is 
\be\label{delo} \delta\psi_1=\psi_2,~~ \delta\psi_2=-\psi_1.\ee
Since this formula is real and the $1+1$-dimensional Euclidean gamma matrices can also be chosen to be real,
we conclude that the boundary fermions of the $2+1$-dimensional topological insulator are real in Euclidean signature.
(A remark made in  section \ref{orch} also applies here, for essentially the same reason: 
on an orientable 2-manifold $Y$, these fermions
can be given a pseudoreal structure as well as a real one.)

Given that the fermions are real, more or less everything we have said about the $2+1$-dimensional superconductor
has an analog for the topological insulator.  On a closed 3-manifold $X$, the sTQFT associated to the topological insulator
has partition function $(-1)^\zeta$, where $\zeta$ is the mod 2 index that must be computed in the space of all fermion components
(meaning that in computing $\zeta$, we have to include both $\psi_1$ and $\psi_2$ or equivalently both charge eigenstates $(\psi_1\pm \i\psi_2)/\sqrt 2$).  If $X$ has a boundary $Y$, then $(-1)^\zeta$ cannot be defined as a topological invariant, without breaking the
time-reversal or reflection symmetry.  We can preserve the symmetry in defining $(-1)^\zeta$ if we use APS boundary conditions,
but the result is not a topological invariant.  It jumps in a way that only makes physical sense if we also include the boundary fermions,
and then the partition function of the combined system is as usual  $|Z_\psi|(-1)^\zeta$.

But is the invariant $(-1)^\zeta$ nontrivial in this situation?  Here we run into an essential novelty.  Because the $\sT$ transformation
that is a symmetry of a topological insulator commutes with the electric charge generator $Q$, it  really does correspond
in relativistic terms to $\sT$ and not to $\sCT$.  In a relativistic theory, there is a $\sCRT$ theorem, so $\sT$ symmetry is equivalent
to $\sCR$ symmetry. In Euclidean signature, it is more natural to work with $\sCR$ (which upon analytic continuation back to Lorentz
signature implies both $\sT$ and $\sCR$).   

In detail,  $\sCR$ is a symmetry that reverses the orientation of space and also {\it anticommutes} with the generator
$Q$ of electric charge.  In terms of gauge theory, the last statement means that the electromagnetic gauge potential $A$ does not
transform as a 1-form under a spatial reflection, but transforms with an extra minus sign. For example, the reflection $\sR:(x_1,x_2,x_3)\to
(-x_1,x_2,x_3)$, combined with charge conjugation $\sC:A\to -A$ gives the combined $\sCR $ transformation
 \be\label{beff}\sCR A_1(x_1,x_2,x_3)=A_1(-x_1,x_2,x_3),~~~\sCR A_i(x_1,x_2,x_3)=-A_i(-x_1,x_2,x_3),~~ i=2,3.\ee

Once one understands that an orientation-reversing symmetry should be combined with charge conjugation,
 it is straightforward to see that the mod 2
index $\zeta$ can be nonzero.  
We start with the two-torus $T^2$ parametrized by $0\leq x_1,x_2\leq 1$ with endpoints identified.  On $T^2$, we place
a gauge field $A$ with 1 unit of magnetic flux: $\int_{T^2}\d x_1\d x_2\, F_{12}=2\pi/e$.  Pick a basis $\psi^\pm_\pm$ for
the components of a Dirac fermion on $T^2$, where the superscript is the $\U(1)$ charge $\pm 1$ and the subscript is the
2d chirality.  Using standard facts about the 2d Dirac equation in a magnetic field, one finds that  $\psi^+_+$ and $\psi^-_-$ have 1 zero-mode each
and $\psi^+_-$,  $\psi^-_+$ have none.    Now consider a reflection $\sR:(x_1,x_2)\to (-x_1,x_2)$.  One can choose $A$ to be invariant
under  the combined operation $\sCR$ (this depends on the fact that $\int_{T^2}F$ is  odd under both $\sC$ and $\sR$ and so is 
$\sCR$-invariant).    The effect of  $\sCR$  is to exchange $\psi^+_+$ with
$\psi^-_-$, so precisely one linear combination of the two zero-modes on $T^2$ is $\sCR$-invariant.  If it were the case that $\sR$
acted freely on $T^2$, we would now consider the quotient of $T^2$ by the $\Z_2$ group generated by $\sR$ to get a 2d example
with a nonzero mod 2 index.  However, this is not the case.  Instead we take a product $T^3=T^2\times S^1$ where $S^1$ is 
a circle parametrized by $0\leq x_3\leq 1$ again with endpoints identified.  We take the gauge field $A$ to be a pullback from $T^2$,
and we define the orientation-reversing symmetry $\sR':(x_1,x_2,x_3)\to (-x_1,x_2,x_3+1/2)$.  Now $\sR'$ does act freely on $T^3$,
and the quotient $T^3/Z_2$, where $\Z_2$ is generated by $\sCR'$, is a 3d example with a nontrivial mod 2 index.

It is probably fairly clear at this point  that whatever we have said about the $D=3$ topological superconductor has an analog for the 
topological insulator.

To conclude, we will consider a system with a conserved charge $Q$ that is {\it odd} rather than even under $\sT$.
In condensed matter physics, $Q$ could be a component of the electron spin, which for some systems is approximately conserved.
A time-reversal symmetry that reverses the only pertinent conserved charge would be called $\sCT$ in relativistic language.
Then in Euclidean signature, one would have to consider an $\sR$ symmetry, rather than $\sCR$, meaning that the 
gauge field $A$ transforms under reversal of orientation as an ordinary 1-form.  In this case, one can show\footnote{In the case
of $\sR$ rather than $\sCR$ symmetry, $A$ is a connection on a $\U(1)$ bundle $\L\to X$, where $X$ is a possibly unorientable
3-manifold with $\pin^+$ structure. ($\sCR$ symmetry would mean that the structure group is $\O(2)$ rather than $\U(1)$.)
 The first Chern class $c_1(\L)$ is Poincar\'e dual to an embedded circle $U\subset X$.  Let
$\h U$ be a tubular neighborhood of $U$ in $X$. Topologically $\h U=U\times D^2$ where $D^2$ is a two-dimensional disc.  Since
we are trying to calculate a topological invariant, we can assume
that in a suitable gauge,
$A=0$  on the complement of $\h U$.
  One can construct an elementary cobordism from $X$ to the disjoint union
$X_1\cup X_2$, where $X_1$ is obtained from $X$ by cutting out $\h U$ and then gluing back in a copy of $\h U$ with $A=0$ (so that
$A=0$ everywhere on $X_1$),
and $X_2$ is obtained by embedding $\h U=U\times D^2$ in $U\times S^2$ (using the obvious embedding $D\subset S^2$, and taking $A=0$ outside $\h U$). Let $\zeta_X$, $\zeta_{X_1}$,
and $\zeta_{X_2}$ be the mod 2 indices on the indicated 3-manifolds.  By cobordism invariance of the mod 2 index, $\zeta_X=\zeta_{X_1}+
\zeta_{X_2}$.   On $X_1$,
$A=0$ so a charged Dirac fermion on $X_1$ is equivalent to 2 Majorana fermions on $X_1$ coupled to gravity only;  they contribute
equally to  $\zeta_{X_1}$, which therefore trivially vanishes.  $X_2$ is orientable, so $\zeta_{X_2}$ vanishes because of the usual antilinear symmetry. So
$\zeta_X=0$.}
that on any three-manifold $X$, and for any gauge field, $\zeta$ is universally 0.  So there is no nontrivial sTQFT with partition
function $(-1)^\zeta$.  This is consistent with the fact that in $D=3$, there is according to standard arguments no analog of a topological insulator for a $\sT$-invariant system with a $\U(1)$ symmetry whose generator $Q$ is assumed to be odd under $\sT$.  In particular, a
 Dirac fermion with a conserved
charge that is odd under $\sT$ can be given a symmetry-preserving bare mass.

\section{Complex Fermions}\label{complex}

\subsection{Overview}\label{overv}

Our basic example with complex fermions will be the topological insulator or superconductor in 3 space or 4 spacetime dimensions.
We have already treated this problem in section \ref{pseudo}, but there we considered only  orientable spacetimes.
Here we discuss what happens if the spacetime is permitted to be unorientable.    
In this case,\footnote{We consider systems with $\sT^2=(-1)^F$ so that the appropriate group is $\Pin^+(3)$ and not $\Pin^-(3)$.  See
Appendix \ref{unor}.} the boundary fermions transform in a complex representation of $\Pin^+(3)$ rather than a pseudreal representation of $\Spin(3)
\cong \SU(2)$.  Their partition function, even at a formal level, is now complex-valued, not real-valued, and we have to consider
anomalies in a complex-valued  path integral of complex fermions.

We start with the topological superconductor, which proves to be the more interesting case.
The discussion proceeds in a more or less familiar way.  First we consider the anomaly of the boundary modes, which are a $d=3$
Majorana fermion.  As usual,
this anomaly can be expressed as a topological invariant of the mapping torus.  For complex fermions in general \cite{Global}, 
the relevant invariant of the mapping torus is $\exp(-\i\pi\eta/2)$,
where $\eta$ is an APS $\eta$-invariant of a Dirac operator in one dimension more -- in our example the Dirac operator of a $D=4$
Majorana fermion. 
  Then we consider the partition function of the sTQFT associated 
to the bulk topological superconductor.  Using a standard characterization of the phase transition to topological
superconductivity in $D=4$, we show that the
partition function of the bulk sTQFT is again $\exp(-\i\pi\eta/2)$.  Like other bulk invariants we have encountered
as sTQFT partition functions, this one cannot be defined with a local, symmetry-preserving boundary condition.  We can define
it in a symmetry-preserving way using APS boundary conditions, but then it is not a topological invariant and again is not satisfactory
by itself.   But the product $|Z_\psi|\exp(-\i\pi \eta/2)$ is physically sensible, as is shown by the Dai-Freed theorem \cite{DF}.  This product should be interpreted as the combined
partition function of the boundary fermions and the bulk sTQFT.

Finally, we consider a question that does not have a close analog in cases studied earlier in this paper.  
This involves
a $D=4$ system whose boundary supports $\nu$ identical gapless Majorana fermions.
At the free fermion level, $\nu$ is an integer-valued invariant, but it has been 
argued \cite{FCV,WS,MFCV,KitTwo} that at the interacting level, $\nu$ is
only an invariant mod 16.  From the present point of view, this is true because 
in $D=4$, $\exp(-\i\pi\eta/2)$ is in general a
$16^{th}$ root of 1, and there is no satisfactory, symmetry-preserving 
definition of $\exp(-\nu\cdot \i\pi\eta/2)$ on a 4-manifold
with spatial boundary, unless $\nu$ is a multiple of 16.  

A somewhat similar explanation of the value $\nu=16$ was proposed in \cite{KTTW} based on a suggestion that the relevant sTQFT
partition function would be a cobordism invariant.   An explicit global anomaly computation \cite{HCR} showed that the boundary
theory with $\nu$ identical Majorana fermions is anomalous unless $\nu$ is an integer multiple of 8.  It is actually rather subtle
what happens for $\nu=8$.  We will show that although $\exp(-\i\pi\eta/2)$ is a $16^{th}$ root of unity in general,
it is an $8^{th}$ root of unity in the case of a mapping torus.  Thus the $\nu=8$ theory does not have a traditional anomaly
that can be detected via the mapping torus construction.    Nevertheless, for $\nu=8$ there is a more subtle problem, analogous
to what we explained in section \ref{signp} for a single Majorana fermion considered on an oriented 3-manifold only: even though
there is no anomaly in the traditional sense, there is also no natural way to specify the sign of the partition function.  In section \ref{signp}, this
was because $(-1)^{\I/2}$ is in general nontrivial, though it equals 1 for a mapping torus; in the present context, the reason is
that the analogous
statement holds for the invariant $\exp(-8\cdot\i\pi\eta/2)$.  

The $\nu=8$ problem has one more interesting property that certainly shows its subtlety.  It actually is possible to define a gapped
symmetry-preserving boundary state without topological order for the $\nu=8$ theory -- that is for the $8^{th}$ power of the $\nu=1$ sTQFT  -- 
but this boundary state is not
compatible with unitarity.

The   $d=3$ topological insulator is also related to the $\eta$-invariant by the same arguments. But in this case, there is no analog of
the integer-valued parameter $\nu$.  Hence the results are less interesting and going to an unorientable spacetime does not reveal
 much structure beyond what we explored in section \ref{pseudo}.

\subsection{Anomaly Of The Boundary Fermions}\label{anombo}

An unorientable 3-manifold $Y$ can always be constructed as a quotient $\h Y/\Z_2$, where $\h Y$ is the oriented double cover of $Y$,
and $\Z_2$ is generated by an orientation-reversing symmetry
$\tau:\h Y\to \h Y$ that obeys $\tau^2=1$.   Suppose that we are given a $\tau$-invariant spin structure $\S$ on $\h Y$.  This means in particular that
we have a notion of fermion field $\h\psi$ on $\h Y$ and an action of $\tau$ on $\h\psi$, obeying $\tau^2=1$.

In this case, we can define a $\pin^+$ structure $\P$ on $Y$, by saying that a fermion field $\psi$ on $Y$ with this $\pin^+$ structure
is a fermion field $\h\psi$ on $\h Y$ that obeys $\tau\h\psi=\h\psi$.  Moreover, every $\pin^+$ structure on $Y$ arises in this way, for some
choice of $\tau$-invariant spin structure $\S$ on $Y$.

But we can also define a ``complementary'' $\pin^+$ structure $\P'$ on $Y$, by saying that a fermion field $\psi'$ on $Y$ with this
alternative $\pin^+$ structure is a fermion field $\h\psi$ on $\h Y$ that obeys $\tau\h\psi=-\h\psi$.  Actually, the relationship between $\P$ and $\P'$
is completely reciprocal and neither of them is preferred.  When are given a geometrical symmetry $\tau:\h Y\to \h Y$ and a spin structure $\S\to
\h Y$ that is $\tau$-invariant up to isomorphism, this does not tell us the sign with which $\tau$ should be taken to act on $\S$.  The choice
of sign determines what we mean by $\P$ and what we mean by $\P'$.

On $\h Y$, we can define the usual hermitian Dirac operator $\D=\i\slashed{D}=\i\sum_k\g^k D_k$.  However, in odd dimensions, the Dirac
operator anticommutes with an orientation-reversing symmetry.   This point was illustrated in eqn. 
(\ref{refact}), and, as explained there, is the reason that a mass term added to the Dirac equation in $2+1$ dimensions
would violate reflection symmetry.   

In particular, $\D$ is odd under $\tau$, so if $\tau\psi=\pm\psi$, then $\tau(\D\psi)=\mp\D\psi$.  Interpreted on $Y$, this
statement means that there is no self-adjoint Dirac operator acting on a fermion field $\psi$ that is a section of $\P$,
or on a fermion field $\psi'$ that is a section of $\P'$.  Rather, the natural Dirac operator on $Y$ maps $\P$ to $\P'$ and $\P'$ to $\P$.
The operators $\D_\P:\P\to \P'$ and $\D_{\P'}:\P'\to \P$ are adjoints, so we can combine them to a self-adjoint operator which in the basis
$\bp \P\cr \P'\ep$ is
\be\label{intr}\D=\bp 0 & \D_{\P'}\cr \D_\P & 0 \ep. \ee
In fact, $\D$ is just the usual self-adjoint Dirac operator of $\h Y$ (for spin structure $\S$), acting on a spinor field $\h\psi$ that has
been reinterpreted as a pair  $\psi, $ $\psi'$ on $Y$.

More fundamentally, the reason that there is no self-adjoint Euclidean signature Dirac operator for $\psi$ or for $\psi'$ is that in Euclidean
signature these are complex fermions,
not real or pseudoreal ones.  As explained in Appendix \ref{wwod}, the
 group $\Pin^+(3)$ has two different two-dimensional representations, which are complex conjugates
of each other.   The fields $\psi$ and $\psi'$ are associated to these two different representations.

Although we cannot write a self-adjoint Dirac operator for $\psi$ or $\psi'$ only, there is no problem with the classical Dirac action
\be\label{terb}\int\d^3x \sqrt g\bar\psi\i\slashed{D}\psi \ee
for $\psi$ (or likewise for $\psi'$) only.  Recall that, as discussed in section \ref{ferp}, the Dirac action
really involves an antisymmetric bilinear form $\DD$ rather than an operator $\D$.  Up on $\h Y$, the relation between the two
is $\DD_{\gamma\beta}=\veps_{\gamma\alpha}\D^\alpha_\beta$, where $\veps_{\gamma\beta}$ is the antisymmetric bilinear form
on the 2-dimensional representation of $\Spin(3)=\SU(2)$.  But this form is odd under $\tau$ (more generally, it is odd under the disconnected
component of $\Pin^+(3)$).  So as $\D^\alpha_\beta$ is also odd under $\tau$, it follows that $\DD_{\gamma\beta}$ is even, and thus
can be defined only for the $\tau$-even fermion $\psi$ or the $\tau$-odd fermion $\psi'$.

All of what we have said has an analog that is probably much more familiar on an orientable manifold of, say, dimension 2.
In 2 dimensions, we can consider chiral fermions $\psi_+$ and $\psi_-$ which in Euclidean signature 
are complex conjugates of each other.  The 2d Dirac operator maps $\psi_+$ to $\psi_-$ and $\psi_-$ to $\psi_+$,
so to write a self-adjoint Dirac operator one has to combine the two chiralities.  But the Dirac action pairs $\psi_+$ with $\psi_+$ and
$\psi_-$ with $\psi_-$, so (modulo questions of anomaly cancellation) it makes sense to consider a theory with one chirality only
-- such as the edge modes of a $2+1$-dimensional chiral  topological superconductor.

Returning to the fields $\psi$ and $\psi'$ on a possibly unorientable 3-manifold,
the Dirac action and path integral for $\psi'$ are the complex conjugates of those for $\psi$, so the 
two partition functions $Z_\psi$ and $Z_{\psi'}$
are complex conjugates.  In fact, concretely $\psi$ and $\psi'$ combine to the fermion $\h\psi$ on $\h Y$, 
so formally $Z_\psi Z_{\psi'}=Z_{\h\psi}$, and we claim that $Z_{\h\psi}\geq 0$.
As usual, we have
$Z_{\h\psi}=|Z_{\h\psi}|\exp(-\i\pi\eta/4)$.  However, as the Dirac operator $\D$ of $\h Y$ is odd under $\tau$, its spectrum is
symmetric under $\lambda \leftrightarrow -\lambda$, so nonzero modes on $\h Y$ do not contribute to $\eta$.  So either
there are no zero-modes, in which case  $\eta=0$ and $Z_{\h\psi}>0$, or there are zero-modes and $Z_{\h\psi}=0$.  

While the product $Z_\psi Z_{\psi'}$ is positive and in particular is completely anomaly-free, $Z_\psi$ and $Z_{\psi'}$ separately
are complex and subject to anomalies.  They have no perturbative anomalies (in general there are no perturbative anomalies in
odd-dimensional spacetimes without boundary), but they have global anomalies.  Indeed, the global anomaly of $Z_\psi$ was computed recently
in \cite{HCR} in a particular example.   This is analogous to the fact that in the more familiar case of the chiral fermions $\psi_+$ and
$\psi_-$ in 2 spacetime dimensions, $Z_{\psi_+}$ and $Z_{\psi_-}$ are anomalous (in this case with perturbative as well as
global anomalies), but the product $Z_{\psi_+}Z_{\psi_-}$ is nonnegative and anomaly-free.  

A general framework to calculate the global anomaly in a complex fermion path integral was described in section 4 of \cite{Global}.
Given a diffeomorphism $\phi$ of $Y$ (or a combined diffeomorphism and gauge transformation in a more general problem), one constructs as in section \ref{trada} a mapping torus $X$.  $X$ is a 4-manifold obtained by gluing together the ends of $Y\times I$, where
$I$ is the interval $0\leq s\leq 1$, via $\phi$.
We endow $X$ with a metric \be\label{dif}\d \ell^2= \d s^2+\epsilon^2 g_s,\ee where $g_s$ is an $s$-dependent familiy of metrics on $Y$
and $\epsilon$ is a suitable small parameter.  In the basis (\ref{intr}), one introduces a new gamma matrix
\be\label{mezzo}\g_s=\bp 1 & 0 \cr 0 & -1 \ep. \ee
Then one defines a Dirac operator on $X$:
\be\label{ezzo}\D_X=\i\g_s\frac{D}{D s}+\frac{1}{\eps} \bp 0 & \D_{\P'}\cr \D_\P & 0 \ep. \ee
Here $\D_Y=\bp 0 & \D_{\P'}\cr \D_\P & 0 \ep$ is the Dirac operator on $Y$ with $s$-dependent metric $g_s$, and the explicit
factor $1/\epsilon$ comes because in (\ref{dif}), we use the metric $\eps^2 g_s$.     
The result of the analysis in \cite{Global} is to show that the change of the partition function $Z_\psi$ under $\phi$ can be expressed
in terms of the $\eta$-invariant of $D=4$ Dirac operator $\D_X$:
\be\label{wezzo}Z_\psi\to Z_\psi \exp(-\i\pi\eta/2). \ee
We will not repeat this calculation in detail and just mention a few points.  Because we are trying to compute a topological invariant,
we can consider a generic $s$-dependent family of metrics $g_s$ such that $\D_Y$ has no level
crossings as a function of $s$.  By also taking $\epsilon$ sufficiently small, one can reduce to the case that the spectrum
of $\D_X$, and therefore its $\eta$-invariant, can be computed by an adiabatic approximation.\footnote{If $\lambda_i(s)$ are the eigenvalues of $\epsilon^{-1}\D_Y$ as a function of $s$ (we assume that these are nondegenerate except for the 2-fold degeneracy that comes
from pseudoreality), the condition for the adiabatic approximation is  that
$\d\lambda_i/\d s$ should be sufficiently small compared to the relevant eigenvalue differences.
     This condition is satisfied for sufficiently small $\epsilon$, assuming there are no level crossings.}
In the adiabatic approximation, the contribution of each eigenvalue of $\D_Y$ to $\exp(-\i\pi\eta/2)$ comes essentially
from the holonomy of its Berry connection.  (The Berry connection for adiabatic evolution of a given eigenstate  is introduced, though not by that name, in the middle of p. 215
of \cite{Global}.)  In the adiabatic approximation, the same holonomy gives the contribution of each eigenvalue to the change of
phase of $Z_\psi$ under $\phi$.  

A consequence of eqn. (\ref{wezzo}) is that, as in previous examples that we have studied, the anomaly is a cobordism invariant,
meaning that it is trivial if $X$ is the boundary of a 5-manifold $Z$ over which the $\pin^+$ structure extends.  
Here we make use of eqn. (\ref{molg}).  On a 5-dimensional $\pin^+$ manifold, the index $\I$ is always even because of pseudoreality, so eqn. (\ref{molg})
implies that if $X$ is a boundary, then
$\eta$ is a multiple of 4 and hence $\exp(-\i\pi\eta/2)=1$.  

The reader who actually consults \cite{Global} and the later paper \cite{DF} in which an important refinement was made will
find that it is assumed that $Y$ has even dimension and $X$ has odd dimension.  The reason for this was simply that, if $Y$
is orientable, complex fermions on $Y$ exist only if its dimension is even.  Applications involving unorientable manifolds
were not envisioned so it was natural to assume $Y$ to have even dimension. 
However, the reasoning in these papers apply equally well to complex fermions in any dimension, even or odd.  

In terms of the applications of the results, an important difference comes from the fact that
 in even dimensions, complex fermions generically give rise to perturbative anomalies.\footnote{In \cite{Global} but not in \cite{DF},
 it is assumed at some points that perturbative anomalies cancel.}  Moreover, if $Y$
has even dimension, then $X$ has odd dimension, and if it is the boundary of some $Z$, then $Z$ again has even dimension. 
 If $\D_X$ is a Dirac operator on $X$ and $\eta$ is its $\eta$-invariant,
then typically $\exp(-\i\pi\eta/2)$ is not a cobordism invariant (or even a topological invariant) because there is a curvature
term in the APS index theorem on $Z$.  (For $Z$ of dimension 4, this curvature term is the term $\hA-P$ of eqn. (\ref{trof}).)
However, on an even-dimensional manifold $Y$, we recover cobordism invariance in the case of complex fermions  whose perturbative anomalies cancel.\footnote{An example is the Standard Model of particle physics, in $D=4$, with gauge group 
$G=\SU(3)\times \SU(2)\times \U(1)$. In Euclidean signature, the fermions
are in a complex but highly reducible representation of $\Spin(4)\times G$. This representation is such that perturbative anomalies
cancel.} This follows from the formula for $\eta$ given by the APS index theorem.
  Here of course  the global anomaly comes from an $\eta$-invariant that receives contributions from all fermion multiplets in a given theory, so in using the
APS index theorem, we  sum over all multiplets.  The curvature term in the APS index theorem on $Z$ is precisely the anomaly polynomial of the fermions, so after summing over multiplets, it cancels in a theory
that is free of perturbative anomalies.  
The APS index theorem then shows that in any theory without perturbative anomalies, in even or odd $D$,  the invariant $\exp(-\i\pi\eta/2)$ that controls
the global anomaly is a cobordism invariant.  

The relationship of $\eta$ to global anomalies is a close cousin or slight refinement of a result that is much better-known:
perturbative anomalies in $D$ dimensions are related to Chern-Simons in $D+1$ dimensions and to a curvature polynomial in $D+2$
dimensions.  In fact the APS index theorem implies (see eqns. (\ref{moj})-(\ref{oj})) a close relationship between 
$\eta$ and Chern-Simons. Usually the distinction between them is important only when one asks rather delicate questions, for example about global anomalies.

\subsection{Bulk Partition Function}\label{bpf}

Here we will compute the bulk partition function of the $3+1$-dimensional topological superconductor.  The strategy should be familiar
from sections \ref{why} and \ref{bulkboundary}. 

A standard description of the phase transition between an ordinary and a topological superconductor in $3+1$ dimensions
is that it occurs when the mass $m$ of
a $D=4$  Majorana fermion $\psi$ passes through 0.  
Even on an unorientable 4-manifold $X$, one can define a hermitian Dirac operator $\D$
(we recall from section \ref{anombo} that this is not true in 3 spacetime dimensions), and its spectrum is doubled by a version of Kramers
doubling, reflecting the fact that the Majorana fermion transforms in a pseudoreal representation of $\Pin^+(4)$.  For 
details on these statements, see Appendices \ref{do} and \ref{fwod}.  The path integral of the Majorana fermion is a Pfaffian which 
can be computed by taking one eigenvalue
from each pair.    We write $m$ for the bare mass of $\psi$ and we regulate the partition function $Z_\psi$ by including a Pauli-Villars
field of mass $\mu$ obeying the same Dirac equation but with opposite statistics.  The regularized partition function is
\be\label{regpar}Z_{\psi,\reg}=\prod_k{}'\,\frac{\lambda_k+\i m}{\lambda_k+\i\mu},  \ee
where as usual $\prod_k'$ is an instruction to take a single eigenvalue from each pair.

There are essentially two cases, depending on whether $m$ and $\mu$ have the same sign or opposite signs.
If $m$ and $\mu$ have the same sign, we get a topologically trivial superconductor.  Indeed, in the limit that $X$ is very large,
the precise values of $m$ and $\mu$ do not matter, only their signs.  If $m$ and $\mu$ have the same sign, we may as well set
$m=\mu$, and then $Z_{\psi,\reg}=1$ is completely trivial.

The interesting case is that $m$ and $\mu$ have opposite signs.  In this case, we may as well set $m=-\mu$, so
\be\label{negpar}Z_{\psi,\reg}=\prod_k{}'\,\frac{\lambda_k-\i\mu}{\lambda_k+\i\mu}. \ee
A shortcut to analyzing this expression is to factor it:
\be\label{wegpar}Z_{\psi,\reg}=\prod_k{}'\,\frac{\lambda_k}{\lambda_k+\i\mu}\cdot \prod_k{}'\,\frac{\lambda_k-\i\mu}{\lambda_k} . \ee
The second factor is the inverse complex conjugate of the first, so they contribute the same phase.  The first factor was analyzed
in section \ref{zelfan}, with a result given in eqn. (\ref{rv}).  In using this result, we have to replace $\upeta$ (computed in the
space of positive charge fermions only) with $\eta/2$ (where $\eta$ is computed using all eigenvalues of the Dirac operator and
the factor of 1/2 reflects the $\prod_k'$ symbol).  This replacement $\upeta\to \eta/2$ should be familiar from eqns.
(\ref{riv}) and (\ref{roov}).  Thus the two factors in eqn. (\ref{wegpar}) each contribute a factor $\exp(-\i\pi\eta/4)$, and  the final
result for the partition function of the bulk sTQFT of the topological superconductor is $\exp(-\i\pi\eta/2)$.

As a check, let us verify that if $X$ is orientable, this answer agrees with the result $(-1)^{\I/2}$ that was claimed in section \ref{signp}.
If $X$ is orientable, we can define the chirality operator $\bar\g$ (eqn. (\ref{chiro})) and the operation $\psi\to\bg\psi$
ensures that nonzero eigenvalues of $\D$ are paired under $\lambda\leftrightarrow-\lambda$.  (The spectrum also has
even multiplicity for each $\lambda$ because of pseudoreality,  so nonzero modes are in quartets $\lambda,\lambda,-\lambda,-\lambda.$) Pairs
$\lambda,-\lambda$ do not contribute
to $\eta$, so for $X$ orientable, $\eta$ receives contributions only from zero-modes.  Let $n_+$ and $n_-$ be the number
of zero modes of positive or negative chirality.  Each zero-mode contributes $+1$ to $\eta$ (see eqn. (\ref{uv})), so $\eta=n_++n_-$
and $\exp(-\i\pi\eta/2)=\exp(-\i\pi (n_++n_-)/2)$.  On the other hand $\I=n_+-n_-$, so $(-1)^{\I/2}=(-1)^{(n_+-n_-)/2}$.  These
are equal as $n_+$ and $n_-$ are even.  For a more direct explanation, the reader can return
to eqn. (\ref{negpar}) and see that a quartet of eigenvalues $\lambda,\lambda,-\lambda,-\lambda$ makes no contribution but
a pair of zero-modes contributes a factor of $-1$, just as in $(-1)^{\I/2}$.

Note that in the course of the argument, we have shown that if $X$ is orientable, then $\eta$ is an even integer.   But if $X$
is orientable and has an orientation-reversing symmetry, then $n_+=n_-$ and so $\eta$ is a multiple of 4.  

\subsection{Anomaly Cancellation And The Dai-Freed Theorem}\label{comb}

As usual, there is a problem in defining the partition function $\exp(-\i\pi\eta/2)$ of the bulk sTQFT on a 4-manifold $X$
with boundary.  There does not exist a local, symmetry-preserving, self-adjoint boundary condition for the 4d Dirac operator in
Euclidean signature that could be used to give a satisfactory definition of $\eta$.  We can use nonlocal APS boundary
conditions to define $\eta$ in a symmetry-preserving fashion.  But $\exp(-\i\pi\eta/2)$ defined this way is not a topological
invariant and only makes sense physically when combined with gapless modes on the boundary.

In section \ref{anombo}, we explained that the usual boundary fermions of the topological superconductor are complex fermions
with a complex path integral when formulated on an unorientable 3-manifold $Y$.  Moreover, according to the general
global anomaly formula for complex fermion path integrals \cite{Global}, the anomaly involves $\exp(-\i\pi\eta/2)$
where as in the last paragraph, $\eta$ is the $\eta$-invariant of a $D=4$ Majorana fermion.  

As in other cases that we have studied, what is physically sensible is the product
\be\label{durf}|\Pf(\DD)|\exp(-\i\pi\eta/2), \ee
where $\Pf(\DD)$ is the Pfaffian of the boundary fermions and $\eta$ is the bulk $\eta$-invariant computed with APS boundary
conditions.  Indeed, the assertion that this product is physically sensible is essentially the content of what we will call
the Dai-Freed theorem \cite{DF}, which is a refinement of the general global anomaly formula.

We will not attempt to describe the proof of the Dai-Freed theorem, but we will make a few remarks that may be helpful.
First of all, the theorem is actually phrased in \cite{DF} in a more abstract way.  To explain this, we should explain that the
usual mathematical viewpoint about anomalies is to say that although $\Pf(\DD)$ cannot be naturally defined as a complex
number, or as a complex-valued function of the metric on $Y$, there is always a unitary complex line bundle over the space
of metrics on $Y$ (or metrics plus gauge fields in a more general case) such that $\Pf(\DD)$ is canonically defined as a section
of this line bundle.  This line bundle is called the Pfaffian line bundle and we will denote it as $\PF(\DD)$.  The Dai-Freed theorem
as stated in \cite{DF} asserts that $\exp(\i\pi\eta/2)$ is naturally-defined as a section of $\PF(\D)$.  An equivalent statement
is that the product $\Pf(\DD)\exp(-\i\pi\eta/2)$ (where the two factors are sections of inverse line bundles) is well-defined as a complex number.  However, as long as we restrict to the
locus on which the boundary Dirac operator has no zero-modes, the line bundle $\PF(\DD)$ has a natural trivialization. Relative
to this trivialization, the Pfaffian $\Pf(\DD)$ is positive and can be replaced by its absolute value, and $\exp(-\i\pi\eta/2)$ is likewise
a well-defined complex number,\footnote{We do not need to discuss the meaning of $\exp(-\i\pi\eta/2)$
when the boundary Dirac operator does have zero-modes, since in that case $\Pf(\DD)=0$.  As explained in \cite{DF},
this case is subtle because the definition of APS boundary conditions is subtle when the boundary fermions have
zero-modes.} with the usual definition (\ref{roov}).

Finally, we should explain in concrete terms the meaning of the statment that the product $|\Pf(\DD)|\exp(-\i\pi\eta/2)$ is physically 
sensible.  For real or pseudoreal fermions, a key point was that $|\Pf(\DD)|(-1)^\zeta$ or $|\Pf(\DD)|(-1)^{\I/2}$ behaves
sensibly when the boundary Dirac operator
 develops a zero-eigenvalue.  This happened because the sign of $(-1)^\zeta$ or $(-1)^{\I/2}$ jumps
whenever the path integral of the boundary fermions would be expected to change sign.    For complex fermions, we are not
dealing with sign changes, because $\Pf(\DD)$ is not naturally real.  But it is still essential to understand what
happens when zero-modes develop. 

Near a point at which the boundary Dirac operator develops a pair of zero-modes, the associated bilinear form $\DD$,
in its canonical form, has a $2\times 2$ block
\be\label{zumo}\bp 0 & \lambda \cr -\lambda & 0\ep, \ee
with small $\lambda$.  Here $\lambda$ is a complex number and to make it vanish we must vary two parameters in
the metric and/or gauge field on $Y$.  For example, there may be two parameters $u_1,u_2$ such that
\be\label{umo}\lambda=u_1+\i u_2. \ee
Equivalently, setting $u_1+\i u_2=\varrho e^{\i\varphi}$, we have $\lambda=\varrho e^{\i\varphi}$.  We expect the path integral
to be proportional to $\lambda$ for $\lambda\to 0$.  So for  the formula $|\Pf(\DD)|\exp(-\i\pi\eta/2)$ for the combined
bulk and boundary partition function to be physically sensible near $\lambda=0$, $\exp(-\i\pi\eta/2)$ must not be smooth
near $\lambda=0$.  On the contrary, it must be proportional to $e^{\i\varphi}$, which is ill-defined at $\lambda=0$.  Concretely,
this happens because of the way APS boundary conditions depend on $\varphi$.  APS boundary conditions depend on which modes
of the Dirac operator on $Y$ have negative eigenvalues.  For very small (positive) $\varrho$, the Dirac operator on $Y$ has low-lying
modes, and which linear combinations of them are eigenmodes with negative eigenvalue depends on $\varphi$.

\subsection{The Case of $\nu$ Bands}\label{bands}

At the free fermion level, the topological superconductor is characterized by an integer-valued invariant $\nu$.
Let us recall how $\nu$ is defined.

On the boundary, we consider $n_1$ Majorana fermions  that transform under $\sT$
as 
\be\label{trt}\sT\psi(-t,x_1,x_2)=\g_0\psi(t,x_1,x_2),\ee
and $n_2$ that transform with the opposite sign:
\be\label{utrt}\sT\psi(-t,x_1,x_2)=-\g_0\psi(t,x_1,x_2).\ee
A pair of Majorana fermions $\psi,\psi'$ that transform with opposite signs can have a $\sT$-conserving
bare mass $\bar\psi\psi'$.  So up to $\sT$-invariant perturbations,  only the difference $\nu=n_1-n_2$
 is relevant.

In bulk, the basic object, as explained in section \ref{bpf}, is a Majorana fermion $\psi$ of mass $m$ that has a regulator
of mass $\mu$.  The interesting case is that $m$ and $\mu$ have opposite signs.  In section \ref{bpf}, we took $\mu>0$,
$m<0$ and arrives at $\exp(-\i\pi\eta/2)$ as the sTQFT partition function.  With the opposite signs, $\mu<0$ and $m>0$,
the same derivation would give $\exp(\i\pi\eta/2)$.  In bulk, $\nu$ is defined as the difference between the number of
$\mu>0,\,m<0$ pairs and the number of $\mu<0,\,m>0$ pairs. The bulk partition function is $\exp(-\nu\i\pi\eta/2)$.

At the free fermion level, $\nu$ is an integer-valued invariant, but it is known \cite{KitTwo}
that when $\sT$-conserving interactions are taken into account, $\nu$ is a topological invariant only mod 16.

It was pointed out in
 \cite{KTTW} that for a 4-dimensional $\Pin^+$ manifold, $\exp(-\i\pi\eta/2)$ is always  \cite{G,Stolz} a 
 $16^{th}$ root\footnote{Moreover,
 all $16^{th}$ roots do occur.  For $X=\RP^4$, one has $\exp(-\i\pi \eta/2)=\exp(\pm 2\i\pi /16)$, where the sign
 depends on the choice of $\pin^+$ structure. For an introduction to these matters, see Appendix \ref{feta}.}   of 1.
 Accordingly, the product of $\nu$ copies of the sTQFT whose partition function
 is $\exp(-\i\pi\eta/2)$ is trivial if and only if $\nu$ is an integer multiple of 16.  This was offered in \cite{KTTW} as an explanation
 of why even at the interacting level, $\nu$ is well-defined mod 16.  We have put those considerations on a firmer footing
 by showing in section \ref{bpf} that the partition function of the basic sTQFT associated to the $D=4$ topological superconductor
 is indeed $\exp(-\i\pi \eta/2)$.  
 
 \subsection{Special Behavior For $\nu=8$}\label{zeb}
 
It was shown in \cite{HCR} that, at least in a special case, anomalies associated to the mapping torus detect the value of $\nu$
mod 8 but not mod 16.  We can now explain why this is true.  First of all, $\exp(-\i\pi \eta/2)$ is a cobordism invariant of 
a $\pin^+$ 4-manifold $X$.  Moreover, it generates the group of cobordism invariants, so any other $\U(1)$-valued cobordism
invariant is a power of this one.  In particular, taking $w_4$ to represent the fourth Stieffel-Whitney class, $(-1)^{w_4}$ is
a nontrivial $\U(1)$-valued cobordism invariant so it must be a power of $\exp(-\i\pi \eta/2)$.  The power must be 8 since
$(-1)^{w_4}$ is valued in $\{\pm 1\}$:
\be\label{reffo}(-1)^{w_4}=\exp(-8\i\pi \eta/2). \ee
For example, for $X=\RP^4$, one has $w_4=1$, $\exp(-\pi \i\eta/2)=\exp(\pm 2\i\pi /16)$, with 
the sign depending on the $\pin^+$ structure.

Thus the $\nu=8$ case of a topological superconductor has $(-1)^{w_4}$ for its bulk partition function.  
The Stieffel-Whitney classes are defined without choosing a spin or pin structure, so unlike what happens for  other values of
$\nu$, the $\nu=8$ topological superconductor does not depend on the $\pin^+$ structure of the four-manifold $X$.   There is a full-fledged
TQFT with partition function $(-1)^{w_4}$.   

For a rank 4 real vector bundle -- such as the tangent bundle of a 4-manifold -- $w_4$ is the same as the mod 2 reduction of the
Euler class.  So an alternative statement is that the $\nu=8$ theory has $(-1)^{\chi(X)}$ as its bulk 
partition function,\footnote{We started with $w_4$ rather than $\chi$ because Stieffel-Whitney numbers such as
$w_4$ are always cobordism invariants, which made it easier to explain the relationship to $\exp(-\i\pi \eta/2)$. Likewise, the relationship of $\chi$
to $w_4$ is the simplest way to explain that $(-1)^{\chi(X)}$ is a cobordism invariant.} where $\chi(X)$
is the Euler characteristic of $X$.  

Now we can explain the result found in \cite{HCR}, and see that it is general.  If $X$ is a mapping torus, meaning
that it is total space of a fiber bundle $X\to S^1$, the fiber being a 3-manifold $Y$, then $\chi(X)=0$.  One way to prove
this is to observe that in general, the Euler characteristic is multiplicative in fibrations.  So $\chi(X)=\chi(Y)\chi(S^1)$.  But
this vanishes, since $\chi(S^1)=0$.

In short then,  we are in a similar situation to what we found in section \ref{signp} for the same problem on an orientable manifold.
If $\nu$ is a multiple of 8, this ensures that
the  theory of $\nu$ identical Majorana fermions on a 3-manifold $Y$ has no traditional global anomaly -- 
detectable via the invariant $\exp(-\nu\i\pi \eta/2)$ of a mapping torus.  But to give a natural definition of
the sign of the partition function for a system of $\nu$ identical $2+1$-dimensional  Majorana fermions, with no need for ``anomaly inflow'' from a 4-manifold $X$, one wants $\exp(-\nu\i\pi \eta/2)$
to equal 1 for all four-manifolds, not just mapping tori.  For this, we need $\nu$ to be a multiple of 16.  

The fact that the bulk state for $\nu=8$ has partition function $(-1)^{\chi(X)}$ may seem, at first sight, to offer a way
to define a symmetry-preserving gapped boundary state without topological order.  
After all, the Euler characteristic of a manifold with boundary
is perfectly well-defined and has natural cobordism properties.    
In the TQFT with partition function $(-1)^{\chi(X)}$, one can define a symmetry-preserving
boundary state by saying that the partition function on a 4-manifold $X$ with any spatial boundary $Y$ is $(-1)^{\chi(X)}$.
This boundary state satisfies cutting and gluing axioms of TQFT, but it is not compatible with reflection positivity or unitarity.
For a simple counterexample, take $Y$ to be a 3-sphere and $X$ a 4-ball with boundary $Y$.  Then $(-1)^{\chi(X)}=-1,$
but because $X$ can be built by gluing two identical pieces as in fig. \ref{dodotwo} of section \ref{rpo}, reflection positivity
would require that the partition function for $X$ should be positive.

\subsection{The $d=3$ Topological Insulator On An Unorientable Spacetime}\label{topsupun}

Now we very briefly consider the $3+1$-dimensional  topological insulator on a possibly unorientable spacetime.
Everything that we have said about the topological superconductor has a fairly obvious analog,
with one exception.  In contrast to the integer-invariant $\nu$ of the topological superconductor,
the topological insulator has only a $\Z_2$-valued invariant, even at the free fermion level.   Indeed, we have already
written in eqn. (\ref{worx}) a $\sT$-conserving mass term for a pair of $2+1$-dimensional  Dirac fermions $\psi_1,\psi_2$ that transform the
same way under $\sT$.
So $\sT$ symmetry only protects mod 2 the number of massless charged Dirac fermions.  

We therefore expect that it will turn out that $\exp(-\i\pi \eta/2)$ is always $\pm 1$ for the topological insulator,
even on an unorientable 4-manifold.  To understand this statement, we first have to recall that the symmetry
that protects the SPT state of a topological insulator is, in relativistic terminology,\footnote{$\sC$ and $\sCT$, which exchange electrons
with positrons, are never true symmetries in condensed matter physics.} $\sT$ and not $\sCT$.  As we discussed in section
\ref{welfin}, this means that in Euclidean signature, we should consider the symmetry $\sCR$; in other words, a symmetry
that reverses the orientation of spacetime also acts as charge conjugation.

Now let $X$ be a possibly unorientable 4-manifold and $\h X$ its oriented double cover.  The invariant $\exp(-\i\pi \eta/2)$
on $X$ that we want to understand is the $\eta$-invariant of a Dirac operator acting on  fermion fields 
$\psi^\pm _\pm$ where the superscript is the $\U(1)$ charge and the subscript is the chirality.  There is an equivalent
definition on $\h X$ in which one considers only  fermion fields $\h\psi^+_\pm$ of positive
$\U(1)$ charge but any chirality: the Dirac spectrum on $\h X$ computed  from eigenvalues of the Dirac operator
acting on $\h\psi^+_\pm$ only is the same as the Dirac spectrum on $X$ in the full set of fields $\psi^\pm_\pm$.  Indeed, $X$
is the quotient of $\h X$ by a $\sCR$ symmetry, so the fields $\psi^+_+$ and $\psi^-_-$ on $X$ combine to $\h\psi^+_+$
on $\h X$, and similarly $\psi^+_-$ and $\psi^-_+$ on $X$ combine to $\h\psi^+_-$ on $\h X$.  So instead of computing
an $\eta$-invariant by summing over the spectrum of a Dirac operator on $X$ acting on the full set of fields $\psi^\pm_\pm$,
we can sum over the spectrum of a Dirac operator on $\h X$ acting on $\psi^+_\pm$ only.

As $\h X$ is orientable, the Dirac spectrum on $\h X$ is symmetric under $\lambda\leftrightarrow -\lambda$.  This follows
as usual from considerations of chirality.  But a pair of eigenvalues $\lambda,-\lambda$ does not contribute to the $\eta$-invariant.
So in computing $\eta$ on $\h X$, we can consider zero-modes only.  The space of zero-modes of the Dirac operator
on $\h X$ is even-dimensional because of the usual antilinear symmetry of the Dirac equation, so $\eta$ is an even integer.
Consequently, $\exp(-\i\pi \eta/2)=\pm 1$, as we aimed to show.

\section{The Majorana Chain}\label{maj}

At a microscopic level, the time-reversal operation that holds in the real world obeys $\sT^2=(-1)^F$, where $(-1)^F$ is
the operator that counts fermions mod 2.  But there are some situations in condensed matter physics in which it is natural
to consider a time-reversal symmetry that obeys $\sT^2=1$.  This usually happens either because spin is unimportant
or because the symmetry under consideration is really the product $\sRT$ of time-reversal with a reflection of one space coordinate.
(In nature $(\sRT)^2=1$, and $\sRT$ can act as time-reversal on a thin layer or long chain that is localized at the fixed point
set of $\sR$.)

In the context of SPT phases of matter, there is a very interesting example in which a time-reversal symmetry is assumed with
$\sT^2=1$.  This is the Majorana chain in 1 space dimension \cite{FK}.

A boundary of such a system is 0-dimensional, that is a point, and we will begin there.  In this paper, we have built in the
condition $\sT^2=(-1)^F$ by taking our Lorentz signature gamma matrices to be real matrices obeying
 $\{\g_\mu,\g_\nu\}=2\eta_{\mu\nu}$,
where $\eta_{\mu\nu}=\diag(-1,1,1\dots,1)$.   Then we take $\sT$ to act by \be\label{horz}\psi(t,\vec x)
\to \pm \g_0\psi(-t,\vec x),\ee
 ensuring that $\sT^2\psi=-\g_0^2\psi=-\psi$.  Since $\g_i^2=+1$ for $i>0$,
it follows that a spatial reflection squares to $+1$.  In Euclidean signature, this leads to a $\pin^+$ structure.

In $0+1$ dimensions, we only need one gamma matrix $\g_0$.  To get the most economical possible theory,
we take $\g_0^2=+1$, rather than $-1$ as in all our discussion so far.  Then we can represent $\g_0$ by the number $1\times 1$
matrix 1 (or $-1$), and it is possible to consider a 1-component real (Majorana) fermion field $\psi$ with action
\be\label{dolb}I=\frac{i}{2}\int \d t \,\psi\frac{\d}{\d t}\psi. \ee
We take $\sT$ to act on such a field by 
\be\label{olb}\sT\psi(t)=\pm \psi(-t), \ee
with some choice of the sign, so that $\sT^2=1$.  A pair of fermions $\psi$, $\psi'$ that transform with opposite signs can have a $\sT$-invariant
``mass'' term $i\psi\psi'$, so as in section \ref{bands}, the natural invariant is the difference $\nu$ between the number of
fermions transforming under $\sT$ with a $+$ sign, and the number transforming with a $-$ sign.  At the free fermion level,
$\nu$ is an integer invariant, but  \cite{FK}  at the interacting level, $\nu$ is only an invariant mod 8.
(This statement means that 
a $0+1$-dimensional system consisting of 8 real fermions all transforming under $\sT$ with the same sign
can be gapped by a suitable $\sT$-invariant four-fermion coupling.)

When $\nu$ is not a multiple of 8, this system has a subtle ($\nu$-dependent) anomaly in the realization of the symmetries
$\sT$ and $(-1)^F$.  We will not repeat this story  here, and instead we will ask how the system with $\nu$ boundary fermions
can naturally appear on the boundary of a $\sT$-conserving system in $1+1$ dimensions.  The choice $\g_0^2=+1$ that
we have made in setting up this problem corresponds to a $\Pin^-$ structure, and once we have assumed a $\Pin^-$
structure on the $0+1$-dimensional boundary of spacetime, we must also (if we want the Dai-Freed theorem or a similar
tool to be available) assume a $\Pin^-$ structure on the $1+1$-dimensional
bulk.  So we consider fermions with $1+1$-dimensional gamma matrices that obey $\g_0^2=1=-\g_1^2$, or more covariantly
$\{\g_\mu,\g_\nu\}=-2\eta_{\mu\nu}$.  In Euclidean signature, this becomes $\{\g_\mu,\g_\nu\}=-2\delta_{\mu\nu}$, where the
minus sign corresponds to a $\Pin^-$ structure.  ($2\times 2$ matrices satisfying $\{\g_\mu,\g_\nu\}=-2\delta_{\mu\nu}$ exist
but cannot be chosen to be real.  The fact that a bare mass is possible, as discussed in the next paragraph,
makes it clear that a Majorana fermion with $\sT^2=1$ is pseudoreal in Euclidean signature.)

In contrast to the case $\sT^2=-1$, a 2d Majorana fermion with $\sT^2=1$
can have a $\sT$-invariant bare mass.  In Lorentz signature, the $\sT$-invariant Dirac equation can be written
\begin{equation}\label{sx}\left(\g^0\partial_0+\g^1\partial_1-m\bar\g\right)\psi=0, \ee
where $m$ is real and $\bar \g=\g^0\g^1$ is the chirality operator.  Actually, this equation is $\sT$-invariant for either sign in
$\{\g_\mu,\g_\nu\}=\pm 2\eta_{\mu\nu}$, but if one writes the equation in Hamiltonian form $\i\partial_t\psi=H\psi$
for some linear operator $H$, one sees that $H$ is only hermitian for the case of a $-$ sign.  So that is the case for
which this mass term is physically sensible.  

After generalizing to curved spacetime and continuing to Euclidean signature, eqn. (\ref{sx}) becomes $\left(\slashed{D}-\i m\bg \right)\psi=0$. But it is convenient
to multiply by $\bg$ and write the equation as
\be\label{mirth}\left(\D'+\i m\right)\psi=0.  \ee
Here $\D'=\bg \slashed{D}$ is a hermitian and reflection-symmetric Dirac operator.  Because it is reflection symmetric,
it makes sense on any 2-manifold, possibly unorientable, that is endowed with a $\Pin^-$ structure.  (See Appendix \ref{do} for
more on this.)

The bulk sTQFT of the Majorana chain is constructed by taking a fermion field $\psi$ of mass $m$, together with a Pauli-Villars
regulator field of mass $\mu$.  As in section \ref{bpf}, the interesting case is that $m$ and $\mu$ have opposite signs.
For $\mu>0$, $m<0$, the phase of the partition function is $\exp(-\i\pi \eta/2)$, by the familiar argument.  This is the partition function of the bulk
sTQFT associated to the Majorana chain. 
 It is always an $8^{th}$ root of 1; it equals $\exp(\pm 2\i\pi /8)$ for $\RP^2$,
where the sign depends on the choice of $\Pin^-$ structure. (See Appendix \ref{feta}.)  Because $\exp(-\pi \i\eta/2)$ is always an $8^{th}$ root of 1,
the bulk sTQFT of the Majorana chain is trivial for $\nu$  multiple of 8.  (This was argued in \cite{KTTW} based on an assumption
of cobordism invariance.)

As usual, if $\nu$ is not a multiple of 8, the bulk partition function $\exp(-\nu\i\pi \eta/2)$ cannot be defined as a topological invariant
in a symmetry-preserving fashion on a possibly unorientable 2-manifold with boundary.  However, the Dai-Freed theorem
applies in this situation and says that the product of the path integral of $\nu$ boundary fermions times $\exp(-\nu\i\pi \eta/2)$
is well-defined.  This is the path integral of the combined system consisting of the bulk sTQFT and the boundary fermions.

\appendix
\section{Spinors In Riemannian Geometry}\label{dirram}

\subsection{Spinors On Orientable Manifolds}\label{spinor}

To introduce spinors and the Dirac equation on a Riemannian manifold $W$ of dimension\footnote{In our applications, $W$ is usually
either the spacetime $X$ of dimension $D$ or its boundary $Y=\partial X$ of dimension $d=D-1$.} $n$,
it is useful to first pick a ``vielbein,'' that is an orthonormal basis of tangent vectors $e_a$, $a=1,\dots,n$.  We denote
as $e_a^i$, $i=1,\dots,n$, the components of $e_a$ in an arbitrary local coordinate system $x^i=\{x^1,\dots,x^n\}$ on $W$.
The statement that the $e_a$ are orthonormal means that, if $g_{i j}$ is the metric tensor of $W$, then
\begin{equation}\label{orth} e_a^i e_b^jg_{ij}=\delta_{ab}.\end{equation}
Equivalently,
\begin{equation}\label{north} \delta^{ab}e_a^ie_b^j =g^{ij}. \end{equation}
Indices of $e_a^i$ are raised and lowered with $\delta_{ab}$ and $g_{ij}$ and in particular
$e_i^a$ is the inverse matrix to $e_a^i$. 
``Flat space'' gamma matrices are $2^{[n/2]}\times 2^{[n/2]}$ matrices $\g_a, \,a=1,\dots,n$,
obeying 
\begin{equation}\label{worth}\{\g_a,\g_b\}=2\delta_{ab}.\end{equation}
The corresponding ``curved space'' gamma matrices
are  $\g_i=e_i^a\g_a$, obeying
\begin{equation}\label{zorth}\{\g_i,\g_j\}=2g_{ij}.\end{equation}

While the orientation-preserving rotation group $\SO(n)$ acts on the tangent space to $W$ at a given point $p\in W$, the corresponding
group that acts on spinors on $W$ is a double cover of $\SO(n)$ that is known as $\Spin(n)$.  Its generators are
\begin{equation}\label{torth}\Sigma_{ab}=\frac{1}{4}[\g_a,\g_b]. \end{equation}
The covariant derivative of a spinor field $\Psi$ is defined
as
\begin{equation}\label{porth}\frac{D}{D x^i}\Psi=\left(\frac{\partial}{\partial x^i}+\omega_i^{ab}\Sigma_{ab}\right)\Psi,\end{equation}
where $\omega_i^{ab}$ is the Levi-Civita connection on the tangent bundle of $W$.  
So  we can define a hermitian Dirac operator:
\begin{equation}\label{dirac}\D=\i\sum_{i=1}^n\g^iD_i. \end{equation}

This construction depended on the choice of a vielbein $\{e_a\}$, which in general can only be made locally.  To define a global
Dirac operator, one glues together local descriptions.  A down-to-earth way to proceed is to cover $W$ with small open sets $W_\x$, $\x=1,\dots,s$,
picking a vielbein $e_{a\,\x}$, $a=1,\dots,n$, on each $W_\x$.  On $W_\x\cap W_\y$, since $e_{a\,\x}$ and $e_{a\,\y}$ are
both orthonormal bases of the tangent bundle of $W$, they are related by an orthogonal transformation $M_{\x \y}$: $e_\x=M_{\x \y}
e_\y$, or in more detail
\begin{equation}\label{toroth}e_{a\,\x}=\sum_b M_a{}^b{}_{\x \y} e_{b\,\y}. \end{equation}
(Here $M_a{}^b{}_{\x \y}$ are the matrix elements of the $n\times n$ orthogonal matrix $M_{\x \y}$.)
If $W$ is orientable, as we assume in this section,
 we can pick the local vielbeins $e_\x$ such that $M_{\x \y}$ is valued in $\SO(n)$ (rather than $\O(n)$).
  The transition matrices $M_{\x \y}$ obey a consistency condition: in triple
overlaps $W_\x\cap W_\y\cap W_\z$, we have
\begin{equation}\label{horth} M_{\x \y}M_{\y \z}M_{\z \x}=1.\end{equation}

Once a local vierbein is picked in an open set $W_\x$, we can define a spinor field $\Psi_\x$ (with components $\Psi_{\alpha \,\x}$,
$\alpha=1,\dots, 2^{[n/2]}$) in this open set.  The covariant
derivatives and Dirac equation for $\Psi_\x$ are as  in eqns. (\ref{orth}) and (\ref{dirac}).  On intersections $W_\x\cap W_\y$, we compare
$\Psi_\x$ and $\Psi_\y$ via
\begin{equation}\label{compare}\Psi_\x =\h M_{\x \y}\Psi_\y,\end{equation}
where $\h M_{\x \y}$ is a $2^{[n/2]}\times 2^{[n/2]}$ matrix that is obtained by ``lifting'' $M_{\x \y}$ to the spinor representation.
Thus $\h M_{\x \y}$ is an element of $\Spin(n)$, the double cover of $\SO(n)$ associated to spin.  Because the sign with which
a given element of $\SO(n)$ acts on spinors is not uniquely determined, the  lift from $M_{\x \y}$
to $\h M_{\x \y}$ is only uniquely determined up to sign.  For consistency, the signs must be chosen so that in triple overlaps
\begin{equation}\label{ompare}\h M_{\x \y}\h M_{\y \z}\h M_{\z \x}=1. \end{equation}

A choice of signs satisfying this condition is called a spin structure on $W$; once a spin structure is picked, we can piece together the local
descriptions and globally study
fermions and the Dirac equation on $W$.    In dimension $\geq 4$, a given $W$ may not admit a spin structure because it may be impossible to satisfy
eqn. (\ref{ompare}); in this case, spinors cannot be defined on $W$
(an example is $W=\Bbb{CP}^2$).
 This possibility will not play a primary role in the present paper. We simply restrict our attention
to spacetimes on which spinors can be defined.

More important for our purposes is the fact that a given $W$ may admit more
than one spin structure.   If we do find a choice of signs consistent with (\ref{compare}), then a different choice $\h M'_{\x \y}
=\h M_{\x \y} (-1)^{c_{\x \y}}$, for some $\Z_2$-valued function $c_{\x \y}$, obeys the same condition if
\be\label{yero}(-1)^{c_{\x \y}}(-1)^{c_{\y \z}}(-1)^{c_{\z \x}}=1\ee
for all $\x,\y,\z$.  We want to impose on $c$ a gauge equivalence $c_{\x\y}\cong c_{\x\y}+d_\x-d_\y$
for any $\Z_2$-valued function $d_\x$, since in equation (\ref{compare}) the change in $\h M_{\x\y}$ resulting from such a change
in $c_{\x\y}$ can be absorbed in redefining the local spinors by $\Psi_\x\to (-1)^{d_\x}\Psi_\x$.  Given the 
condition (\ref{yero}) and the gauge-invariance just stated, $c_{\x\y}$ defines an element of the group $H^1(W,\Z_2)$.  
Any two spin structures on $W$ differ by ``twisting'' by an element of this group.  

For an elementary example of this, take $W$ to be an $n$-torus $T^n$, with flat metric.  Parallel transport of tangent vectors on $W$ is
completely trivial: the Levi-Civita connection on tangent vectors is 0.  However, in defining spinors on $W$, we are free to say that a spinor 
changes sign under parallel transport around a noncontractible loop $\ell\subset T^n$.  As there are $n$ independent loops to consider,
there are $2^n$ choices of sign in defining spinors on $T^n$, and these are the spin structures on $T^n$.  The group $H^1(T^n,\Z_2)$
is isomorphic to $\Z_2^n$ and labels these choices of sign.  (For $W=T^n$, but not in general, there is a distinguished spin structure,
namely the ``trivial'' one in which spinors are parallel transported with no sign changes.  As a result, in this example, the set
of spin structures can be canonically identified with the group $H^1(W,\Z_2)$.)

A manifold $W$ endowed with a choice of spin structure is called a spin manifold. (Orientability is built into the notion of a spin manifold,
since in the starting point, we took the transition functions of the tangent bundle to be valued in $\SO(n)$.)
 If $W$ is a spin manifold, the matrices
$M_{\x\y}$, understood as $2^{[n/2]}\times 2^{[n/2]}$ matrices in the spin representation of $\Spin(n)$, are transition functions for a vector bundle $\S\to W$.  A fermion field on $W$ is a section of this bundle. We sometimes refer to $\S$ as the spin structure.  If $\S$
is one spin structure, then any other spin structure is $\S\otimes \rho$, where $\rho\to W$ is a real line bundle of order 2 (defined
in the above construction by the transition functions $(-1)^{c_{\x\y}}$).

For even dimension $n$, 
the irreducible representation of the Clifford algebra is unique up to isomorphism. This can be proved by writing $n$
gamma matrices in terms of $n/2$ pairs of creation and annihilation operators, which have a unique irreducible representation. But for
even $n$, the irreducible representation of the Clifford algebra is not irreducible as a representation of $\Spin(n)$;
it  decomposes as the sum of
two representations of $\Spin(n)$ of positive or negative chirality.  For odd $n$, matters are different.  Up to isomorphism,
there is only one spinor representation of $\Spin(n)$, but there are two inequivalent representations of the Clifford algebra.
The two representations differ by $\g_i\to -\g_i$, which preserves the anticommutation relations of the Clifford algebra and
does not affect the $\Spin(n)$ generators $\Sigma_{ij}=\frac{1}{4}[\g_i,\g_j]$, but gives an inequivalent representation of the
Clifford algebra. The two representations differ by the sign of the product $\g_1\g_2\cdots \g_n$ -- which commutes with the Clifford
algebra and so is a $c$-number ($\pm 1$ or $\pm \i$, depending on $n$) in an irreducible representation of the Clifford algebra.  
For example, in 3 dimensions, the two representations
of the Clifford algebra correspond in a locally Euclidean frame to $\g_i=\pm \sigma_i$, and thus the gamma matrices obey
\be\label{zoffo}\g_i\g_j=g_{ij}\pm\i\epsilon_{ijk}\g^k, \ee
where $\epsilon_{ijk}$ is the Levi-Civita tensor.  In this formula, $\pm\epsilon_{ijk}$ can be regarded as a choice of orientation of $W$.
Thus the two representations of the Clifford algebra are associated to the two possible orientations of $W$.  The Dirac operator
$\D=\i\sum_k\g^k D_k$ is odd under $\g_k\to -\g_k$, and so the sign of the Dirac operator depends on the orientation of $W$.
The $\eta$-invariant of the Dirac operator changes sign if one changes the sign of this operator (thereby changing the sign of
all of its eigenvalues), and this is why the $\eta$-invariant is odd under parity, that is under reversal of orientation.

\subsection{Fermions On An Unorientable Manifold}\label{unor}

Now we consider the unorientable case.
On an unorientable $n$-manifold $W$, the transition matrices $M_{\x\y}$ of the tangent bundle of $W$
 are valued in $\O(n)$ rather than $\SO(n)$, and the corresponding
matrices $\h M_{xy}$ that act on spinors must  similarly take values in a double cover of $\O(n)$.  There are two choices of this
double cover,\footnote{Here we consider Majorana fermions, that is fermions coupled to gravity only, as is appropriate physically for a topological superconductor.
In the presence of the $\U(1)$ gauge symmetry of electromagnetism, one can combine a spatial reflection with a gauge transformation, and define a symmetry
for which $\sR^2$ acts on electrons with an arbitrary phase.  Mathematically, this is related to a possible generalization from spin and pin structures to $\mathrm{spin}_c$
and $\mathrm{pin}_c$ structures.  This generalization is useful in analyzing topological states of matter, but will not be relevant in the present paper.}
depending on whether a spatial reflection $\sR$ acting on spinors satisfies $\sR^2=1$ or $\sR^2=-1$.  For most physical
applications, the appropriate choice is $\sR^2=1$.
 The double cover of $\Spin(n)$ that arises if $\sR^2=1$ is called $\Pin^+(n)$.  (If we choose $\sR^2=-1$,
 we get a double cover of $\Spin(n)$ that is called $\Pin^-(n)$.)
 
 A reflection symmetry in Euclidean signature can be continued to either a reflection or a time-reversal symmetry in Lorentz signature,
 depending on whether the reflection acts on the spacetime coordinate that is analytically continued in  the change of signature.
 This process is subtle and, as is well-known, $\sT$ is antilinear in quantum field theory while $\sR$ is linear.  Related to this,
 $\sC$ (charge conjugation) enters in the continuation, so $\sR$ in Euclidean signature can be continued to $\sCT$ in Lorentz signature,
 and $\sCR$ can be continued to $\sT$.  If $\sR$ (or $\sCR$) squares to $+1$ on fermions, then $\sCT$ (or $\sT$) squares to $-1$,
 and vice-versa. 

To describe concretely the group $\Pin^+(n)$, we work on $\R^n$ and ask how a reflection that reverses the sign of one coordinate acts
on spinors.    Let $w$ be a unit vector and consider the reflection $x^i \to x^i -2w^i  w\cdot x$.
Setting $\g\cdot w=\g_i  w^i $, the reflection acts on spinors by 
\be\label{bel}\psi\to \g\cdot w\, \psi.\ee  
These transformations square to $+1$, and adjoining them to the $\Spin(n)$ transformations generated by
$\Sigma_{ab}=\frac{1}{4}[\g_a,\g_b]$ gives $\Pin^+(n)$.  One way to describe $\Pin^-(n)$ is simply to take the given reflection to act by
$\psi\to \i \g\cdot w\psi$, an operation that squares to $-1$.

To define a $\pin^+$ structure on $W$, one lifts the transition functions $M_{\x\y}$ of the tangent bundle from $\O(n)$ to $\Pin^+(n)$-valued
functions  $\h M_{\x\y}$.
In dimension $\geq 2$, there is a possible obstruction to this (for example, $\RP^2$ admits no $\pin^+$ structure).
If $W$ does admit a $\pin^+$ structure, this structure is not necessarily unique; just like spin structures, $\pin^+$ structures
differ by twisting by an element of $H^1(W,\Z_2)$.  A manifold with a choice of $\pin^+$ structure is called a $\pin^+$ manifold.

If $W$ is a $\pin^+$ manifold, the transition functions $\h M_{\x\y}$, understood as before as
$2^{[n/2]}\times 2^{[n/2]}$ matrices, 
define a vector bundle $\P\to W$.   We can define a spinor field $\psi$ that is a section of this bundle and the Dirac
operator $\D=\i\slashed{D}$ can act on $\psi$.    But some of its properties are different from the
orientable case, as we have seen in the main text and will discuss further in Appendix \ref{do}.

If $\P\to W$ is one $\pin^+$ structure, then just as in the spin case, any other $\pin^+$ structure is $\P\otimes \rho$ for some
real line bundle $\rho\to W$.  However, there is a new ingredient when $W$ is unorientable.  There is then a canonical real line
bundle $\veps\to W$, namely the orientation bundle: the holonomy of $\veps$ around a closed loop $\ell\subset W$ is $-1$
if the orientation of $W$ is reversed in going around $\ell$, and otherwise it is $+1$. $\P'=\P\otimes \veps$ is a new $\pin^+$ structure on $W$ that we call the $\pin^+$ structure complementary to $\P$.  Concretely, the transition functions $\h M'_{\x\y}$ of $\P'$
are obtained from the transition functions $\h M_{\x\y}$ of $\P$ by reversing the sign of $\h M_{\x\y}$ whenever it is in the
orientation-reversing component of $\Pin^+(n)$.

Everything that we have just said for $\pin^+$ has an immediate analog for $\pin^-$ except that the obstruction to a $\pin^-$ structure
begins in dimension 4.  

\subsection{Dirac Operators}\label{do}

Consider a reflection in $\R^n$:
\be\label{porz}\sR\psi(x_1,x_2,\dots,x_n)=\g_1\psi(-x_1,x_2,\dots,x_n).\ee
It is not difficult to see that such a reflection anticommutes with the Dirac operator
\be\label{orzo} \D=\i\slashed{D}=\i\sum_{j=1}^n\g^j D_j. \ee

More generally, on a spin manifold $\h W$, any orientation-reversing symmetry anticommutes with $\D$.  
Suppose that $\h W$ has an orientation-reversing symmetry $\tau$ that acts freely and obeys $\tau^2=1$.
Then we can pass from $\h W$ to its quotient $W=\h W/\Z_2$.  
If in addition  $\tau$ acts as a symmetry of the spin bundle $\S\to \h W$, still obeying $\tau^2=1$, then
$W$ is endowed with two $\pin^+$ structures $\P$ and $\P'$ that
can be defined as follows.  A section $\psi$ of $\P\to W$
is a section $\h\psi$ of $\S\to \h W$ that obeys
\be\label{milox} \tau\h\psi=\h\psi, \ee
and a section $\psi'$ of $\P'\to W$ is a section $\h\psi$ of $\S\to \h W$ that obeys
\be\label{ilox}\tau\h\psi=-\h\psi.\ee
Since the sign with which $\tau$ acts on $\S$ depends on an arbitrary choice, the relationship between $\P$ and $\P'$ is completely
symmetrical.  

Consider any point $p\in \h W$ and a path $\h\ell$ from $p$ to $\tau(p)$.  After dividing by $\tau$, $\h\ell$
projects to a closed loop $\ell\subset W$ around which the orientation of $W$ is reversed.  The relative minus sign in eqns.
(\ref{milox}) and (\ref{ilox}) means that the monodromies around $\ell$ of the $\pin^+$ bundles $\P$ and $\P'$ differ in sign.
Thus $\P$ and $\P'$ are complementary $\pin^+$ bundles in the sense of Appendix \ref{unor}.

Every $\pin^+$ manifold $W$ arises in this construction, with $\h W$ being the oriented double cover of $W$.

Now suppose that $\psi$ is a fermion field on $W$ valued in a particular $\pin^+$ structure $\P$.
From what we have said, it is clear that the usual Dirac operator $\D=\i\sum_k\g^k D_k$ cannot be defined as a self-adjoint
operator acting on $\psi$.  On the covering space $\h W$, $\D$ anticommutes with $\tau$, so it maps sections of $\P$ to 
sections of $\P'$ and vice-versa. 

For odd dimension $n$, there is no way to remedy this situation and there is no self-adjoint Dirac operator acting on a $\P$-valued
field $\psi$.  The upshot of this is that the corresponding path integral $Z_\psi$ can be defined as the Pfaffian of a complex-valued
antisymmetric bilinear form, but not as the determinant of a self-adjoint operator. Accordingly it is complex-valued, and in general
it can be affected by a global anomaly involving a complex number of modulus 1, not just a real number
$\pm 1$.  This issue was explored in the main text in section \ref{anombo}.  

For even $n$, matters are different and there is always a self-adjoint Dirac operator acting on a fermion field valued in any given 
$\pin^+$ structure $\P$.  This was assumed in the main text at several points and here we will explain the details.  First
we start on the orientable manifold $\h W$ with spin structure $\S$.  As $\h W$ is orientable, one can define a chirality operator
$\bg$ that acts on $\S$ (and leads to a decomposition 
$\S=\S_+\oplus \S_-$, where $\S_+$ and $\S_-$ are the bundles of positive or negative
chirality spinors).  Assuming that we want $\bg^2=1$, the proper definition of $\bg$ depends slightly on whether $n$ is congruent
to 0 or 2 mod 4.  For example, in 2 dimensions (and similarly in $4k+2$ dimensions for any $k$), one
defines 
\be\label{toroz}\bg=\frac{\i}{2!}\epsilon^{ij}\g_i\g_j. \ee
In 4 dimensions (and similarly in $4k$ dimensions for any $k$), one omits the factor of $\i$ and defines
\be\label{orz}\bg=\frac{1}{4!}\epsilon^{ijkl}\g_i\g_j\g_k\g_l. \ee

On an orientable manifold $\h W$ of even dimension, instead of the usual Dirac operator $\D=\i\slashed{D}$, we can use the alternate
Dirac operator 
\be\label{elf} \D'=\bg \slashed{D}. \ee
$\D'$ is self-adjoint, like $\D$, and the two operators have the same spectrum, since they are conjugate.  Indeed, if $U=(1-\i\bg)/\sqrt 2$, then 
\be\label{welf}\D'=U\D U^{-1}.\ee  This depends on the fact that $\bg$ commutes with the action of $\Spin(n)$ on the spinor representation,
and thus  commutes with the $\Spin(n)$ connection that is hidden in the definition of $\D$.

So on $\h W$, we could equally well use $\D$ or $\D'$.
However, if we want to descend to the unorientable manifold $W=\h W/\Z_2$, where $\Z_2$ is generated by the orientation-reversing
isometry $\tau$, then it is better to use $\D'$.  Indeed, since $\bg$ and $\D$ are both odd under $\tau$, $\D'$ is even, so it descends
to a perfectly good self-adjoint operator acting on the $\pin^+$ bundle $\P$ of $W$.

In the main text, we sometimes write simply $\D$ rather than $\D'$ for the self-adjoint Dirac operator of an even-dimensional $\Pin^+$ manifold $W$.
A justification for this is that in fact up on $\h W$, what we mean by $\D$ or by  $\D'$ depends on the choice of representation of the Clifford 
algebra.
We can define new gamma matrices
\be\label{werm} \g_i'=U\g_iU^{-1}   \ee
Concretely
\be\label{zerm}\g_i'=-\i\bg\g_i=\begin{cases}-\eps_{ij}\g^j & {\mathrm{if}} ~n=2\cr
                                                                       \frac{\i}{3!}\epsilon_{ijkl}\g^j\g^k\g^l& {\mathrm{if}} ~n=4 .\end{cases}\ee
In terms of the new gamma matrices, $\D'$ takes the standard form
\be\label{termo}\D'=\i\sum_k\g'{}^kD_k . \ee
(The covariant derivatives $D_k$ take the same form in old or new gamma matrices.)

However,  if we use the new gamma matrices, then the transformation law (\ref{bel}) under a reflection
(and therefore under any orientation-reversing element of $\Pin^+(n)$) is modified.  It becomes
\be\label{yrf}\psi\to \i\bg'\g'\cdot w\,\psi. \ee
(Here $\bg'$ is the chirality operator constructed from the new gamma matrices; it  coincides with $\bg$.)
Actually, this is the most common choice in particle physics in 4 spacetime dimensions.\footnote{For example,
see eqn. (2.33) of \cite{BD}, which however is written with an arbitrary phase for Dirac fermions, and also
is a formula for the transformation of a fermion under parity (a reflection of all three spatial coordinates, not just one).}
  The reason that we started with
(\ref{bel}) instead, apart from the fact that it is fairly standard mathematically, is that it leads for many purposes to a formalism
that works uniformly in all dimensions.  It is hard to find conventions for fermions that are convenient for all purposes.

\section{Examples In Low Dimension}\label{examples}

We will look more closely at  low-dimensional examples that are important in this paper.

\subsection{Two Dimensions}\label{twod}

To describe spinors on $\R^2$,
we only need two gamma matrices, and we can pick real $2\times 2$ gamma matrices
$\g_1=\sigma_1$, $\g_2=\sigma_3$.  The group $\Spin(2)$ is generated by $\Sigma=\frac{1}{2}\g_1\g_2$.
Abstractly, $\Spin(2)$ is a copy of $\SO(2)$, with the spinors as the fundamental real 2-dimensional representation. But, because the eigenvalues of $\Sigma $ are $\pm \i/2$ (not $\pm \i$),
$\Spin(2)$ is a double cover of the original $\SO(2)$ that acts by rotations on $\R^2$.

To get $\Pin^+(2)$, we include the reflections $x^i\to x^i-2w^i w\cdot x$, acting by
\be\label{pes} \psi\to \gamma\cdot w\, \psi. \ee
This gives a real, 2-dimensional representation of $\Pin^+(2)$.  The group $\Pin^+(2)$ is abstractly isomorphic
to $\O(2)$, though it is a double cover of the usual $\O(2)$ that acts by rotation and reflection of $\R^2$.

Naively, one might think that one can define a second real, 2-dimensional representation of $\Pin^+(2)$
in which a reflection acts by
\be\label{nes}\psi\to -\gamma\cdot w\psi. \ee
Shortly we will see that something like this does work in 3 dimensions, but in 2 dimensions the two representations
are conjugate under $\psi\to \g_1\g_2\psi$, which represents a $\pi$ rotation of $\R^2$.  

Since the spinor representation of $\Pin^+(2)$ is real, there is an antilinear operation
\be\label{wes}\T\psi =\psi^*,~~\T^2=1 \ee
that anticommutes with the hermitian Dirac operator $\D'$ and ensures that on any 2-manifold, orientable or not,
the spectrum  is symmetric under $\lambda\leftrightarrow -\lambda$.

On an {\it orientable} 2-manifold, we can also define an antilinear operation
\be\label{res}\T'\psi=\bg \psi^*,~~~(\T')^2=-1 \ee
that {\it commutes} with $\D$ (or $\D'$) and ensures that the Dirac eigenvalues all have even multiplicity. 
Here $\bg=\i\g_1\g_2=\sigma_2$ is the chirality operator. 

The reason that both structures exist is that, if we restrict to $\Spin(2)$, then the 2-dimensional spinor representation
is reducible; it decomposes in 1-dimensional representations of positive or negative chirality which are complex conjugates
to each other.   Accordingly, the Dirac operator anticommutes with $\bg$ and we can include it in defining $\T'$.  On an unorientable
2-manifold, the definition of $\T$ makes sense, but $\T'$ can no longer be defined, because $\bg$ is not defined globally.

\subsection{Three Dimensions}\label{wwod}

Now we go to 3 dimensions.  For the $2\times 2$ gamma matrices, we can take the Pauli $\sigma$-matrices
\be\label{zob} \g_a=\sigma_a,~~~a=1,\dots,3.\ee
The double cover of $\SO(3)$ that acts on spinors is $\Spin(3)=\SU(2)$.   Spinors transform in the spin 1/2 representation of $\SU(2)$.
This representation is pseudoreal, since the trivial representation of $\SU(2)$ appears {\it antisymmetrically} in the tensor product $1/2\otimes 1/2$.
Concretely, the invariant antisymmetric tensor on the spin 1/2 representation of $\SU(2)$ is given by the $2\times 2$ Levi-Civita symbol $\veps_{\alpha
\beta}$, $\alpha,\beta=1,2$.
If $\psi$ is a fermion field valued in this representation, then the corresponding ``mass term,'' often written as $\bar\psi\psi$, is
\be\label{mast}(\psi,\psi)=\veps_{\alpha\beta}\psi^\alpha\psi^\beta. \ee 

The group $\Pin^+(3)$ is obtained by adjoining to $\Spin(3)=\SU(2)$ the reflection matrices $\gamma\cdot w$ for $w$ a unit vector in $\R^3$.
Such a reflection matrix is a $2\times 2$ unitary matrix of determinant $-1$.  So $\Pin^+(3)$ is the subgroup of $\U(2)$ consisting of matrices of determinant
$\pm 1$:
\be\label{pinthree}\Pin^+(3)=\{g\in \U(2)|\det g=\pm 1\}. \ee
In contrast to $\Spin(3)$, $\Pin^+(3)$ has two inequivalent 2-dimensional irreducible representations $R$ and $R'$.  If $\psi$ and $\psi'$ are
valued respectively in $R$ and $R'$, then $g\in \Pin^+(3)$ acts on them by
\be\label{inth}\psi \to g\psi,~~  \psi'\to g(\det g)\,\,\psi'. \ee
Thus the difference is precisely that a spatial reflection (or any orientation-reversing transformation) acts on $\psi$ and $\psi'$ with opposite signs.
The representations $R$ and $R'$ are both complex representations, meaning that they do not admit an invariant bilinear form.
Indeed, the mass terms $(\psi,\psi)=\veps_{\alpha\beta}\psi^\alpha\psi^\beta$ and $(\psi',\psi')=\veps_{\alpha\beta}\psi^{'\alpha}\psi^{'\beta}$
are both odd under reflection. Under $\psi\to g\psi$, $\psi'\to g\det g\,\,\psi'$, we have
\be\label{zomb}(\psi,\psi)\to \det g\,\,(\psi,\psi),~~(\psi',\psi')\to \det g\,\,(\psi',\psi'). \ee
But the ``off-diagonal'' mass term
\be\label{minth} (\psi,\psi')=\veps_{\alpha\beta}\psi^\alpha\psi^{'\beta} \ee
that links two fermions $\psi$ and $\psi'$ that transform oppositely under $\sR$ (or $\sT$) is invariant. 

Since there is always an invariant pairing between a representation of a compact group such as $\Pin^+(3)$ and its complex conjugate, the
invariance of the pairing $(\psi,\psi')$ means that the representation $R$ is isomorphic to the complex conjugate of $R'$.  
Indeed, one can show that the complex
conjugates of the matrices $g$ acting on representation $R$ can be conjugated by a unitary transformation
to the corresponding matrices $g\det g$ acting on representation $R'$:
\be\label{winth}\sigma_y \,g^*\sigma_y^{-1}=g\det g. \ee
To show this, one can consider separately the cases that $g=\gamma\cdot w$ is a reflection, and that $g\in \Spin(3)$, with $\det g=1$.
Eqn. (\ref{winth}) is equivalent to $\T g\T^{-1}=g\det g$, where $\T$ is the antilinear operation that we introduce momentarily.

On an {\it orientable} 3-manifold, the spinor representation of $\Spin(3)\cong \SU(2)$ admits the invariant antisymmetric tensor
$\veps_{\alpha\beta}$, as discussed above.  Hence one can define an antilinear operation
\be\label{tork} (\T\psi)^\alpha=\veps^{\alpha\beta}\psi^*_\beta,\ee
where $\psi^*_\beta$ is the complex conjugate of $\psi^\beta$.  It obeys $\T^2=-1$, and commutes with the Dirac operator $\D$.
So there is a version of Kramers doubling: the eigenvalues of $\D$ all have even multiplicity.

We have given a rather intrinsic definition of $\T$, but some readers may also want to see a description in local coordinates.
After picking a vielbein and ``flat space'' gamma matrices $\g_a=\sigma_a$, and writing $\K$ for complex conjugation, one can define
\be\label{zork} \T=\K \sigma_y,~~\T^2=-1.\ee
This operation can be defined using any vielbein, and because it is actually $\SU(2)$-invariant, despite not being written
in a way that makes this manifest, the definition of $\T$ does not depend on the choice of vielbein, as long as we consider
only vielbeins that determine the same orientation on spacetime (i.e., as long as we allow only orientation-preserving transformations
betwen vielbeins).  

A special case of the fact that the spectrum of the Dirac operator has even multiplicity on an orientable 3-manifold is that the mod 2
index of this operator vanishes.   However, $\T$ anticommutes with the gamma matrices and hence with the reflection elements
of $\Pin^+(3)$.  As a result, $\T$ cannot be defined as a transformation of spinor fields on an unorientable 3-manifold $W$.
Going up to the oriented double cover $\h W$, with $W=\h W/\Z_2$, $\T$ can be defined but anticommutes with the orientation-reversing 
generator $\tau$ of $\Z_2$.  So if $\h\psi$ is a spinor field on $\h W$ and $\tau\h\psi=\pm \h\psi,$ then $\tau(\T\h\psi)=\mp \T\h\psi$.
Recalling the definition of the complementary $\pin^+$ structures $\P$ and $\P'$, we can interpret this statement on $W$.
It means that although $\T$ cannot be defined to map sections of $\P$ to themselves or sections of $\P'$ to themselves, it makes sense as a map
from sections of $\P$ to sections of $\P'$ and vice-versa.  This means, in particular, that $\P$ and $\P'$ have the same mod 2 index (not
necessarily 0, as we know from section \ref{inter}).  In fact,
$\T$ transforms a zero-mode of the Dirac operator $\D:\P\to \P'$ to a zero-mode of the adjoint Dirac operator $\D:\P'\to\P$.

\subsection{Four Dimensions}\label{fwod}

In 4 dimensions, we need 4 gamma matrices.  We can take them to be $4\times 4$ matrices, but it is not possible for them to be all
real.  For example, we can take
\be\label{dorf}\g_1=\sigma_1\otimes 1,~~\g_2=\sigma_2\otimes 1,~~\g_3=\sigma_3\otimes \sigma_1, ~~\g_4=\sigma_3\otimes \sigma_3.\ee
With this choice, $\g_2$ is imaginary and the others are real. 

The group $\Spin(4)$ is isomorphic to the product of two copies of $\SU(2)$, which we write as $\SU(2)_\ell\times \SU(2)_r$.
The action of $\Spin(4)$ commutes with the chirality operator $\bg=\g_1\g_2\g_3\g_4$, and accordingly the 4-dimensional
spinor representation of $\Spin(4)$ splits as the direct sum of two 2-dimensional representations with $\bg=\pm 1$.
These are the representations $(1/2,0)$ and $ (0,1/2)$ of $\SU(2)_\ell\times \SU(2)_r$.  

Those representations are both pseudoreal, simply because the 2-dimensional representation of $\SU(2)$ is pseudoreal.
So spinors of $\Spin(4)$ are pseudoreal.  If we denote a field valued in the $(1/2,0)$ representation as $\psi^{\alpha'}$
and one valued in the $(0,1/2)$ representation as $\psi^{\alpha''}$, then there are invariant antisymmetric tensors
$\veps_{\alpha'\beta'}$ and $\veps_{\alpha''\beta''}$ for the $(1/2,0)$ and $(0,1/2)$ representations, respectively.  Each of
these is uniquely determined by $\SU(2)_\ell$ or $\SU(2)_r$ symmetry up to a constant multiple.

To pass from $\Spin(4)$ to $\Pin^+(4)$, we add reflection symmetries $x^i\to x^i-2w^i(w\cdot x)$.  A reflection acts as usual by
\be\label{zim}\psi\to \gamma\cdot w \,\psi.\ee
 Such a reflection exchanges $\SU(2)_\ell$ with $\SU(2)_r$,
anticommutes with $\bg$, and  exchanges $(1/2,0)$ with $(0,1/2)$.  So the representation $(1/2,0)\oplus (0,1/2)$ 
of $\Spin(4)$ is irreducible as a representation of $\Pin^+(4)$.  There is only one such spinor representation of $\Pin^+(4)$,
up to isomorphism, since a sign in eqn. (\ref{zim}) could be removed by conjugation with $\bg$.  

$\Spin(4)$ does not determine a relative normalization between $\veps_{\alpha''\beta''}$ and $\veps_{\alpha'\beta'}$.  However,
a reflection that exchanges $\SU(2)_\ell$ with $\SU(2)_r$ maps a particular choice of $\veps_{\alpha'\beta'}$ to a choice
of $\veps_{\alpha''\beta''}$.  With this choice, the sum $\veps_{\alpha'\beta'}\oplus \eps_{\alpha''\beta''}$ is an invariant
antisymmetric bilinear form $\veps_{\alpha\beta}$ on the 4-dimensional spinor representation of $\Pin^+(4)$.  (Indices
$\alpha,\beta$ can now be of either type $\alpha',\beta'$ or $\alpha'',\beta''$.)

This enables us to define an antilinear operation on spinors on a $\pin^+$ 4-manifold $W$:
\be\label{orf}(\T\psi)^\alpha =\veps^{\alpha\beta}\psi^*_\beta. \ee
Since $\T^2=-1$, the eigenvalues of the self-adjoint Dirac operator $\D'$ on such a manifold have even multiplicity.

The reader might prefer to see $\T$ defined explicitly using a local vielbein.  Using the representation (\ref{dorf}) for the
flat space gamma matrices, the definition is
\be\label{norf}\T=\K\gamma_2\bg. \ee
This commutes with $\Pin^+(4)$, so this definition does not really depend on the choice of the vielbein and makes sense globally.

\subsection{Five Dimensions}\label{fvod}

Naively, in this paper, we do not need to know about spinors and the Dirac operator in 5 dimensions.  But in Appendix \ref{feta},
we will want to know what the APS index theorem says about the 4-dimensional $\eta$-invariant, and for this
one needs to know some facts about 5 dimensions.

In 5 dimensions, one can find a $4\times 4$ representation of the Clifford algebra.  One uses the 4 gamma matrices in eqn. (\ref{dorf}).
along with $\g_5=\sigma_3\otimes \sigma_2$.  Note that of the 5 gamma matrices, 2 are imaginary ($\g_2$ and $\g_5$) and the
others are real.  

Thus the group $\Spin(5)$ has a 4-dimensional spinor representation.\footnote{This representation is pseudoreal in Lorentz signature,
so one needs to take two copies of it to make a sensible theory of fermions.   This is not really relevant for our limited purposes here.}
The group $\Spin(5)$ is actually isomorphic to $\Sp(4)$, and its 4-dimensional spinor representation is simply the fundamental
4-dimensional representation of $\Sp(4)$.  $\Sp(4)$ is defined as the subgroup of $\U(4)$ that preserves a certain 
nondegenerate antisymmetric bilinear form, so in particular the 4-dimensional representation of $\Sp(4)$ is pseudoreal
and spinors of $\Spin(5)$ are pseudoreal.

Concretely, on spinors of $\Spin(5)$, we can define an antilinear operation
\be\label{zanti}\T=\K\g_2\g_5,~~~ \T^2=-1.\ee
Thus on a 5-dimensional spin manifold 
(orientable for the time being), there is a form of Kramers doubling, and all eigenvalues of the Dirac operator
have even multiplicity.

So far all this parallels what happens in 3 dimensions.  Now let us go to the unorientable case.  Reflections are included in the usual
way.  The reflection $x^i\to x^i-2w^i(w\cdot x)$ acts by
\be\label{wanti}\psi\to \pm (\gamma\cdot w)\psi. \ee
As in 3 dimensions, the sign is meaningful and gives two distinct 4-dimensional representations $R$ and $R'$ of $\Pin^+(5)$.
However, there is an important difference.  In 3 dimensions, the representations $R$ and $R'$ are complex representations that
are complex conjugates of each other, but in 5 dimensions, they are each pseudoreal.  To see this, we just observe that the antilinear
operation $\T$ defined in eqn. (\ref{zanti}) commutes with a reflection acting as in eqn. (\ref{wanti}).  So $\T$ can be defined separately
in $R$ and in $R'$, and these representations are both pseudoreal. At an elementary level, the difference from 3 dimensions
is that in 5 dimensions, the number of gamma matrices that are imaginary is even.

As in any odd dimension, there is no self-adjoint Dirac operator acting on a field valued in a particular $\pin^+$ structure; 
rather $\pin^+$ structures come in complementary pairs
$\P$, $\P'$ that are exchanged by the Dirac operator.  We can define the {\it index} of the Dirac operator $\I$ as the number of
zero-modes of $\D:\P\to\P'$ minus the number of zero-modes of the adjoint operator $\D:\P'\to\P$.  This index is 0 on a five-manifold
without boundary, but on a manifold with boundary (and with APS boundary conditions), it can be nonzero, as we will learn in Appendix
\ref{feta}.  However, $\I$ is always even because the antilinear symmetry $\T$ ensures that the number of zero-modes of $\D$
acting on either $\P$ or $\P'$ is even.

\section{The $\eta$-Invariant In Four Dimensions}\label{feta}

In section \ref{bands} and \ref{zeb}, we needed some facts about the $\eta$-invariant in $D=4$.  Analogous facts in $D=2$
were invoked in section \ref{maj}.  Our goal in this appendix is to briefly explain these facts.  (The arguments we give are similar
to the original ones \cite{G}.)

Our first goal is to show that in $D=4$, $\eta$ is always  an integer multiple of $1/4$, so that $\exp(-\i\pi \eta/2)$ is always
a $16^{th}$ root of 1.
The nontrivial case is that $X$ is unorientable; otherwise $\eta$ is a multiple of 2, as remarked at the end of section \ref{bpf}.
 Suppose that
$X$  has  $\pin^+$ structure
$\P$ and  complementary $\pin^+$  structure $\P'=\P\otimes \veps$, where $\veps $ is  a real line
bundle, the orientation bundle of $X$.  We can define $\eta$-invariants $\eta_\P$, $\eta_{\P'}$ for a Majorana
fermion valued in either of these $\pin^+$ structures.

A first useful relation is that
\be\label{simpler}\eta_\P+\eta_{\P'}=0 ~~~{\mathrm {mod}}~4. \ee
The point is that a Majorana fermion on $X$ coupled to $\P$ together with a Majorana fermion on $X$ coupled to $\P'$  
combine together to a Majorana fermion coupled to a spin structure on $\h X$, the oriented double cover of $X$.
So $\eta_\P+\eta_{\P'}=\h\eta$, where $\h\eta$ is the $\eta$-invariant of the Majorana fermion on $\h X$.  Because
$\h X$ is orientable and also has an orientation-reversing symmetry (the quotient by this symmetry being $X$),  $\h\eta$ is
a multiple of 4, as remarked at the end of section \ref{bpf}.

We need one more relation, which is that
\be\label{impler}8\eta_{\P'}=8\eta_\P~~{\mathrm{mod}}~4. \ee
Together, these relations imply that $\eta_\P$ and $\eta_{\P'}$ are both multiples of $1/4$.

To explain eqn. (\ref{impler}), is helpful to generalize the problem a little.  Let $V\to X$ be any real vector bundle.  We consider the Dirac operator $\D_V$
acting on fermions valued in $V$, that is, acting on sections of $\P\otimes V$.  For any $V$, we can define $\eta_{\P\otimes V}$,
the $\eta$-invariant of the Dirac operator acting on such a field.  Here $\eta_{\P\otimes V}$, for any $V$, is a topological
invariant (independent of the metric of $X$ and the connection on $V$) mod 4.   This follows from the APS index theorem,
which in odd dimensions takes the simple form (\ref{molg}).  To apply the theorem without assuming that $X$ is a boundary,
we consider the 5-manifold $Z=X\times [0,1]$ with one metric and gauge field $(g,A)$ at one end and some other $(g',A')$
at the other end.  The theorem then says that the difference of $-\eta/2$ between the two ends  is equal to $\I$, the index
of the Dirac operator on $Z$ with APS boundary conditions.  This index is always even, as explained in section \ref{fvod},
so the difference between the two values of $\eta$ is a multiple of 4.  More directly, what is happening is that as we vary $(g,A)$
on a manifold of even dimensions, $\eta$ is constant except that it jumps by 2 when an eigenvalue passes through 0.
In 4 dimensions, the eigenvalues have even multiplicity because of a version of Kramers doubling, so the jumps are multiplies of 4
and $\eta$ is a topological invariant mod 4.

We can interpret $8\eta_{\P'}$ as $\eta_{\P\otimes V}$ where $V=\veps^{\oplus 8}$ is the direct sum of 8 copies of the real
line bundle $\veps$.  Similarly $8\eta_\P=\eta_{\P\otimes \R^8}$, where $\R^8$ is a trivial real vector bundle of rank 8.
So to establish eqn. (\ref{impler}), it suffices to prove that $\veps^{\oplus 8}$ is trivial, that is isomorphic to $\R^8$. 

To show that a real vector bundle $V$ over a 4-manifold $X$ is trivial, it suffices to show vanishing of Stieffel-Whitney classes
$w_i(V)$, $1\leq i\leq 4$, and also of a certain class $\lambda\in H^4(X,\Z)$ that is subtle to define, but that obeys
$2\lambda=p_1(V)$, where $p_1$ is the first Pontryagin class.   

For the first point, the total Stieffel-Whitney class of $V$ is
$w(V)=w(\veps^{\oplus 8})=w(\veps)^8=(1+w_1(\veps))^8=1$, where in the last step we use the fact that the relevant
binomial coefficients vanish mod 2. So $w_i(V)=0$ for $i\geq 1$.

To show vanishing of $\lambda(V)$, we first write $V=V_0\oplus V_0$, with $V_0=\veps^{\oplus 4}$.  So $\lambda(V)
=2\lambda(V_0)=p_1(V_0)$.  So we are reduced to showing that $p_1(V_0)=0$.  To compute $p_1(V_0)$, we observe
that the complexification of $V_0$ is the direct sum of 4 copies of $\veps_\C=\veps\otimes_\R\C$. In general, if $V_0\to X$
is a real vector bundle whose complexification is $\oplus_i \L_i$ for some complex line bundles $\L_i\to X$, then
$p_1(V_0)=\sum_i c_1(\L_i)^2$.  In the present case, this formula gives $p_1(V_0)=4c_1(\veps_\C)^2$, and this
vanishes because $c_1(\veps_\C)$ is 2-torsion.  The last statement is a general one that holds for $\veps_\C=\veps\otimes_\R\C$
for any real line bundle $\veps$ over a topological space.  

This completes the proof that $\exp(-\i\pi \eta/2)$ is a $16^{th}$ root of 1.  

We also want to show that for $X=\RP^4$, $\exp(-\pi i\eta/2)=\exp(\pm 2\i\pi /16)$, with the sign depending on the choice of 
$\Pin^+$ structure.  We start on the five-torus $T^5$ parametrized by variables $x_1,\dots,x_5$ of period 1.  We give $T^5$
a spin structure in which spinors are periodic in all directions.  The spin bundle $\S$ of $T^5$ has rank 4,
since the spinor representation of $\SO(5)$ is 4-dimensional.

The Dirac operator $\D=\i \slashed{\partial}$ on $T^5$ is self-adjoint, so its index certainly vanishes.  Now we replace
$T^5$ by $T^5/\Z_2$, where $\Z_2$ acts by $\sP:x_i\to -x_i$, $i=1,\dots,5$.  Here $T^5/\Z_2$ is not a manifold, and we will
correct for that in a moment.   We define $\H_+$ and $\H_-$ to be the spaces of spinor fields on $T^5$ that obey,
respectively, $\psi(-x)=\psi(x) $ and $\psi(-x)=-\psi(-x)$.  Because $\D$ anticommutes with $\sP$, it maps $\H_+$ to 
$\H_-$, and vice-versa.  We define the {\it index} $\I$ of $\D$ on $T^5/\Z_2$ to be the number of zero-modes of $\D$
acting on $\H_+$ minus the number of zero-modes of $\D$ acting on $\H_-$.  (These operators are adjoints of each other
since the full Dirac operator acting on $\H_+\oplus \H_-$ is self-adjoint.)

It is straightforward to compute $\I$.  The zero-modes of $\D$ acting on $\H_+$ are the 4-dimensional space of constant
spinors, and there are no zero-modes of $\D$ acting on $\H_-$.  So the index is $\I=4$.

Now, $T^5/\Z_2$ is not a  manifold because $\sP$ has 32 fixed points: the points with coordinates $x_i\in \Z/2$, $i=1,\dots,5$.
To get a manifold, we remove from $T^5$ a small ball around each fixed point to get a manifold $T'$ with boundary.  The quotient of $T'$
by $\sP$ is then
 a manifold $X$ with a boundary that  consists of 
32 copies of $\RP^4$. $X$ acquires two $\pin^+$ structures $\P$, $\P'$ whose sections are spinor fields on $T'$ that are even or
odd under $\sP$.
The Dirac operator
$\D$ maps sections of $\P$ to sections of  $\P'$, and vice-versa. As such it
has an index $\I$. A fairly elementary argument\footnote{ Near one of the fixed points of the action of $\sP$
on $T^5$, we can replace $T^5/\Z_2$ by $\R^5/\Z_2$ with Euclidean metric $\d \ell^2=\d \vec x^2$.  A Weyl rescaling converts this to $\d\h\ell^2=\d \vec x^2\cdot \left(1+{1}/{|\vec x|^2}\right)$.   Near $\vec x=0$, this has the tubelike behavior that we described in introducing APS boundary conditions
(see fig. \ref{longer} of section \ref{aps}).  In this case, the tube is a copy of 
$\RP^4\times \R_+$.  Because of the conformal
invariance of the Dirac equation, a $\Z_2$-invariant solution of the Dirac equation on 
$\R^5$ that is regular at $\vec x=0$
is equivalent to a solution that is square-integrable in the tubelike geometry.  So we can equally well
compute the index $\I$ by counting regular solutions  in the original $T^5$ 
geometry or by counting square-integrable solutions after making a Weyl transformation to 
replace the 32 fixed points
by  tubes.  But the definition of APS boundary conditions  ensures 
that the latter procedure is equivalent to the
index computed on $X $ with APS boundary conditions.}
 using the conformal invariance of the Dirac operator shows that the
Dirac index on $X$, with APS boundary conditions, has the same value $\I=4$ 
found in the last paragraph on $T^5/\Z_2$.
In odd dimensions, as in eqn. (\ref{molg}) the APS index theorem just says that $\eta/2=-\I$, 
where $\I$ is the Dirac index on a manifold $T'$
and $\eta$ is the $\eta$-invariant of its boundary. In the present case, the boundary 
consists of 32 copies of $\RP^4$
(all with the same $\pin^+$  structure, since they are permuted by obvious symmetries).  
So the formula becomes
$16\eta_{\RP^4}=-\I=-4$, or $\eta_{\RP^4}=-1/4$. 

We leave it to the reader to show in a similar way that in 2 dimensions, $\exp(-\i\pi \eta/2)$ 
is an $8^{th}$ root of 1,
and $\eta_{\RP^2}=\pm 1/2$.

Research supported in part by NSF Grant PHY-1314311.  I thank M. F. Atiyah,
G. Burdman, D. Freed, A. Kitaev, S. Ryu, and N. Seiberg for comments and discussions.

\bibliographystyle{unsrt}

\end{document}